%% file: quarkonium-fac.tex
\documentclass[preprint,onecolumn,aps,prd,groupedaddress,nofootinbib,showpacs]{revtex4}
\usepackage{graphicx}
\DeclareGraphicsExtensions{.eps}
\usepackage{amsmath}
\usepackage{bm}
\usepackage{slashed}
\usepackage{epsfig}
\usepackage[hypertex]{hyperref}
\newcommand{\ben}{\begin{eqnarray}}
\newcommand{\een}{\end{eqnarray}}
\newcommand{\nnu}{\nonumber\\}
\newcommand{\bef}{\begin{figure}[!htp]}
\newcommand{\eef}{\end{figure}}

\newcommand{\bea}{\begin{eqnarray}}
\newcommand{\eea}{\end{eqnarray}}

\begin{document}
\title{Heavy Quarkonium Production at Collider Energies:\\ 
Factorization and Evolution}

\author{Zhong-Bo Kang$^1$, Yan-Qing Ma$^2$,
Jian-Wei Qiu$^{2,3}$, and 
George Sterman$^{3}$}

\email{zkang@lanl.gov, yqma@bnl.gov, jqiu@bnl.gov, sterman@insti.physics.sunysb.edu}

\affiliation{Theoretical Division, 
                  Los Alamos National Laboratory, 
                  Los Alamos, NM 87545, USA}
                
\affiliation{$^2$Physics Department,
                Brookhaven National Laboratory,
                Upton, NY 11973-5000, USA}

\affiliation{$^3$C.N.\ Yang Institute for Theoretical Physics,
             Stony Brook University, 
             Stony Brook, NY 11794-3840, USA}

\date{\today}

\begin{abstract}
We present a perturbative QCD factorization formalism for inclusive production of heavy quarkonia of large transverse momentum, $p_T$ at collider energies, including both leading power (LP) and next-to-leading power (NLP) behavior in $p_T$.    We demonstrate that both LP and NLP contributions can be factorized in terms of perturbatively calculable short-distance partonic coefficient functions and universal non-perturbative fragmentation functions, and derive the evolution equations that are implied by the factorization.    We identify projection operators for all channels of the factorized LP and NLP infrared safe short-distance partonic hard parts, and corresponding operator definitions of fragmentation functions.  For the NLP, we focus on the contributions involving the production of a heavy quark pair, a necessary condition for producing a heavy quarkonium.  We evaluate the first non-trivial order of evolution kernels for all relevant fragmentation functions, and discuss the role of NLP contributions.  
\end{abstract}
\pacs{12.38.Bx, 13.88.+e, 12.39.-x, 12.39.St}

\maketitle


\input ./sec1.tex

\newpage
\input ./sec2.tex

\newpage
\input ./sec3.tex
\newpage
\input ./sec4ab.tex
\newpage
\input ./sec4c.tex
\newpage
\input ./sec5.tex

\section*{Acknowledgments}

We thank Geoff Bodwin, Eric Braaten, Sean Fleming, Adam Leibovich and Tom Mehen for helpful discussions.  This work was supported in part by the U. S. Department of Energy under contracts No.~DE-AC52-06NA25396 and DE-AC02-98CH10886, 
and the National Science Foundation under grants No.~PHY-0969739 and -1316617.

\appendix
\input ./appendixa.tex
\input ./appendixb.tex

\input ./appendixc.tex

\bibliographystyle{h-physrev5} 
\bibliography{./references/references}{}

\end{document}

%% file: sec1.tex
\section{Introduction}

Almost forty years since the discovery of the J/$\psi$ \cite{Aubert:1974js,Augustin:1974xw}, the production of heavy quarkonia remains one of the most active and fascinating subjects in strong interaction physics \cite{Brambilla:2010cs,Bodwin:2013nua}.  The inclusive production of a pair of charm or bottom quarks is an essentially perturbative process because the heavy quark mass $m_Q$ is much larger than $\Lambda_{\rm QCD}$, while the subsequent evolution of the pair into a quarkonium is nonperturbative.  Different treatments of the nonperturbative transformation from a heavy quark pair to a bound quarkonium have led to various theoretical models for quarkonium production, most notably, the color singlet model (CSM), the color evaporation model (CEM), and the non-relativistic QCD (NRQCD) model \cite{Brambilla:2004wf,Brambilla:2010cs}.   Among these models, the NRQCD treatment of heavy quarkonium production, proposed in Ref.~\cite{Bodwin:1994jh}, is both the most theoretically sound and phenomenologically successful \cite{Braaten:1996pv,Petrelli:1997ge,Kramer:2001hh,Brambilla:2010cs}.  

With full next-to-leading order (NLO) contributions in powers of $\alpha_s$ and properly fitted NRQCD long-distance matrix elements, theory predictions for inclusive J/$\psi$ and $\Upsilon$ production are generally consistent with experimental data from the Tevatron and the LHC \cite{Ma:2010yw,Butenschoen:2010rq,Wang:2012is,Gong:2013qka}.   However, global fits of data on J/$\psi$ production from various high energy collisions, including $e^+e^-$, lepton-hadron, and hadron-hadron collisions \cite{Butenschoen:2010rq,Butenschoen:2011yh} show slight discrepancies of shape compared to data \cite{Bodwin:2013nua}.    In addition, existing theoretical calculations \cite{Brambilla:2010cs,Butenschoen:2012px,Chao:2012iv,Gong:2012ug,Gong:2013qka} have not been able to explain fully the polarization of high-$p_T$ heavy quarkonia at the Tevatron \cite{Affolder:2000nn,Abulencia:2007us,Acosta:2001gv,Abazov:2008aa} and LHC \cite{Chatrchyan:2012woa}.  Motivated in part by these challenges to existing theory, new approaches based on perturbative QCD factorization \cite{Nayak:2005rt,Nayak:2006fm,Kang:2011zza,Kang:2011mg} and soft-collinear effective theory \cite{Fleming:2012wy,Fleming:2013qu} have been proposed for the systematic study of heavy quarkonium production at collider energies.

In this paper, we follow up our earlier work \cite{Kang:2011zza,Kang:2011mg} and develop an extended QCD factorization formalism beyond the leading power for heavy quarkonium production at large transverse momentum $p_T\gg m_H\gg \Lambda_{\rm QCD}$ in hadronic collisions (or at a large energy $E\gg m_H$ in $e^+e^-$ collisions).  In this approach, we first expand the cross section in a power series of $1/p_T^2$, and argue that the first two terms of the expansion can be factorized systematically  into infrared safe short-distance partonic functions in convolution with  universal long-distance matrix elements.  The relevant matrix elements for quarkonium production then take the form of generalized fragmentation functions, for which we derive a set of evolution equations that mix single-parton and heavy quark-pair states.   The importance of  the evolution of quark pair states was suggested originally by calculations in the color singlet model.

Although it was proposed shortly after the discovery of J/$\psi$, the CSM is still a valuable tool for heavy quarkonium production, since it has practically no free parameters once the heavy quarkonium wave function at the origin is fixed by data on heavy quarkonium decay \cite{Einhorn:1975ua,Ellis:1976fj,Carlson:1976cd,Chang:1979nn,Berger:1980ni,Baier:1981uk,Baier:1983va}.  In addition, the CSM is actually a special case of NRQCD \cite{Brambilla:2010cs}.  Nevertheless, at leading order (LO) in powers of strong coupling constant, $\alpha_s$, the CSM cross section falls off as $1/p_T^8$ and  is more than two orders of magnitude smaller than the Tevatron data on J/$\psi$ production.   A complete CSM calculation at NLO and an estimated contribution at next-to-next-to-leading order (NNLO) to hadronic heavy quarkonium production became available in \cite{Campbell:2007ws,Artoisenet:2007xi,Artoisenet:2008fc}.   It was found, surprisingly, that at large $p_T$, the NLO contribution is more than a factor of 10 larger than the CSM LO result,  for $p_T$ values characteristic of Tevatron data.  Although still far below collider data, it behaves as $1/p_T^6$ rather than $1/p_T^8$.    As we shall see in the next section, this correction is due to heavy quark pairs produced at short distances, which convert radiatively to color singlet configurations.    Estimates of the NNLO contributions suggest further significant enhancements in the CSM cross section over the NLO result.     

In contrast to the NLO enhancement found in its CSM sector, in full NRQCD factorization,  supplemented by leading-power evolution, quarkonium production  at high $p_T$ is dominated by single gluon production at short distances, $\sim {\cal O}(1/p_T)$, beginning at LO.   The gluon then fragments into a heavy quark pair only at a much later time $\sim  {\cal O}(1/(2m_Q))$ in the pair rest frame.  The bound state quarkonium forms over even larger time scales, of order $1/(m_Qv)$, with $v$ a typical relative velocity in the pair rest frame.   This subprocess has the leading power partonic production rate of $1/p_T^4$.   It is largely responsible for the NRQCD prediction that heavy quarkonia produced at large $p_T$ are dominated by transverse polarization \cite{Brambilla:2010cs}, which has not, however, been supported by all existing data \cite{Bodwin:2013nua}.   

Part of the motivation for this study is to explore the possibility that with the large phase space available for producing a heavy quark pair with high $p_T$ at collider energies, the production of the heavy quark pair at the ``last minute", that is, at ${\cal O}(1/(2m_Q))$, may not be the whole story.   Heavy quark pairs could be produced directly at the hard collision of the distance scale of ${\cal O}(1/p_T)$, indeed, at any time between ${\cal O}(1/p_T)$ and ${\cal O}(1/(2m_Q))$.  As illustrated by the CSM at NLO, prompt heavy quark pairs contribute to the cross section at next-to-leading power (NLP), $1/p_T^6$, but generally, pairs produced from gluon evolution at intermediate scales $2m_Q<\mu<p_T$ contribute (through evolution) at an intermediate level, typically $m_Q^2/(p_T^4\mu^2)$.  The phenomenology of this evolution requires an analysis beyond leading power in $p_T$.

Heavy quark pair production at short distances gives the relevant NLP term in the $1/p_T^2$ expansion for the production of heavy quarkonia.  Like power corrections to other observables \cite{Qiu:1990xy,Qiu:1998ia}, the factorized form of this term can be proportional to either twist-4 fragmentation functions to a heavy quarkonium or twist-4 parton correlation functions of the colliding hadrons.  We assume that the fragmentation of a heavy quark pair of the correct quark flavor should be much more likely to produce a heavy quarkonium than the fragmentation of other multi-parton states, and in this paper we focus only on those power corrections involving the production of a heavy quark pair, and their fragmentation into a physical quarkonium.  A consistent treatment of NLP factorization within this framework requires us to derive evolution equations for the factorization scale dependence of these new fragmentation functions.  

With the new factorization formalism, including evolution, we effectively organize the production process into three stages based on the dynamics at three different energy scales: $p_T$, $m_Q$ and $\Lambda_{\rm QCD}$.    Specifically, the three stages are: (1) production of a single parton (the first term in the $1/p_T^2$ expansion) or a heavy quark pair (the second term in the $1/p^2_T$ expansion) at the distance scale $1/ p_T$ (or $1/E$), (2) evolution between $1/p_T$ and $1/2m_Q$, which includes the transformation of single partons to heavy quark pairs as well as the resummation of powers of $\ln(p_T^2/m_Q^2)$ for both single partons and quark pairs, and (3) formation of the quarkonium between times $1/(2m_Q)$ and $1/(m_Qv)$.  

Prompt pair production in the stage (1) can be calculated systematically  in perturbative QCD (pQCD) order-by-order in $\alpha_s$ and included in the short-distance functions of the factorization formalism described in Sec.\ \ref{sec:fac} below. The evolution and resummation of logarithms in stage (2) is carried out by solving a closed set of evolution equations for the fragmentation functions of single partons and of  heavy quark pairs to produce a heavy quarkonium.   The operator definitions of these fragmentation functions are derived in Sec.~\ref{sec:frag}.  Their evolution equations and corresponding evolution kernels at the first non-trivial order in $\alpha_s$ are derived in Sec.~\ref{sec:evolution}.   The stage of hadronization, (3), which is essentially nonperturbative, may be treated via NRQCD.  In this sense, the formalism that we develop is fully consistent with NRQCD, although it does not directly address the question of NRQCD factorization \cite{Brambilla:2010cs}.

The predictive power of this new factorization formalism relies on the perturbative calculations of the short-distance functions and the evolution kernels, and our knowledge of the universal fragmentation functions at an input scale $\mu_0$.  With the operator definitions of all fragmentation functions, given in Sec.~\ref{sec:frag}, the factorization formalism provides a unique prescription to calculate all short-distance functions and evolution kernels order-by-order in powers of $\alpha_s$ up to a freedom to choose the factorization scheme.   In a companion paper \cite{KMQS-hq2}, we present calculations of short-distance functions for all partonic production channels at LO in powers of $\alpha_s$.  With the evolution kernels calculated in this paper, we still need the fragmentation functions at the input scale $\mu_0$ in order to make  numerical predictions and comparison with data.  

The input fragmentation functions are non-perturbative, and  in principle,  can be extracted from fitting experimental data.   As explained in our companion paper \cite{KMQS-hq2}, however, we should be able to provide a good estimate of the input fragmentation functions by using NRQCD factorization, since input fragmentation functions to heavy quarkonia can have a large perturbative scale, $\mu_0 \gtrsim 2m_Q$, which is well separated from the soft scales responsible for the binding.  Although there is as yet no formal proof for the NRQCD factorization, the clear separation of momentum scales for the input fragmentation functions provides a good justification for using the formalism as a reasonable conjecture.\footnote{In Refs.\ \cite{Ma:2013yla,Ma:2014eja}, model fragmentation functions have been calculated following this approach for the fragmentation of quark pairs into $S$- and $P$-wave quarkonia.}  Our conclusions and summary are given in Sec.~V.

%% file: sec2.tex
\section{The perturbative QCD factorization formalism}
\label{sec:fac}

We briefly summarize the fundamentals of the NRQCD factorization applied to heavy quarkonium production at collider energies, and review how higher order corrections to its perturbative short-distance functions in singlet channels produce power enhancements relative to the behavior of the singlet channel at LO  \cite{Campbell:2007ws,Artoisenet:2007xi,Artoisenet:2008fc}.   We then argue that cross section for producing a heavy quarkonium at large transverse momentum $p_T\gg m_H$ at collider energies can be expanded as a power series of $m_H^2/p_T^2$, and that the leading power term and the first subleading power terms can be perturbatively factorized into infrared safe short-distance functions (``hard parts") in convolution with nonperturbative but universal long-distance fragmentation functions.  The short-distance hard parts can be systematically calculated as a power series in $\alpha_s(p_T)$.

\subsection{The NRQCD factorization of heavy quarkonium production at high $p_T$}
\label{subsec:csm}

The NRQCD factorization approach to heavy quarkonium production \cite{Bodwin:1994jh} expresses the inclusive cross section for the direct production of a quarkonium state $H$  as a sum of ``short-distance'' coefficients times NRQCD long-distance matrix elements (LDMEs), 
\begin{equation}
\sigma^H(p_T,m_Q) = \sum_{[Q\bar{Q}(n)]} \hat{\sigma}_{[Q\bar{Q}(n)]}(p_T,m_Q,\Lambda)
 \langle 0| {\cal O}_{[Q\bar{Q}(n)]}^H(\Lambda)|0\rangle \, .
\label{eq:nrqcd-fac}
\end{equation}
Here $\Lambda\sim {\cal O}(m_Q)$ is the ultraviolet cut-off of the NRQCD effective theory.  The short-distance coefficients $\hat{\sigma}_{[Q\bar{Q}(n)]}$ are perturbatively calculated in powers of $\alpha_s$, and are essentially the process-dependent perturbative QCD cross sections to produce a $Q\bar{Q}$ pair in various color, spin, and orbital angular momentum states ${[Q\bar{Q}(n)]}$ (including the parton distributions of incoming hadrons).   The LDMEs are nonperturbative, but, universal, representing the probability for a $Q\bar{Q}$ pair in a particular state, ${[Q\bar{Q}(n)]}$ to evolve into a heavy quarkonium.  The sum over the ${[Q\bar{Q}(n)]}$ states is organized in terms of powers of the pair's relative velocity $v$, an intrinsic scale of the LDMEs. For J/$\psi$ production, for example, current production phenomenology mainly uses four NRQCD LDMEs, corresponding to the $c\bar{c}$-pair produced in ${^3}S_1^{[1]}$, ${^1}S_0^{[8]}$, ${^3}S_1^{[8]}$, and ${^3}P_J^{[8]}$ states, respectively, where the superscript $[1]$ (or $[8]$) refers to a color singlet (or octet) heavy quark pair.  The color singlet model and color evaporation model can be thought as a truncation of and a special approximation to the NRQCD approach, respectively \cite{Brambilla:2010cs,Bodwin:2005hm}. 

In the production of the heavy quark pair that evolves into a heavy quarkonium, the heavy quark mass, $m_Q\gg \Lambda_{\rm QCD}$, regulates the perturbative final-state collinear logarithmic behavior.  The NRQCD factorization formalism is an effective field theory approach to separate the long-distance soft physics at the scale $m_Q v$ and below from the short-distance hard physics at the scale of $m_Q$ and larger.   However, when $p_T\gg m_H$, the perturbative functions in Eq.~(\ref{eq:nrqcd-fac}) will have calculable powers of $\ln(p_T^2/m_Q^2)$, which should be resummed systematically.  Furthermore, for the production of certain spin-color $[Q{\bar Q}(n)]$ states, new partonic production channels only open up beyond LO in $\alpha_s$.  As we shall see below, some of these channels can be enhanced by powers of $p_T/m_Q$ compared to their leading order estimates.  For simplicity, we discuss the CSM, as a special case of NRQCD \cite{Brambilla:2010cs} and as an example to illustrate power enhancements at higher orders.   These considerations will motivate an expansion of the cross section for the production of a heavy quark pair in powers of $m_Q/p_T$ first, before expanding coefficient functions in powers of $\alpha_s$.

\bef
\includegraphics[width=0.25\textwidth]{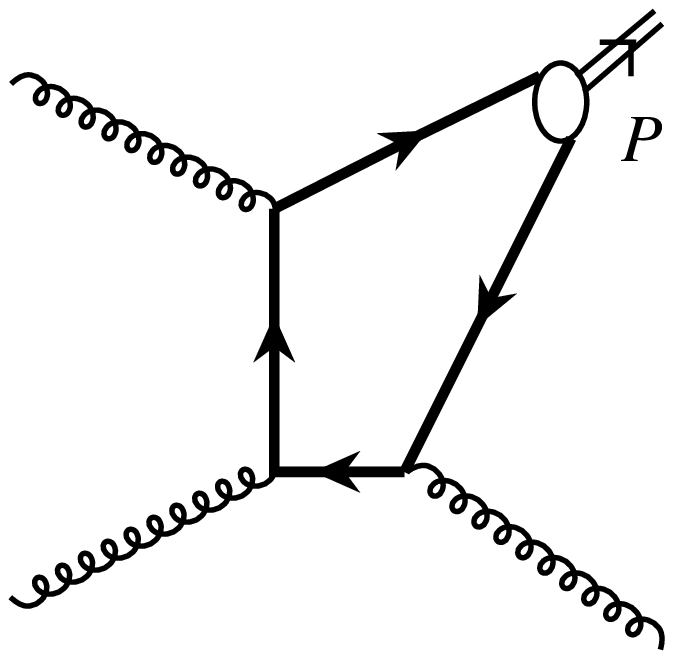}
\caption{Sample lowst-order diagram for
  heavy quarkonium production from the gluon-gluon fusion channel in the CSM.} 
\label{fig:lo_gg}
\eef

In the CSM, quarkonia are formed only from color-singlet, low invariant mass heavy quark pairs, produced perturbatively with the same quantum numbers as the bound states in question.  At LO in $\alpha_s$, the perturbative partonic cross section in a hadronic collision is given by the $2\to 3$ partonic subprocess, $gg\to Q\bar{Q}(P)g$, that produces a pair of color singlet heavy quarks at high transverse momentum $p_T$, as shown in Fig.~\ref{fig:lo_gg}, where heavy quarkonium momentum is defined in the light-cone coordinate as
\ben
P^\mu & \equiv &
\left(\frac{m_T}{\sqrt{2}}\, e^{y},\, \frac{m_T}{\sqrt{2}}\, e^{-y},\, {\bf{p}}_T\right)
\label{eq:momentumP}
\een
with rapidity $y$ and $m_T=\sqrt{m_H^2+p_T^2}$, and $p_T=\sqrt{{\bf p}_T^2}$ in the lab frame.  
For the discussion in this paper, it is more convenient to work in a frame in which the heavy quarkonium has no transverse component as $P^\mu = (P^+, P^-,{\bf 0}_T)$ with
$P^+ = \Big [m_T\,\cosh y+\sqrt{p_T^2+m_T^2\,\sinh^2 y}\; \Big]/\sqrt{2}$ and 
$P^- = \Big[ m_T\,\cosh y-\sqrt{p_T^2+m_T^2\,\sinh^2 y}\; \Big]/\sqrt{2}$ expressed in terms of 
the rapidity and transverse momentum in the lab frame.
In order to produce a color singlet, spin-1 non-relativistic $Q\bar{Q}$ pair at this order, the spinor trace of the heavy quark pair is contracted by the projection operator \cite{Bodwin:1994jh},
\begin{equation}
{\cal P}(^3S_1) \propto 
{\cal C}^{[1]}_{ij}\,
\gamma\cdot \epsilon^{\mu}(P)\,
(\gamma\cdot P/2 +m_Q)\, ,
\label{eq:proj_s1}
\end{equation}
where $\epsilon^\mu(P)$ is the polarization vector for the spin-1 heavy quark pair, 
and ${\cal C}^{[1]}_{ij}=1/N_c \delta_{ij}$ with the superscript ``$[1]$'' indicating a color singlet,  
$N_c=3$ for the SU($N_C$) color of QCD, 
and $i,j=1,2,N_c$ the color indices of the heavy quark and antiquark.
Since the final-state gluon has to balance the transverse momentum of the produced heavy quark pair, both quark propagators of the Feynman diagram in Fig.~\ref{fig:lo_gg} have to be off-shell by the order of $p_T$. With the projection operator in Eq.~(\ref{eq:proj_s1}), the fermion trace does not give an invariant that grows with $p_T$, and the LO cross section in the CSM behaves as $1/p_T^8$, falling  much faster than the generic $1/p_T^4$ behavior of leading power $2\to 2$ partonic cross sections.  Phenomenologically, the LO contribution in the CSM has the wrong $p_T$ shape for the J/$\psi$ transverse momentum distribution at collider energies, and a normalization which can be more than two orders of magnitudes smaller than the high-$p_T$ Tevatron and LHC data \cite{Brambilla:2004wf,Ma:2010yw,Butenschoen:2010rq,Gong:2013qka}. 

\bef
\includegraphics[width=0.25\textwidth]{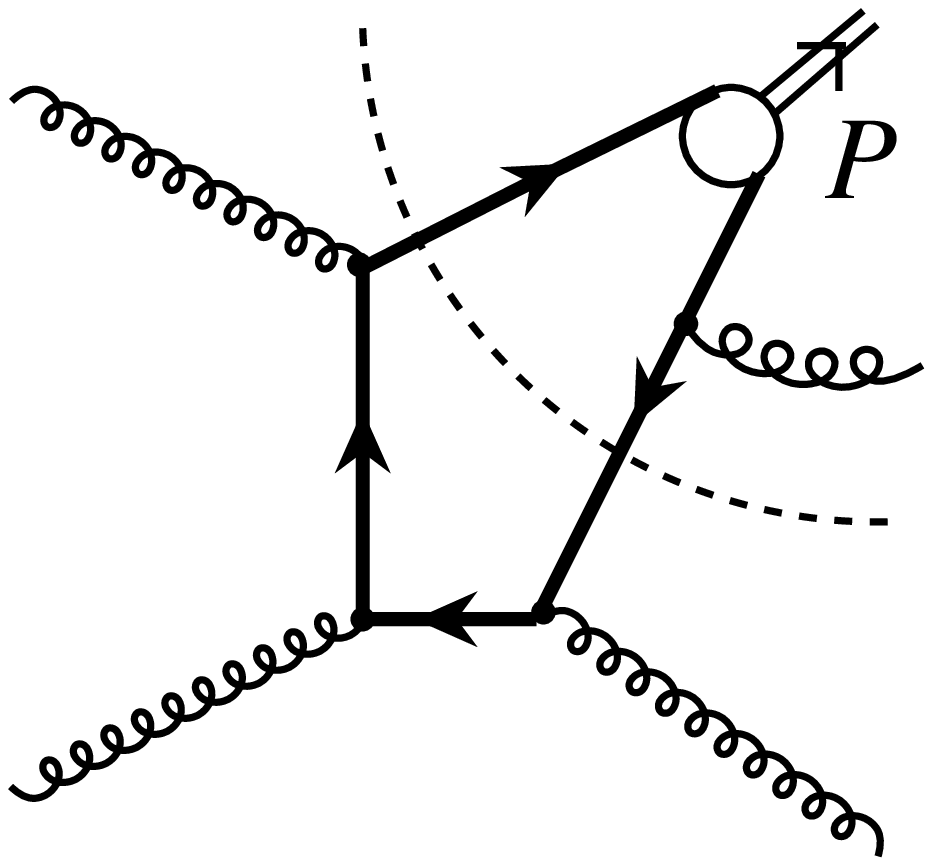}
\hskip 0.5in
\includegraphics[width=0.25\textwidth]{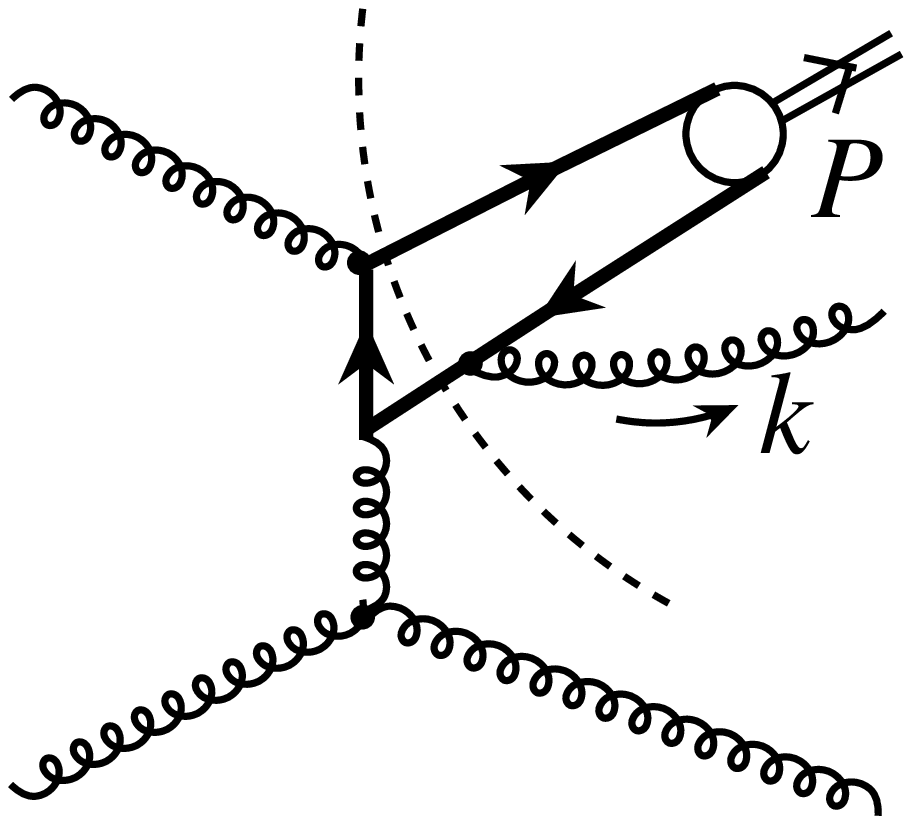}
\hskip 0.5in
\includegraphics[width=0.25\textwidth]{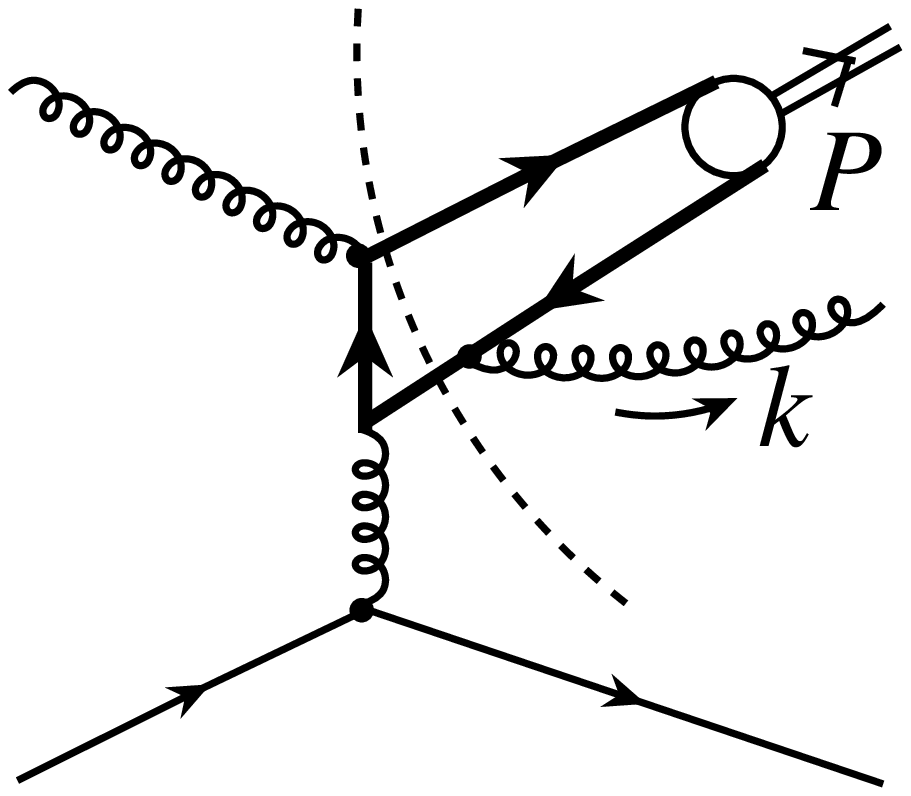}
\caption{Sample NLO Feynman diagrams for heavy quarkonium production at high $p_T$ in hadronic collisions.}
\label{fig:nlo}
\eef

At NLO, real-gluon radiative contributions to the cross section in CSM come from $2\to 4$ Feynman diagrams, as shown in Fig.~\ref{fig:nlo}, where in addition to the heavy quark pair, there are two light partons (or another pair of heavy quarks in the case of associated production) in the final state, while virtual contributions come from the interference between the LO diagram in Fig.~\ref{fig:lo_gg} and its one-loop corrections.  The additional light parton in the final-state in Fig.~\ref{fig:nlo} allows the production of a color octet heavy quark pair at distance scale  $1/p_T$, and opens up a large phase space for the pair to neutralize its color to become a spin-1, color singlet at  much later times, up to the order of $1/m_Q$.  In addition, the heavy quark pair produced at the distance scale of $1/p_T\ll 1/m_Q$ is relativistic and can be in various relativistic spin states before the pair converts itself into the non-relativistic spin-1, color singlet state by radiating additional gluon(s) at a later time.  

The contribution from various relativistic spin states of the heavy quark pair can be separated by a Fierz transformation to decompose the quark spinor trace, as indicated in Fig.~\ref{fig:nlo} by a dashed line.  Like the LO case, the two off-shell propagators needed to produce a heavy quark pair at large $p_T$ give a factor $1/p_T^8$.  However, at this order the heavy quark spinor trace in the numerator can now produce a $p_T^2$ enhancement.   Such factors are isolated by a vector, $\gamma\cdot P$ or an axial vector, $\gamma_5\gamma\cdot P$ spin projection as shown in the figure.   In this way, the NLO contribution can gain a $p_T^2/m_Q^2$ enhancement compared to the LO contribution, and become much larger than the LO term at high $p_T$.  It is this power enhancement that is mainly responsible for the factor of ten enhancement discovered by explicit calculations at NLO in the CSM \cite{Campbell:2007ws,Artoisenet:2007xi,Artoisenet:2008fc}.

\bef
\includegraphics[width=0.25\textwidth]{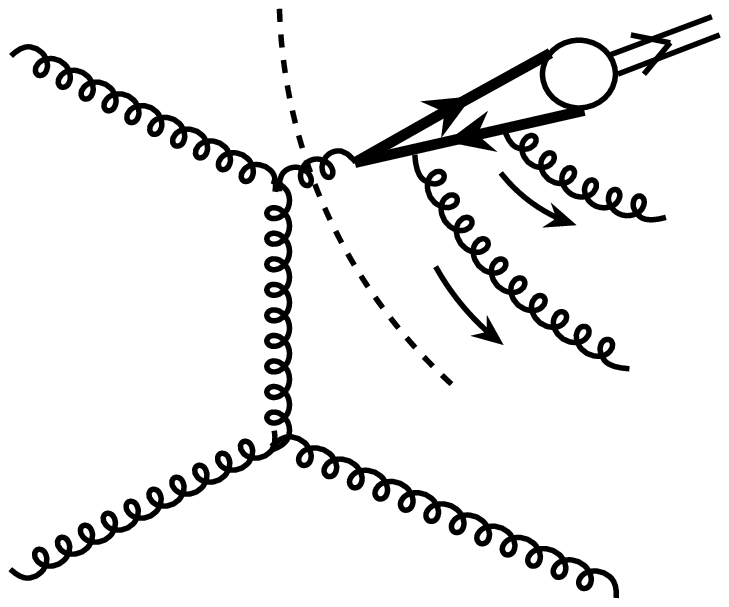}
\hskip 0.5in
\includegraphics[width=0.25\textwidth]{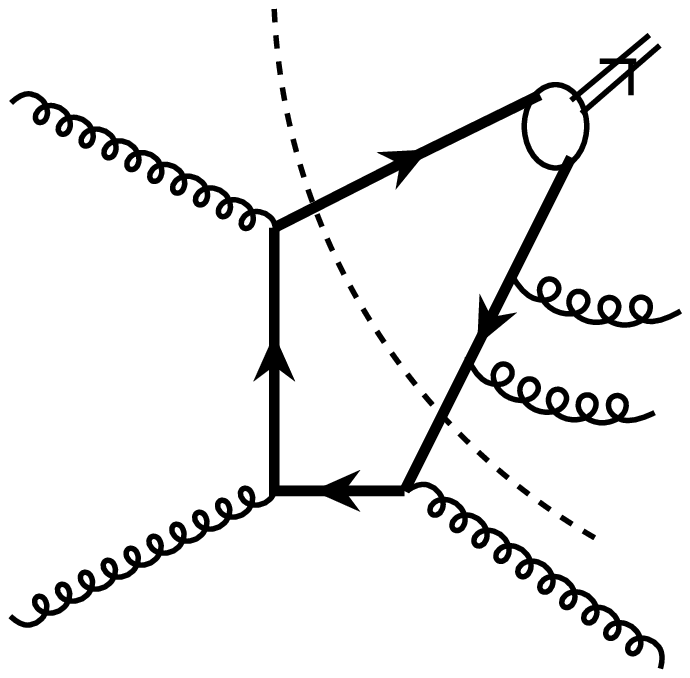}
\caption{Sample NNLO Feynman diagrams having the $p_T^2/m_Q^2$ power enhancement over the LO contribution to heavy quarkonium production.}
\label{fig:nnlo}
\eef

There is no complete NNLO calculation available for heavy quarkonium production at high $p_T$ in the CSM.  The perturbative contribution at this order receives real contributions from the $2\to 5$ Feynman diagrams with one active heavy quark pair.  With an additional parton in the final-state, this contribution can potentially gain two powers of $p_T^2/m_Q^2$ enhancement over the LO from the type of diagram on the left in Fig.~\ref{fig:nnlo}, and one power from the diagram on the right, respectively.  Since the perturbative production rate of a single parton at high $p_T$ already has the strongest $p_T$ behavior at $1/p_T^4$, we do not expect additional power enhancements from contributions beyond NNLO.  

In summary, in these CSM examples, the lowest order in $\alpha_s$ for $\hat{\sigma}_{[Q\bar{Q}(n)]}$ is not always consistent with  the leading power in $p_T$ when $p_T\gg m_Q$.  Large enhancements in the CSM from higher order calculations, even at high $p_T$, suggest that we need to supplement the simplest perturbative expansion in the powers of $\alpha_s$ for the CSM, and by implication for the NRQCD factorization of Eq.\ (\ref{eq:nrqcd-fac}), to take into account radiation from heavy quark pairs produced at short, and intermediate, time scales.

We propose to expand the production cross section of heavy quarkonia at high $p_T$ in powers of $1/p_T$ first, when $p_T\gg m_H$, and only then to expand perturbatively factorizable hard parts  in powers of $\alpha_s$.   In the remainder of this section, we argue that the cross section for producing a heavy quarkonium at large transverse momentum at collider energies can be expanded as a power series of $1/p_T^2$, and that the leading power term and the first subleading power terms can be perturbatively factorized into infrared safe short-distance hard parts in convolution with nonperturbative but universal long-distance fragmentation functions \cite{Kang:2011zza,Kang:2011mg}.

\bef
\includegraphics[width=0.4\textwidth]{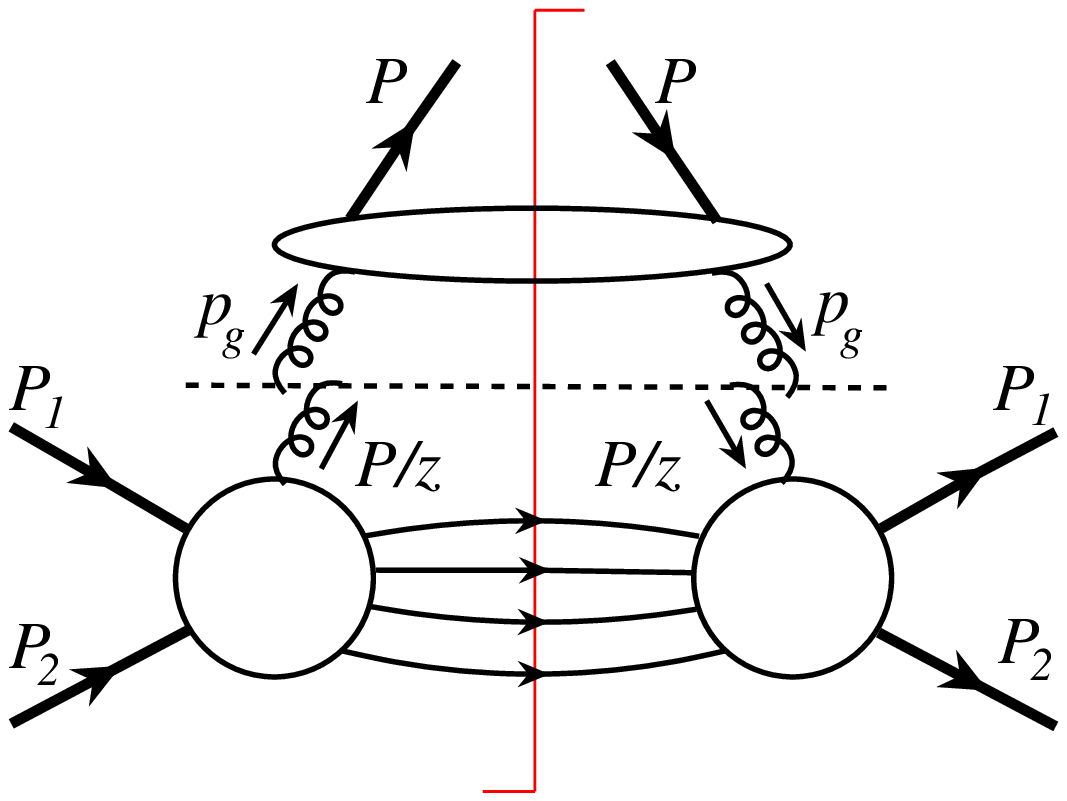}
\hskip 0.8in
\includegraphics[width=0.4\textwidth]{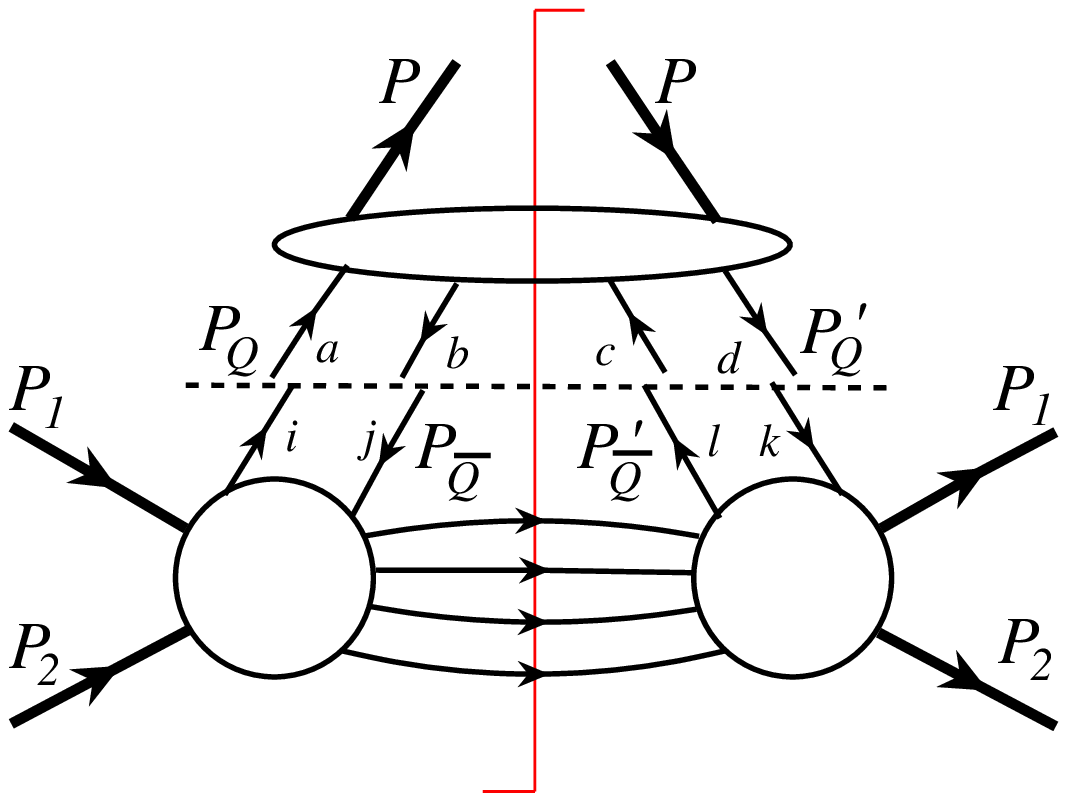}
\caption{These generic Feynman diagrams represent alternative heavy quarkonium production processes, via the production of a single parton, here a gluon, (left) and a heavy quark pair (right) at short distance.}
\label{fig:fac1}
\eef

\bef
\includegraphics[width=0.4\textwidth]{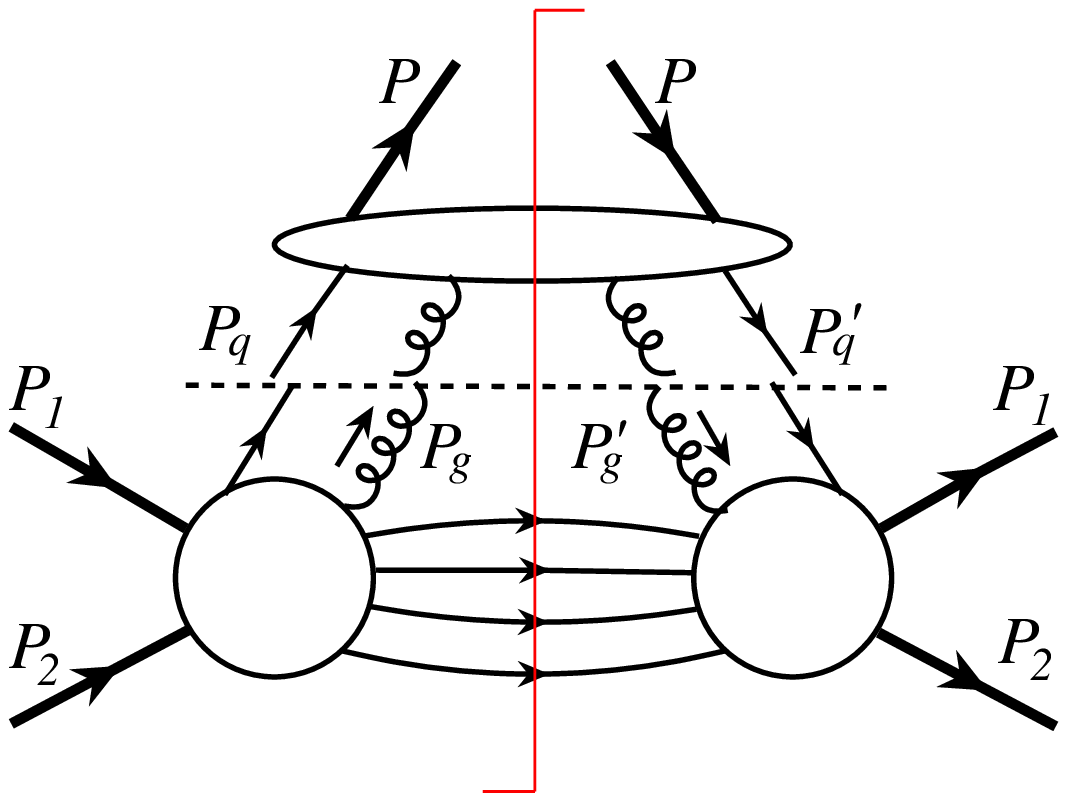}
\hskip 0.8in
\includegraphics[width=0.4\textwidth]{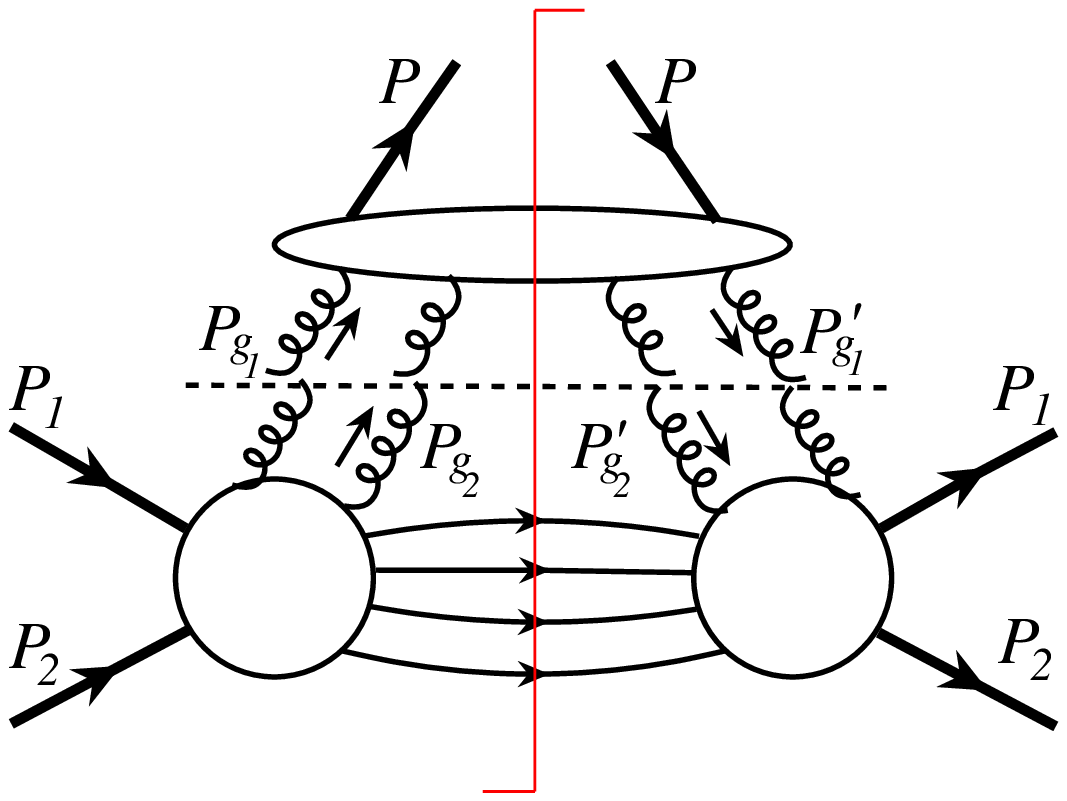}
\caption{Examples of additional production channels that contribute to the first subleading power term in the $1/p_T^2$ expansion of heavy quarkonium production cross sections.
}
\label{fig:fac1more}
\eef

\subsection{Fragmentation and the factorization formula}

Producing a pair of heavy quarks is a necessary condition for producing a heavy quarkonium.   In terms of $1/p_T^2$ expansion, the first two power contributions to the cross section of heavy quarkonium production at high $p_T$ can be presented in terms of the sample diagrams in Fig.~\ref{fig:fac1}.  These figures are shown in cut diagram notation, in which the amplitude and complex conjugate are combined into a forward scattering diagram and the final state is identified by a vertical line.  The diagram on the left represents the leading power term in $1/p_T^2$ expansion and can be interpreted as the perturbative production of a single parton (say a gluon of momentum $p_g$) at the short-distance scale  $1/p_T$.   This parton fragments into a heavy quarkonium at a much later time ($\sim 1/m_H$).
The diagram on the right represents a first subleading power term in the $1/p_T^2$ expansion and corresponds to the production of a heavy quark pair at the short distance scale, which then fragments into a heavy quarkonium.  
In addition, there are $1/p_T^2$ power suppressed contributions to the production of a single active parton in hadronic collisions, the diagram on the left in Fig.~\ref{fig:fac1}.    These include contributions from short-distance collisions involving a single parton from one incoming hadron and two partons from the other, the latter being represented by twist-4 multi-parton correlation functions \cite{Qiu:1990xxa,Qiu:1990xy}.   Other power-suppresssed  terms in the $1/p_T^2$ expansion include the production of a pair of collinear and on-shell light partons, such as those shown in Fig.~\ref{fig:fac1more} plus other combinations and interferences.   With all contributions up to the first subleading power in $1/p_T^2$, we have the corresponding factorization formula \cite{Kang:2011zza,Kang:2011mg},
\begin{eqnarray}
\label{eq:pqcd_fac}
E_P\frac{d\sigma_{A+B\to H+X}}{d^3P}(P)
&\approx &
\sum_{f} \int \frac{dz}{z^2} \, D_{f\to H}(z;m_Q) \, 
E_c\frac{d\hat{\sigma}_{A+B\to f(p_c)+X}}{d^3p_c}\left(p_c=\frac{1}{z}\, p\right) 
\nonumber\\
\ \nonumber\\
&\ &  \hspace{-10mm} +\ 
\sum_{[Q\bar{Q}(\kappa)]} \int \frac{dz}{z^2} \, du\, dv\
{\cal D}_{[Q\bar{Q}(\kappa)]\to H}(z,u,v;m_Q)
\nonumber\\
&\ & \hspace{-5mm}
\times \
E_c\frac{d\hat{\sigma}_{A+B\to [Q\bar{Q}(\kappa)](p_c)+X}}{d^3p_c}
\left(
P_Q=\frac{u}{z}\, p,P_{\bar{Q}}=\frac{\bar{u}}{z}\, p,
P'_Q=\frac{v}{z}\, p,P'_{\bar{Q}}=\frac{\bar{v}}{z}\, p\right)
\nonumber\\
\ \nonumber\\
&\ &  \hspace{-10mm} +\ 
\sum_{[ff']\neq[Q\bar{Q}(\kappa)]} \int \frac{dz}{z^2} \, du\, dv\
{\cal D}_{[ff']\to H}(z,u,v;m_Q)
\nonumber\\
&\ & \hspace{-5mm}
\times \
E_c\frac{d\hat{\sigma}_{A+B\to [ff'](p_c)+X}}{d^3p_c}
\left(
P_1=\frac{u}{z}\, p,P_2=\frac{\bar{u}}{z}\, p,
P_3=\frac{v}{z}\, p,P_4=\frac{\bar{v}}{z}\, p\right)
\, ,
\nonumber\\
\label{eq:pqcd_fac0}
\end{eqnarray}
where $p^\mu = P^\mu(m_H=0)$ in a frame in which the heavy quarkonium moves along $z$-axis, 
defined in and below Eq.~(\ref{eq:momentumP}).  In this expression,
the renormalization scale $\mu$ and the factorization scale $\mu_F$ are suppressed, 
 $\sum_f$ indicates a sum over all parton flavors, $f=q,\bar{q},g$, including heavy flavors with $m_Q\ll p_T$, 
while $\sum_{[Q\bar{Q}(\kappa)]}$ runs over both color and 
spin states of heavy quark pairs $[Q\bar{Q}(\kappa)]$, which will be specified below,
and, finally, $\sum_{[ff']}$ runs over all twist-4 four-parton states excluding those already
included in $\sum_{[Q\bar{Q}(\kappa)]}$.  
The variables $z$, $u$, and $v$, with $\bar{u}=1-u$ and $\bar{v}=1-v$, in Eq.~(\ref{eq:pqcd_fac0}) 
are light-cone momentum fractions defined as,
\begin{equation}
z \equiv \frac{p^+}{p_c^+}
\end{equation}
for the single parton fragmentation term,
\begin{eqnarray}
z &\equiv& \frac{p^+}{P^+_Q + P^+_{\bar{Q}}} = \frac{p^+}{P^{'+}_Q + P^{'+}_{\bar{Q}}}
=\frac{p^+}{p_c^+}\, ,
\nonumber \\
u &\equiv &  z\, \frac{P_Q^+}{p^+} \, , \quad\quad 
\bar{u} \equiv  z\, \frac{P_{\bar{Q}}^+}{p^+} \, , \quad\quad 
v \equiv  z\, \frac{{P'_Q}^{+}}{p^+} \, , \quad\quad 
\bar{v} \equiv  z\, \frac{{P'_{\bar{Q}}}^{+}}{p^+} \, ,
\label{eq:zzetadef}
\end{eqnarray}
for the heavy quark pair fragmentation term, and similarly, 
\begin{eqnarray}
z &\equiv& \frac{p^+}{P^+_1 + P^+_2} = \frac{p^+}{P^{+}_3 + P^{+}_{4}}
=\frac{p^+}{p_c^+}\, ,
\nonumber \\
u &\equiv &  z\, \frac{P_1^+}{p^+} \, , \quad\quad 
\bar{u} \equiv  z\, \frac{P_{2}^+}{p^+} \, , \quad\quad 
v \equiv  z\, \frac{P_3^{+}}{p^+} \, , \quad\quad 
\bar{v} \equiv  z\, \frac{P_{4}^{+}}{p^+} \, ,
\label{eq:zzetadef4f}
\end{eqnarray}
for the other twist-4 fragmentation terms.
The superscript ``$+$'' in these definitions indicates the momentum component 
along the light-cone ``$+$'' direction in a frame where the heavy quarkonium momentum 
has only ``+'' component without ``$-$'' and ``$\perp$'' components.  
Here, we assume that $m_H/p^+ \ll 1$.
In this frame, the heavy quarkonium momentum ${p}^\mu ={p}^+ \bar{n}^\mu$ 
with a light-cone vector $\bar{n}^\mu=(1,0,\bf{0}_\perp)$.
The light-cone components of a general 4-dimensional momentum are $k^\mu=(k^+,k^-,\bf{k}_\perp)$,  
where $k^\pm\equiv (k^0\pm k^z)/\sqrt{2}$.  
The ``+'' component of momentum $p$ can be projected out by another light-cone vector 
${n}^\mu=(0,1,\bf{0}_\perp)$ as $p^+ = p\cdot n$ with $n\cdot \bar{n}=1$ and $n^2=\bar{n}^2=0$.  
Although the total momentum of the heavy quark pair is the same for both the scattering amplitude 
and its complex conjugate, the individual heavy quark momentum in the amplitude does not 
have to be the same as the heavy quark momentum in the complex conjugate amplitude.  
That is, $u$ does not have to be the same as $v$, as defined in Eq.~(\ref{eq:zzetadef}).  
The range for the momentum fractions $u$ and $v$, and $\bar{u}$ and $\bar{v}$, 
is $0$ to $1$.    

Although there is non-trivial interference in the momentum fractions of the heavy quarks, there is no interference between two-quark and single-gluon states, of the sort shown in Fig.\ \ref{fig:absent-mix}, which might suggest a correction suppressed by only a single power of $p_T$.
\bef
\centerline{\includegraphics[width=0.5\textwidth]{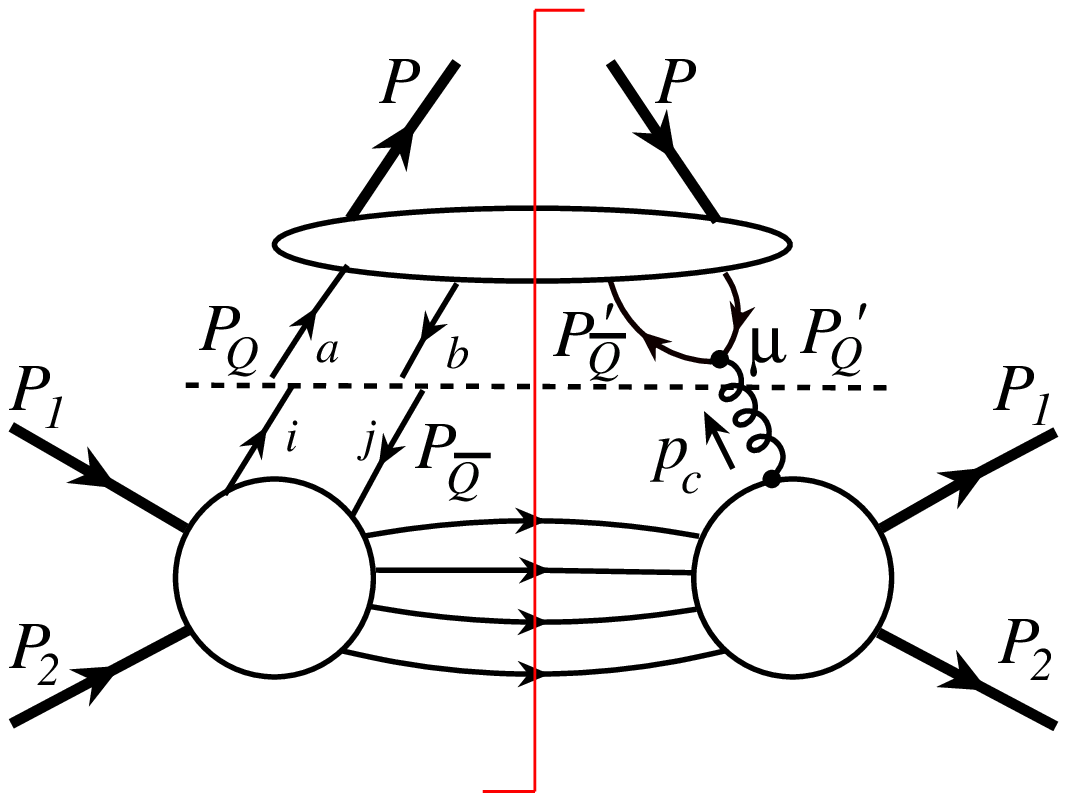}}
\caption{Interference between quark pair and single gluon states.   As explained in the text, this momentum configuration is actually part of the standard quark pair fragmentation in Fig.\ \ref{fig:fac1} because the pole associated with the gluon of momentum $p_c$ is cancelled by the heavy fermion loop.
}
\label{fig:absent-mix}
\eef
As we shall see in Sec.\ \ref{sec:frag} below Eq.\ (\ref{eq:spin-normal}), however, after the internal integrations of the heavy quark subdiagram are carried out, the only vectors that can couple to the gluon of momentum $p_c$ at the vertex above the dashed line  are orthogonal to the physical polarizations of that gluon.   As a result, the gluon pole at $p_c^2=0$ is cancelled, and the heavy quark loop couples to the hard scattering through a contact term on the right of the cut.   The momentum space regions associated with this diagram are then absorbed naturally into the heavy quark fragmentation term in the factorized cross section.

The cross sections $d\hat{\sigma}_{A+B\to f(p_c)+X}$,
$d\hat{\sigma}_{A+B\to [Q\bar{Q}(\kappa)](p_c)+X}$ and
$d\hat{\sigma}_{A+B\to [ff'](p_c)+X}$ in Eq.~(\ref{eq:pqcd_fac0}) 
include all information on the incoming states.
When A and B are hadrons, the $d\hat{\sigma}_{A+B\to f+X}$ in Eq.~(\ref{eq:pqcd_fac0}) 
includes the leading power contribution from collisions of one active parton 
from each colliding hadron, which are proportional to convolutions with 
parton distribution functions (PDFs) at factorization scale $\mu_F$.
They also include the first $1/p_T^2$ suppressed contribution from collisions 
involving one parton from one colliding hadron and two from the other,
given by convolutions of a PDF from one hadron with a twist-4 four-parton correlation function
from the other \cite{Qiu:1990xxa,Qiu:1990xy}, as noted above.  
The $D_{f\to H}(z;m_Q)$ are fragmentation functions for off-shell partons of flavor $f$
to produce a quarkonium state $H$ of momentum $p$ \cite{Collins:1981uw,Nayak:2005rw}. 
The $m_Q$-dependence of these functions indicates that 
the quarkonium state $H$ is a bound state of heavy quarks of mass $m_Q$.  
The functions ${\cal D}_{[Q\bar{Q}(\kappa)]\to H}(z,u,v;m_Q)$ are generalized fragmentation functions 
defined below for a state consisting of a relativistic heavy quark pair $[Q\bar{Q}(\kappa)]$ 
to fragment into the same quarkonium state $H$.  
The ${\cal D}_{[ff']\to H}(z,u,v;m_Q)$ represents all twist-4 fragmentation functions,
excluding those involving a heavy quark pair, ${\cal D}_{[Q\bar{Q}(\kappa)]\to H}(z,u,v;m_Q)$, 
which are already included in the second term on the right-hand-side of Eq.~(\ref{eq:pqcd_fac0}).

In this paper, we neglect all contributions involving twist-4 multi-parton correlation functions of colliding hadrons, because these contributions are expected to be suppressed by $\Lambda_{\rm QCD}^2/p_T^2$ 
and small even at moderate $p_T\gg m_H$, compared to the leading power term in Eq.~(\ref{eq:pqcd_fac0}).
For the NLP contribution, we keep only the fragmentation contribution from heavy quark pairs, the middle term on the right side of Eq.~(\ref{eq:pqcd_fac0}).
This is because only this term can produce a pair of heavy quarks with correct quark flavor, which we assume
has a much larger probability to fragment into a heavy quarkonium
than that of a light gluon at the LP, so that it can compensate in part the power suppression of $1/p_T^2$ 
of the hard parts to provide comparable contributions to the total production rate at moderate $p_T$.
That is, we neglect the third term in Eq.~(\ref{eq:pqcd_fac0}) because  
we expect that the creation of a heavy quarkonium from
fragmentation functions of twist-4 light-parton states is suppressed at least by $\alpha_s^2(m_Q)$ in comparison with the fragmentation from a heavy quark pair with correct quark flavor.
We leave detailed estimates of the size of neglected terms to future work.

It is natural to suppose that the contribution to the heavy quarkonium production comes from 
the region of phase space where the quark and the antiquark have the same momentum, 
$P_Q=P_{\bar{Q}}$ or $u=v=1/2$, which is preferred by the quarkonium wave function.  
However, for a pair produced at very short distance $\sim 1/p_T$, 
which is of a size too small for a physical quarkonium to be formed, 
the quark and antiquark clearly do not need to have precisely the same momentum.  
Their relative momentum changes while they are evolving to a lower momentum scale by radiation.   
It is the non-perturbative fragmentation functions at or near the input momentum scale ($\sim 2m_Q$) 
that are proportional to the wave function of the produced heavy quarkonium, 
which strongly suppresses those configurations where the heavy quark and antiquark 
have a large relative momentum fraction.  
That is, at hadronic scales, the $[Q\bar{Q}]$-fragmentation function is peaked in the region where 
$u=v=1/2$, and vanishes when $u$ and $v$ approach $0$ or $1$.

If the fragmentation functions fall sufficiently fast when $u,v\to 0$ or $1$, 
one could make additional approximation to the factorization formalism in Eq.~(\ref{eq:pqcd_fac0}) 
by setting $u=v=1/2$ in the perturbative hard part $d\hat{\sigma}_{A+B\to [Q\bar{Q}(\kappa)]+X}$, 
and derive the following approximate factorization formula,
\begin{eqnarray}
E_P\frac{d\sigma_{A+B\to H+X}}{d^3P}(P)
&\approx &
\sum_{f} \int \frac{dz}{z^2}\, D_{f\to H}(z;m_Q)\, 
E_c \frac{d\hat{\sigma}_{A+B\to f(p_c)+X}}{d^3p_c}(p_{c}=p/z)
\nonumber\\
&+&
\sum_{[Q\bar{Q}(\kappa)]} \int \frac{dz}{z^2} \, 
D_{[Q\bar{Q}(\kappa)]\to H}(z;m_Q)\, 
\label{eq:pqcd_fac}\\
&\ & \hskip 0.4in
\times
E_c\frac{d\hat{\sigma}_{A+B\to [Q\bar{Q}(\kappa)](p_c)+X}}{d^3p_c}
(P_{Q}=P_{\bar{Q}}=P{'}_{Q}=P{'}_{\bar{Q}}=p/2z)
\,  \, ,
\nonumber
\end{eqnarray}
where we did not list the subleading power terms that are neglected in this paper, and where 
the integrated $[Q\bar{Q}(\kappa)]$-fragmentation functions are given by,
\begin{equation}
D_{[Q\bar{Q}(\kappa)]\to H}(z;m_Q) 
\equiv \int du\, dv\ {\cal D}_{[Q\bar{Q}(\kappa)]\to }(z,u,v;m_Q)\, , 
\label{eq:frag_approx}
\end{equation}
which, like the leading power single parton fragmentation functions, depends only on the total momentum fraction $z$ of the pair, carried by the observed heavy quarkonium $H$.  Without knowing exactly how fast the pair fragmentation functions fall when $u$ and $v$ move away from 1/2, however, we do not make this approximation in the calculations presented in this paper.

The validity of the perturbative QCD factorization formalism in Eqs.~(\ref{eq:pqcd_fac0}) (or(\ref{eq:pqcd_fac})) requires the suppression of quantum interference between the dynamics above and below the dashed lines in Fig.~\ref{fig:fac1} (and in Fig.~\ref{fig:fac1more}).  That is, the dominant contributions of partonic processes in Fig.~\ref{fig:fac1} should necessarily come from the phase space where the fragmenting partons (the gluon in the diagram on the left, and the heavy quark pairs on the right) are forced to their mass shells, and are consequently long-lived compared to the time scale of the hard collision below the dashed line.  The figures illustrate how these regions arise.
The momentum of the single active parton (the gluon), $P_g$ on the left diagram in Fig.~\ref{fig:fac1}, is forced to $P_g^2 \sim 0$ at the boundary of phase space. 
Similarly, in the diagram on the right in Fig.~\ref{fig:fac1} (and those in Fig.~\ref{fig:fac1more}) the limit of low invariant mass for the $Q,\bar Q$ pair is at the boundary in phase space, and in this region the loop momentum flowing between the two lines is pinched between mass shell singularities.   This happens on both sides of the cut, that is, in the amplitude and, independently, in the complex conjugate amplitude.   It is the contribution of this, nonleading region that is summarized in the power suppressed terms in Eq.\ (\ref{eq:pqcd_fac0}) for both heavy quarks of Fig.\ \ref{fig:fac1} and the light partons of Fig.\ \ref{fig:fac1more}.  

We can illustrate the pinch of loop momenta in Fig.\ \ref{fig:fac1} by labeling the heavy quark and antiquark momenta in the amplitude as, $P_Q= P/2 + q$ and $P_{\bar{Q}}= P/2 - q$, respectively.   The integral over $q$ then takes the form,
\begin{equation}
{\cal M} \propto \int \frac{d^4 q}{(2\pi)^4} {\rm Tr}
\left[\hat{H}(P,q,Q) \frac{- \gamma\cdot (P/2 - q)+m_Q}{(P/2-q)^2 - m_Q^2 + i\varepsilon}
       \hat{D}(P,q) \frac{\gamma\cdot (P/2 + q)+m_Q}{(P/2+q)^2 - m_Q^2 + i\varepsilon}
\right]\, ,
\label{eq:amplitude0}
\end{equation}
where $\hat{H}(P,q,Q)$ represents the production of the heavy quark pair with a hard scale $Q$, 
$\hat{D}(P,q)$ represents the fragmentation of the pair, and $q$ is the relative momentum of the pair in the amplitude, which does not have to be the same as the relative momentum of the pair in the complex conjugate amplitude.  If the total momentum of the heavy quark pair is dominated by the $P^+$ component in a frame in which the heavy quarkonium is moving in the $+z$ direction, we can identify the relevant perturbative contribution to the integration of $q$ in Eq.~(\ref{eq:amplitude0}) by examining the pole structure of its $q^-$ integration.  From the denominators in Eq.~(\ref{eq:amplitude0}), we have
\begin{eqnarray}
q^- &=&
\frac{q_\perp^2 - 2m_Q^2 (q^+/P^+)}{P^+ + 2q^+} - i\varepsilon \theta(P^+ + 2q^+)
\to \frac{q_\perp^2}{P^++2q^+} - i\varepsilon \, ,
\nonumber\\
q^- &=&
- \frac{q_\perp^2 + 2m_Q^2 (q^+/P^+)}{P^+ - 2q^+} + i\varepsilon \theta(P^+ - 2q^+)
\to - \frac{q_\perp^2}{P^+-2q^+} + i\varepsilon\, .
\label{eq:poles}
\end{eqnarray}
These two denominators pinch the $q^-$ integral so long as we are away from the region  $q^+\rightarrow \pm \ P^+/2$, where one of the pair carries all the momentum, and the other is at rest.   We shall assume that this region is strongly suppressed for producing a bound quarkonium, and that $P^+\gg m_Q$ (that is, $p_T\gg m_Q$ in the Lab frame).  In any case, it is clear from Eq.~(\ref{eq:poles}) that the contributions from the diagram on the right in Fig.~\ref{fig:fac1} are forced into the region of phase space where the heavy quark and antiquark are both close to their mass-shells, and are factorized from the short-distance hard-scattering process.   We must still argue, of course, that this factorization is respected by higher orders in the perturbative expansion.

The predictive power of the factorization formula in Eq.~(\ref{eq:pqcd_fac0})  relies on our ability to do systematic perturbative calculations of the short-distance partonic hard parts in powers of $\alpha_s$ and of the evolution kernels for the scale dependence of these fragmentation functions, as well as on the universality of the fragmentation functions.  The accuracy of the perturbative calculations and the strength of their predictive power depends on the stability of the perturbative expansion in powers of $\alpha_s$, and the approximation of neglecting terms that are even higher powers in the $m_H^2/p_T^2$ expansion.   

An important feature of the perturbative QCD factorization formalism in  Eq.~(\ref{eq:pqcd_fac0})  is that the short-distance partonic hard parts should not depend on the details of the quarkonium states they produce.  Therefore, we can extract the short-distance partonic hard parts in Eq.~(\ref{eq:pqcd_fac0}) perturbatively order-by-order in powers of $\alpha_s$ by applying the same factorization formula to the production of partonic states, $H\ =\ g, q, \bar{q}, [Q\bar{Q}(\kappa)]$.  When  $H$ in Eq.~(\ref{eq:pqcd_fac0})  is a partonic state, both the cross section to produce the partonic state on the left of the equation and the fragmentation functions to the partonic state on the right of the equation can be systematically evaluated by calculating Feynman diagrams order-by-order in powers of $\alpha_s$, with a  regularization for  the collinear singularity when $m_Q/p_T \to 0$.  Since the short-distance partonic hard parts on the right side of Eq.~(\ref{eq:pqcd_fac0})   are infrared safe, the partonic cross sections on the left and the fragmentation functions on the right of the factorized equation  share the same collinear divergences, if any, order-by-order in $\alpha_s$.  That is, factorization requires that the perturbative fragmentation functions to a given partonic state should absorb all collinear singularities of the cross section for that partonic state.  

\subsection{Arguments for Factorization}

We now go on to give a justification of the factorization formula in Eq.~(\ref{eq:pqcd_fac0}).   In our discussion, we will revisit several arguments given in Ref.\ \cite{Qiu:1990xy}, where a general analysis for factorization of power-suppressed corrections in hadron-hadron scattering was introduced.

The development of factorized cross sections begins with an examination of amplitudes, most conveniently in terms of the general properties of pertubation theory diagrams \cite{Sterman:1978bi}.    We begin by noting that although factorization is simplest to formulate when all particles are massless, the presence of masses does not by itself require a reformulation of factorized cross sections \cite{Qiu:1990xy}.   Particle mass dependence can be incorporated consistently both into hard scattering functions and long distance parton distributions and fragmentation functions \cite{Collins:1998rz,Mitov:2006xs}.   New, power-suppressed corrections to factorization in inclusive and seminclusive cross sections are associated not with fixed mass scales, such as $m_Q$, but with long-distance, nonperturbative effects.  Power suppressed mass corrections in $m_Q^2/p_T^2$ can be included in  the partonic short distance functions by keeping the heavy quark mass term of pQCD Lagrangian as a new local interaction term. 

A convenient classification of potential sources of power corrections requires an analysis of regions of momentum space where integrals are forced near or to the mass shell.    In \cite{Sterman:1978bi} these are referred to as pinch surfaces, an example of which we have seen in Eq.\ (\ref{eq:amplitude0}).    In principle, any pinch surface is associated either with the leading power factorized cross section, or with a power-suppressed (and in general) factorized correction to the leading-power factorized form.  In a hard-scattering cross section, the contributions of any region can be put into a factorized form, but this form will only be useful if its components have some features of universality.   For example, in Eq.~(\ref{eq:pqcd_fac0})  the ``new" quark pair fragmentation function is independent of the choice of initial state, $|A,B\rangle$.   It is possible to estimate the overall power behavior by using the techniques of \cite{Sterman:1978bi}, and we will not reproduce all these arguments here.   
 
Because any cross section, and in particular a single-particle inclusive cross section, is effectively the sum over all possible such regions of momentum space, a cross section $d\sigma_{A+B\to C(p)+X}/dp_T$ for hadron $C$ of momentum $p$ can in principle be thought of a sum of factorized terms, each with its characteristic dependence on $p_T$.     Of these, the leading power term is of special interest.   In cut diagram notation \cite{Libby:1978qf},  the relevant pinch surfaces are illustrated on the left-hand side of Fig.\ \ref{fig:fragmentation_gluon}.   This figure and other figures in this section represent pinch surfaces in physical gauges.  The situation in covariant gauge is slightly more complex, but the basic conclusions are unchanged.

For Fig.\ \ref{fig:fragmentation_gluon}, and indeed all pinch surfaces \cite{Sterman:1978bi}, on-shell lines correspond to physically realizable processes involving free, classical propagation and local interaction of partons with finite momenta.   Such processes can also be dressed in all possible ways by a ``cloud" of soft partons, represented by $S$ in the figure.    In the leading configuration, one parton from each of the two hadrons collides to initiate the hard scattering, while the spectators of each hadron move into the final state, interacting with each other along the way.   These sets of mutually-collinear particles are sometimes referred to as ``incoming jets", $J_1$ and $J_2$ in the figure.  The hard scattering produces a parton of momentum $p_g=p/z$, labelled as a gluon in the figure, with $0<z<1$, and this parton radiates a jet, $J_{C/g}$, of collinear partons, some of which eventually emerge into the final state, including the observed hadron $C(p)$.   In individual perturbative diagrams, soft gluons are attached to the incoming as well as outgoing ``jets".

\bef
\begin{minipage}[c]{2.0in}
   \includegraphics[width=0.9\textwidth]{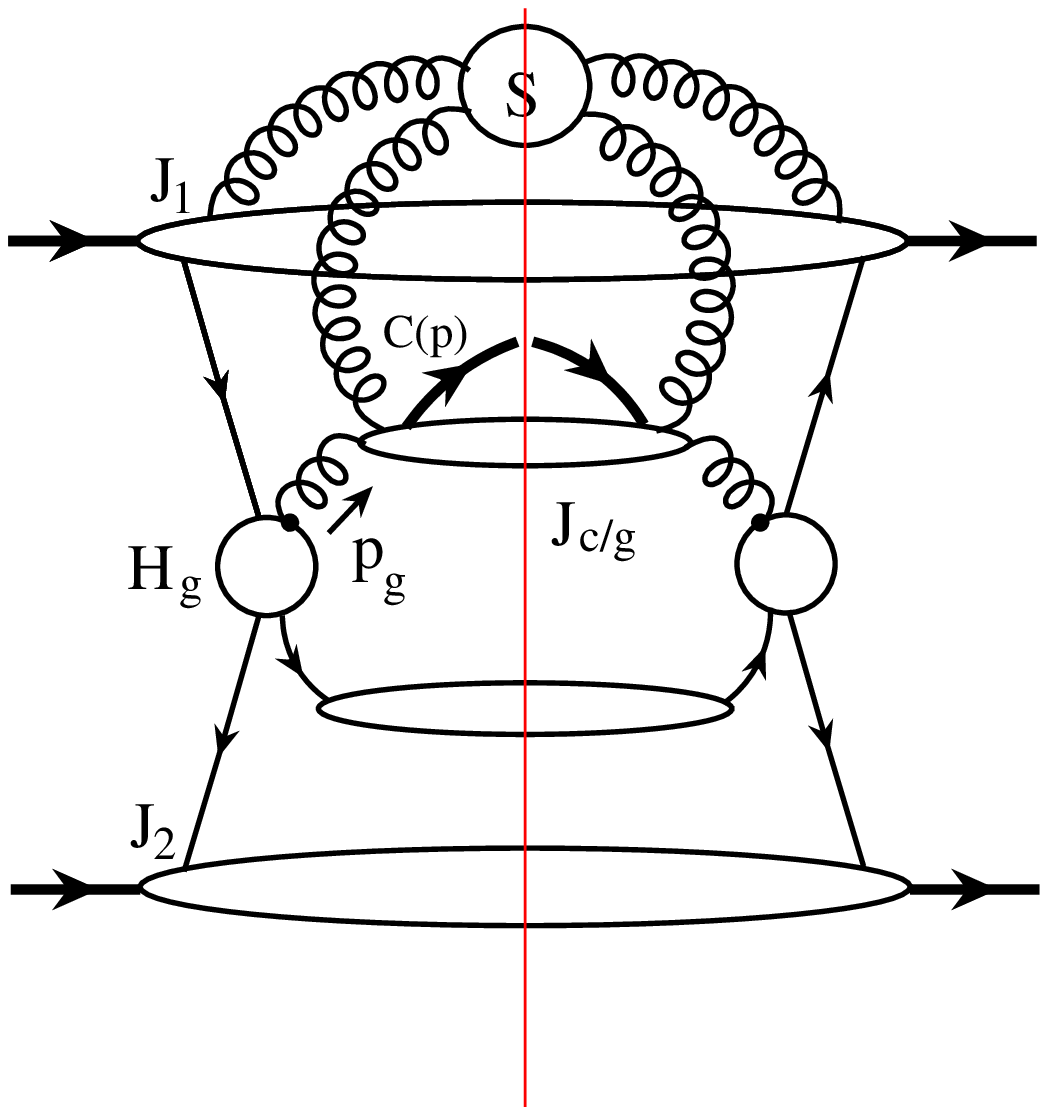}
\end{minipage}
$\ =\ $
\begin{minipage}[c]{2.0in}
   \includegraphics[width=0.9\textwidth]{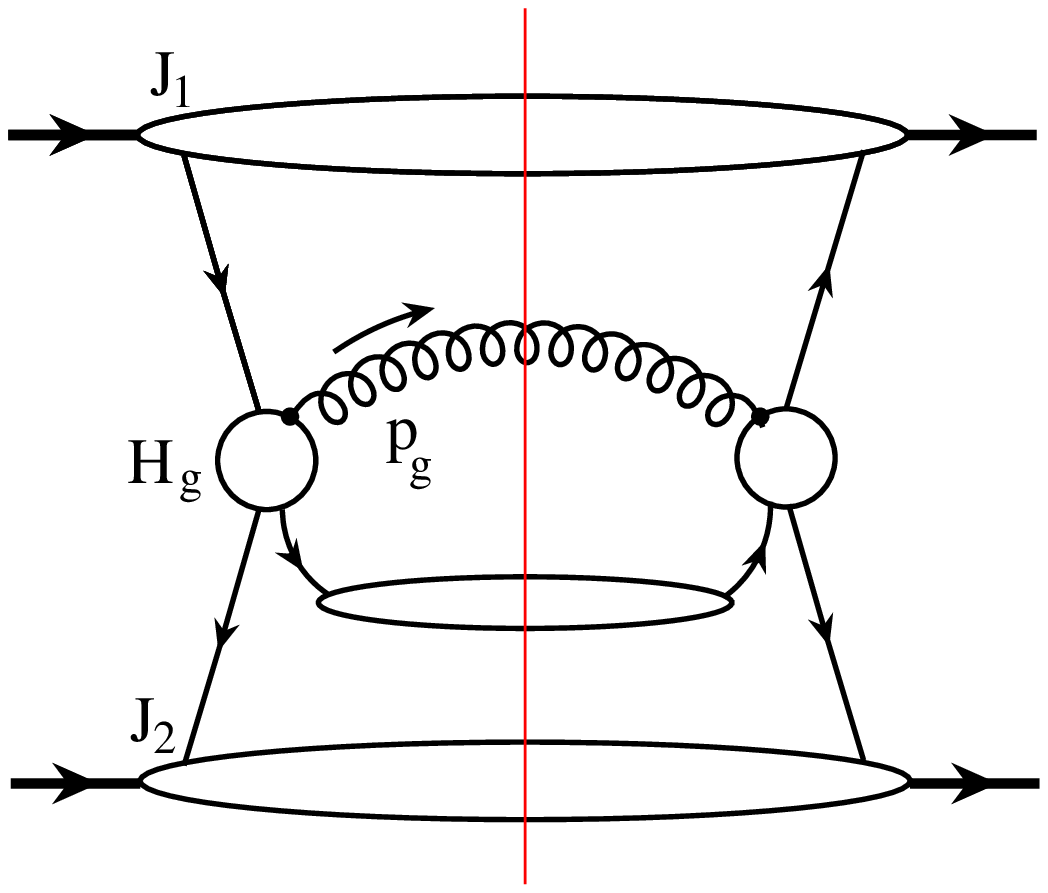}
\end{minipage}
$\ \otimes \ $\hskip 0.1in
\begin{minipage}[c]{1.3in}
   \includegraphics[width=0.9\textwidth]{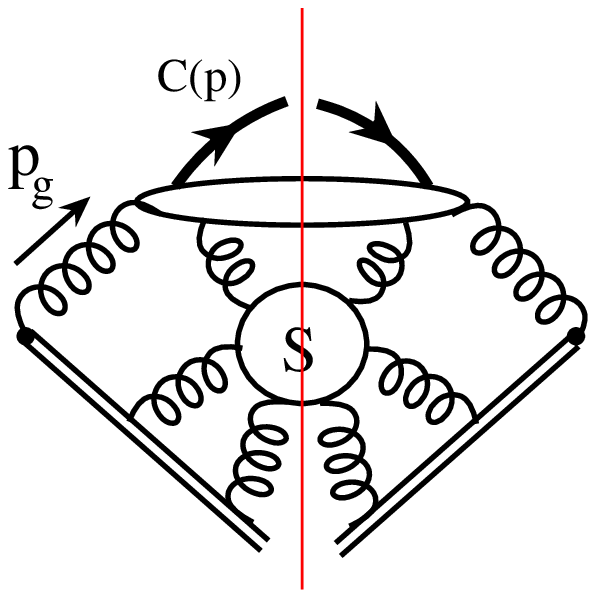}
\end{minipage}
\caption{Leading power pinch surface corresponding to single-parton fragmentation.}
\label{fig:fragmentation_gluon}
\eef

The right-hand side of Fig.\ \ref{fig:fragmentation_gluon} shows the result of factorization, which can be demonstrated in various way, but which is described for leading-power fragmentation in some detail in Refs.\ \cite{Nayak:2005rw,Nayak:2005rt}.  At leading power, the fragmentation function $D_{g\to C(p)}$ decouples from the rest of the process, leaving behind the cross section $\tilde \sigma$ for the production of a parton of momentum $p_g=p/z$, schematically as,
$$
\sigma_{1+2\to C(p)+X} = \widetilde{\sigma}_{1+2\to g+X} \otimes D_{g\to C(p)}\, .
$$ 
In general, the coupling of all soft gluons to the fragmenting jet is represented (again to leading power) by eikonal (Wilson) lines, as indicated in the figure.   Conventionally, these are chosen in a direction $\bar p$, opposite to the particle momentum, $p$, but this is a matter of convention.   In $\tilde \sigma$, soft gluons cancel in the inclusive sum over states, and the incoming jets are organized into parton distributions.  As argued in Refs.~\cite{Nayak:2005rw,Nayak:2005rt}, all leading pinch surfaces take this form.

Nonleading powers in the cross section are associated with nonleading pinch surfaces, where lines are forced to the mass shell at pinch surfaces, but where the integrals from these regions are suppressed by extra inverse powers of $p_T$ relative to leading power.       They must still correspond to physically-realizable processes, however, and can be classified as power-suppressed corrections to the leading factorization of Fig.\ \ref{fig:fragmentation_gluon}.  Figure \ref{fig:Other_PC_rv} illustrates several basic possibilities, which we have already encountered in connection with Eq.\ (\ref{eq:pqcd_fac}) above.  Additional partons may be attached to the hard part within the incoming (or other) jet, as indicated by $T$ in the figure.  Extra soft gluons may attach directly to the hard scattering, as indicated by $U$.    The power counting of Ref.\ \cite{Libby:1978qf} shows that pinch surfaces involving $T$ are nonleading by $1/Q^2$ in any inclusive hard scattering with scale $Q$ for unpolarized cross sections, where $Q$ is the hard scale (in our case, $p_T$).    Such corrections are examples of factorization at $1/Q^2$ into multiparton matrix elements, as discussed in Ref.\ \cite{Qiu:1990xy}.   In the same reference, it was shown that pinch surfaces involving soft lines directly attached to the hard part are suppressed by $1/Q^4$ in general.    
\bef
\includegraphics[width=0.4\textwidth]{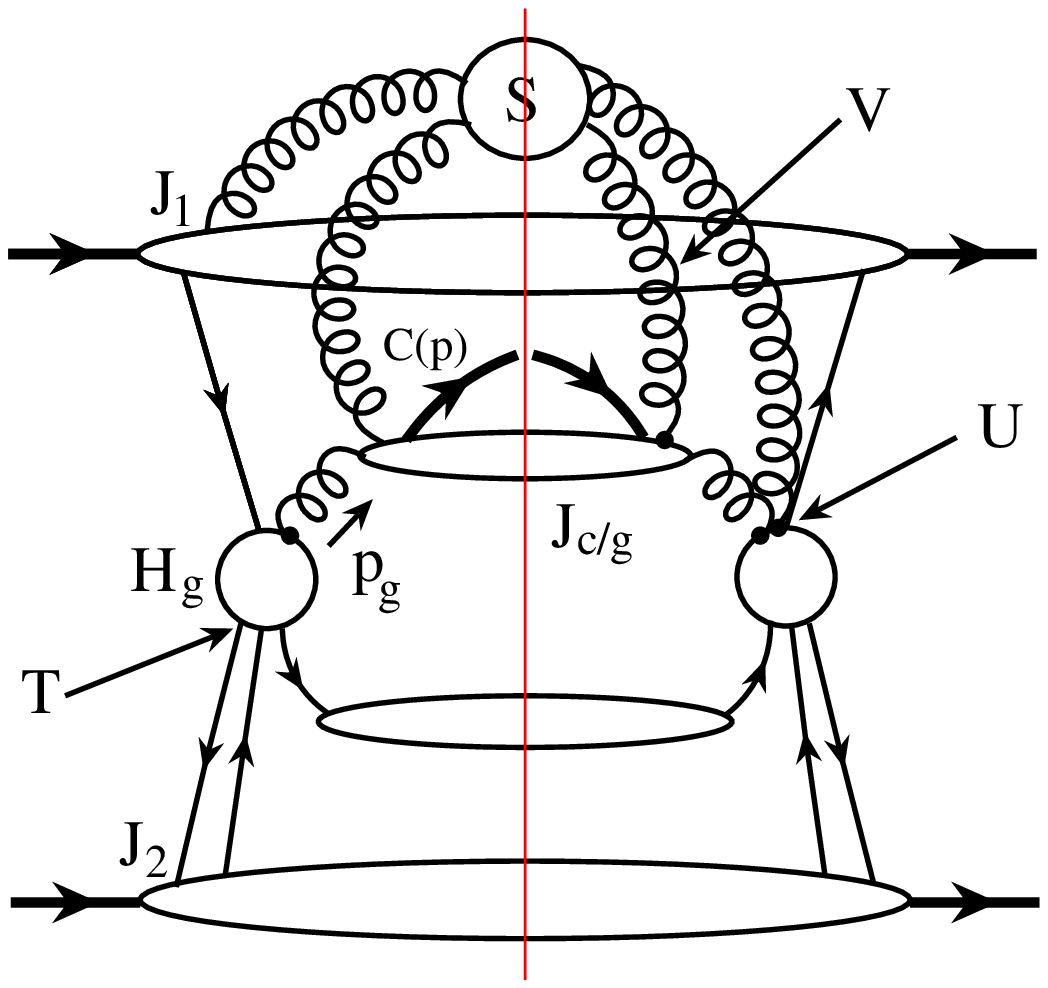}
\caption{Nonleading power pinch surfaces associated with additional initial-state partons (T) 
and soft gluons attached to the hard scattering (U).}
\label{fig:Other_PC_rv}
\eef

 Finally, soft lines may attach the leading fragmenting jet to other jets at the pinch surface, as indicated by $V$, but now with the leading-power couplings that lead to the factorized result of Fig.\ \ref{fig:fragmentation_gluon} removed.   
 In this case, we must analyze these pinch surfaces from the point of view of Ref.\ \cite{Qiu:1990xy}.    We consider a region where all lines in the jet $J_{C/g}$, which contains the final-state parton, are off-shell by some fixed squared mass, call it $M^2$.   Let us take the jet's momentum as approximately $p^\mu_g=p^\mu/z\sim Q\, \bar{n}^\mu$, where $\bar{n}^\mu$ is a light-like vector.   To recall our notation, we write for any jet line $P^\mu \sim Q\, \bar{n}^\mu + (M^2/Q) n^\mu + p_T$, where $p_T^2\sim M^2$ and where $\bar n\cdot n=1$, $\bar n^2 = n^2=0$.    This is a standard scaling of jet-like momentum, as developed in \cite{Sterman:1978bi}.   Soft lines that flow into the jet must have $n$-components of order $M^2/Q$ for the jet lines to remain off-shell by order $M^2$.    We consider first, as in \cite{Qiu:1990xy}, the ``soft central region", where all components of soft momenta are of the same order.  Once the leading terms are removed, the first nonleading contribution is suppressed by one order of soft momentum, so that the contributions is down by order $M^2/Q$ relative to leading power.    But this is not the end of the story.   In this case, the conditions necessary to factor the soft momenta from all other jets (as on the right of Fig.\ \ref{fig:fragmentation_gluon}) are satisfied \cite{Collins:1989gx}, and, as shown in \cite{Qiu:1990xy}, these gluons cancel when we sum over all cuts of the diagrams that are consistent with fixed attachments of soft lines to the fragmenting jet $J_{C/g}$.\footnote{Specifically, we neglect the $\bar n^\mu$ component of the soft gluon momenta in the lines of the frgamenting jet, where they are negligible.   We then integrate the remainder of the diagram over these light cone components.   The cancellation occurs in the resulting sum over final states.}   Remainders are now of order $M^4/Q^4$.    For a more general analysis, it is sometimes necessary to study another region of soft momenta, the so-called ``Glauber region", in which the transverse momenta of soft gluons increase to order $M$, while light-cone components remain at order $M^2/Q$.   This region, however, requires that the soft momenta attach only to spectator lines of the incoming jets \cite{Qiu:1990xy}.   Otherwise, light cone momentum components are not pinched, and one or both may be increased to order $M$.   In the case at hand, it is the $n$-component that is free to be deformed to order $M$, taking the jet lines off-shell to order $QM$, and thus away from the pinch surface.   We conclude that nonperturbative effects found in this way are suppressed to order $1/Q^4$.
 
 It is worth noting at this point the relationship of the above discussion of power corrections to the treatment of power corrections associated with nuclear dependence in Refs.\ \cite{Luo:1994np} and \cite{Wang:2001ifa}.   Reference \cite{Luo:1994np} was concerned with transverse momentum broadening, while Ref.\   \cite{Wang:2001ifa} created the influence of soft rescatterings on fragmentation. In the terminology we have introduced above, both involved an analysis of corrections of the type $T$ in  Fig.\ \ref{fig:Other_PC_rv}, and identified effects suppressed by order $1/Q^2$.   Both are proportional to multi-parton matrix elements in an initial-state hadron, times a perturbative hard-scattering function.    We anticipate these matrix elements to be of order $\Lambda_{\rm QCD}^2$, so that they are relatively modest in their effects, and can be neglected for most hard scattering phenomenology, except when enhanced, for example, by nuclear sizes \cite{Luo:1994np,Wang:2001ifa}.   
 
 Of course, as we have noted above in connection with Fig.\ \ref{fig:fac1more}, power-suppressed pinch surfaces involving more than a single parton at the hard scattering are possible for final state jets as well as the incoming hadrons.  We generaly expect these to be small, as in the incoming case, but matrix elements involving the hadronization of heavy quark pairs into heavy quarkonia may be an exception.   We have in mind a role for the color singlet matrix elements of NRQCD.   Of particular interest are quark pairs in a color singlet configuration with a total transverse momentum $p_T$ or greater.   These appear in pinch surfaces associated with the production of a heavy pair at short distances, which may be created in a singlet configuration, or may evolve into a singlet configuration.   Such surfaces are illustrated in Fig.\ \ref{fig:fragmentation_QQbar}, and include the right-hand side of Fig.\ \ref{fig:fac1}.
 
\bef
\begin{minipage}[c]{2.0in}
   \includegraphics[width=0.9\textwidth]{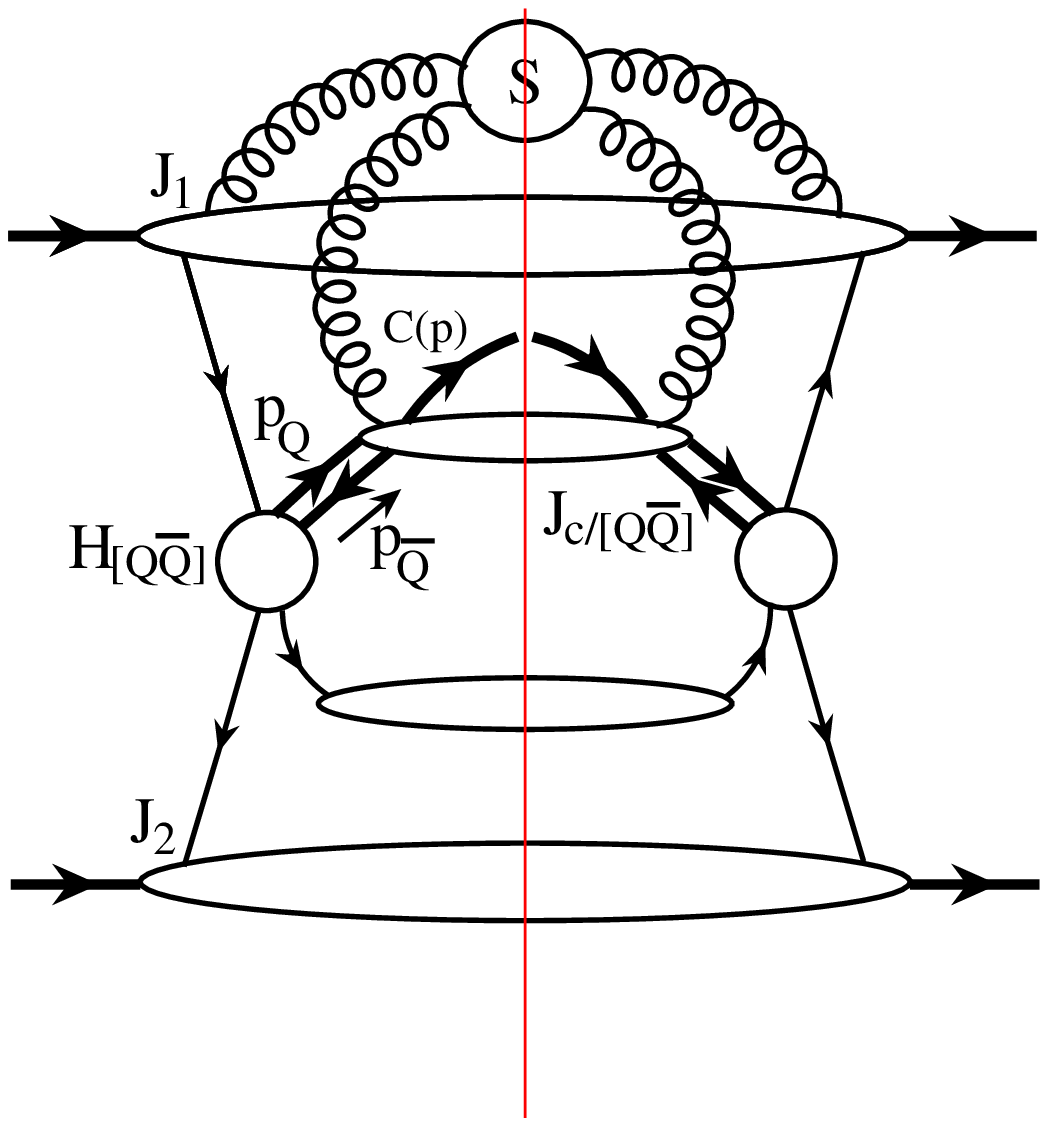}
\end{minipage}
$\ =\ $
\begin{minipage}[c]{2.0in}
   \includegraphics[width=0.9\textwidth]{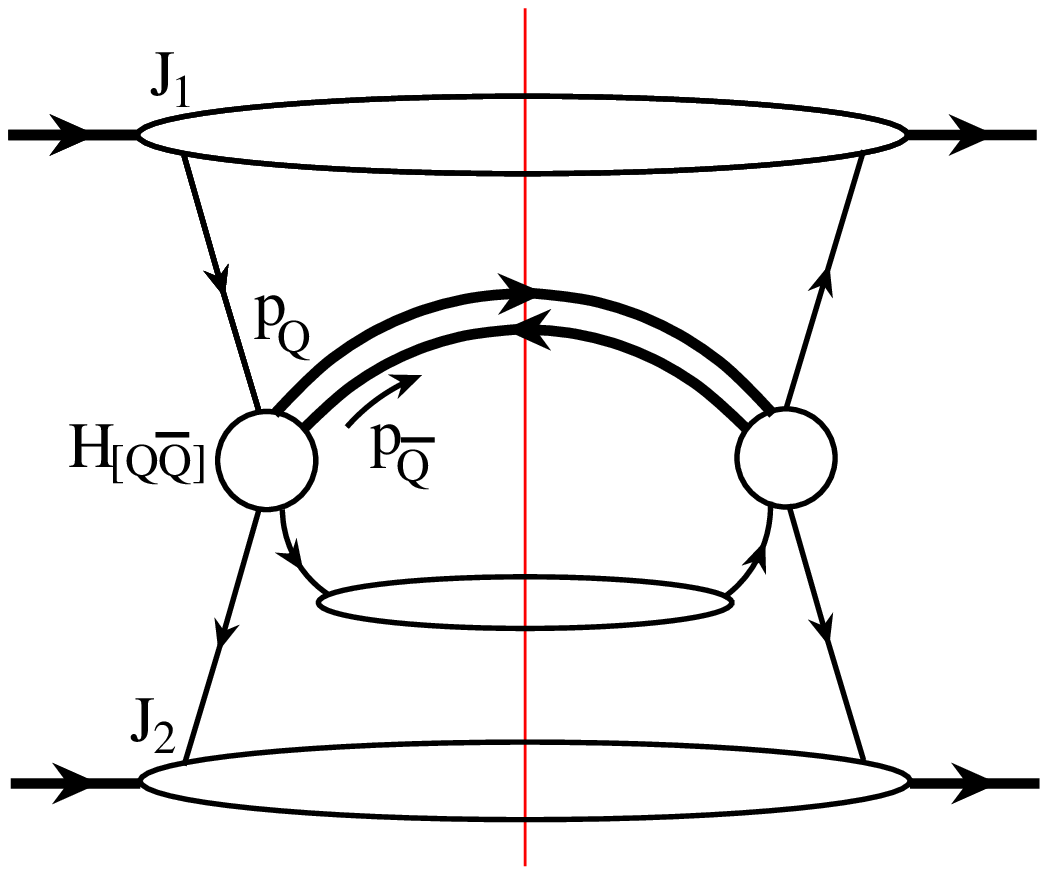}
\end{minipage}
$\ \otimes \ $ \hskip 0.1in
\begin{minipage}[c]{1.3in}
   \includegraphics[width=0.9\textwidth]{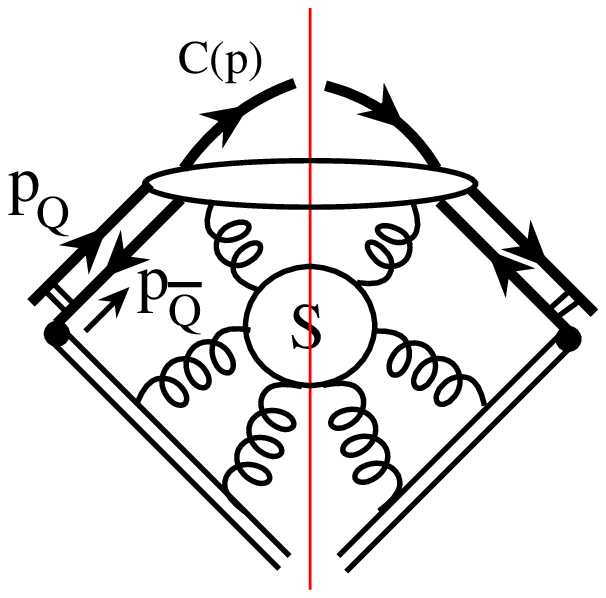}
\end{minipage}
\caption{Nonleading pinch surface representing the production of a pair at short distances.}
\label{fig:fragmentation_QQbar}
\eef
The leading behavior of such surfaces is suppressed by $1/p_T^2$ relative to the overall leading power that of the surfaces shown in Fig.\ \ref{fig:fragmentation_gluon}, simply because two partons rather than one attach to the hard scattering \cite{Sterman:1978bi}.   (This is the same $p_T$-suppression as for item $T$ in Fig.\ \ref{fig:Other_PC_rv}.)    At the same time, if the nonperturbative dynamics for these processes is greatly enhanced relative to the fragmentation functions of individual partons, and here we have in mind the fragmentation of gluons by color octet matrix elements in NRQCD,  they may be competitive, or even dominate for a range of $p_T$.    As indicated by the figure, the same factorization properties analyzed in Refs.\ \cite{Nayak:2005rw,Nayak:2005rt} for single-parton fragmentation apply to the fragmentation of a jet initiated by two partons,
\begin{equation}
\sigma_{1+2\to [Q\bar{Q}](\to C(p)+X')+X} = \widetilde{\sigma}_{1+2\to [Q\bar{Q}]+X} 
\otimes {\cal D}_{[Q\bar{Q}]\to C(p)}\, ,
\end{equation}
in this case heavy quarks have a total transverse momentum that is greater than or equal to $p_T\gg m_Q$.

Recalling again that nonleading pinch surfaces enter the cross section additively, we recognize the momentum configurations of Fig.\ \ref{fig:fragmentation_QQbar} as the source of the heavy pair correction term in Eq.~(\ref{eq:pqcd_fac0}).  This requires us to introduce and analyze these multi-parton fragmentation functions, in the spirit of similar analyses in Ref.\ \cite{Luo:1994np} and \cite{Wang:2001ifa} for higher-twist matrix elements for incoming hadrons.    In the following, we will formulate some of the basic properties of these new two-parton fragmentation functions, ${\cal D}_{[Q\bar{Q}]\to C(p)}$, 
when the hadron $C(p)$ is a heavy quarkonium $H$.

%% file: sec3.tex
\section{The fragmentation functions}
\label{sec:frag}

Determination of the perturbative short distance functions in Eq.~(\ref{eq:pqcd_fac0}) requires 
the perturbative calculations of both the partonic cross sections on the left of the equation, and partonic fragmentation functions on the right.   The factorization assures that the partonic cross sections and the fragmentation functions share the same non-perturbative long-distance dynamics.  The precise forms of the partonic hard parts in Eq.~(\ref{eq:pqcd_fac0})  depend on the operator definitions of the fragmentation functions.  The operator definition for a single parton of flavor $f$ to fragment into hadrons was first introduced by Collins and Soper \cite{Collins:1981uw}.   This definition was further discussed in the context of heavy quarkonium production in Ref.~\cite{Nayak:2005rt,Nayak:2005rw}.   In this section, we derive operator definitions of fragmentation functions, ${\cal D}_{[Q\bar{Q}(\kappa)]\to H}(z,u,v;m_Q)$, for a heavy quark pair $[Q\bar{Q}(\kappa)]$ of quantum number $\kappa$ including both color and spin of the pair, to fragment into a heavy quarkonium $H$.

To derive explicit operator definitions for the heavy quark pair fragmentation functions, we need to identify the leading power contribution to heavy quarkonium production in the heavy-quark pair fragmentation channel, as sketched in the right diagram in Fig.~\ref{fig:fac1}.  We first perform a collinear expansion of all the momenta of the the active heavy quark and antiquark in the partonic part (below the dashed line) to reduce the four-dimensional momentum convolutions to one-dimensional convolutions of light-cone momentum fractions.  Then we factorize color and spin indices between the partonic hard parts that produce the pair and the fragmentation functions that describe the transformation of the produced pair to a bound quarkonium.  From this process, we drive explicit cut-vertices defining the $[Q\bar{Q}(\kappa)]$-fragmentation functions, and the corresponding projection operators to define the partonic hard part for producing the pair.

We write the heavy quark pair fragmentation channel of heavy quarkonium production in Fig.~\ref{fig:fac1} as 
\begin{eqnarray}
d\sigma_{AB\to [Q\bar{Q}]\to H(P)} 
&=&
\int \frac{d^4p_c}{(2\pi)^4} \frac{d^4q_1}{(2\pi)^4} \frac{d^4q_2}{(2\pi)^4}
\bigg[
\hat{\cal H}_{AB\to [Q\bar{Q}]}(P_Q, P_{\bar{Q}}, P'_Q, P'_{\bar{Q}})
\label{eq:QQxsection} \\
&\ & \hskip 1.4in
\times
\hat{\cal T}_{[Q\bar{Q}] \to H(P)}(P_Q, P_{\bar{Q}}, P'_Q, P'_{\bar{Q}}; P)
\bigg]
dPS(P)\, ,
\nonumber
\end{eqnarray}
where $\hat{\cal H}$ and $\hat{\cal T}$ represent the process below and above the dashed line 
in Fig.~\ref{fig:fac1}, respectively.
The dashed line in the figure represents color and spinor traces 
between $\hat{\cal H}$ and $\hat{\cal T}$,
and $dPS(P)$ denotes the differential phase space of the observed final-state.  
The function $\hat{\cal H}$ includes all necessary initial-state factors needed for calculating 
the cross section, including parton distributions if $A$ and $B$ are hadrons.
In Eq.~(\ref{eq:QQxsection}), heavy quark and antiquark momenta are expressed 
in terms of three independent momenta, $p_c$, $q_1$, and $q_2$,
\begin{eqnarray}
P_Q = \frac{p_c}{2} + q_1\, ,
& \quad\quad &
P_{\bar{Q}}  = \frac{p_c}{2} - q_1\, ,
\nonumber\\
P'_Q  = \frac{p_c}{2} + q_2\, ,
& \quad\quad &
P'_{\bar{Q}}  = \frac{p_c}{2} - q_2\, ,
\label{eq:QQmomentum}
\end{eqnarray}
where $p_c=P_Q + P_{\bar{Q}}=P'_Q + P'_{\bar{Q}}$ is the total momentum of the pair, 
while $q_1=(P_Q - P_{\bar{Q}})/2$ and $q_2=(P'_Q - P'_{\bar{Q}})/2$ 
represent the relative momenta of the quark and antiquark in the amplitude and 
 complex conjugate  amplitude, respectively.
When the light-cone component of physically observed quarkonium momentum, $p^+$ (or $p_T$ in the Laboratory frame) is  much larger than heavy quark mass, $m_Q$, and larger than the typical virtuality of active quark and antiquark in $\hat {\cal T}$, as discussed in the last section, we can approximate the contribution to the cross section in Eq.~(\ref{eq:QQxsection}) by expanding the momenta of heavy quarks and antiquarks in $\hat{\cal H}$ along the direction of the observed heavy quarkonium, and obtain the leading power term as 
\begin{eqnarray}
d\sigma_{AB\to [Q\bar{Q}]\to H(P)} 
&\approx &
\int dz\, du\, dv
\bigg[
\hat{\cal H}_{AB\to [Q\bar{Q}]}(\hat{P}_Q, \hat{P}_{\bar{Q}}, \hat{P}'_Q, \hat{P}'_{\bar{Q}})
\nonumber\\
&\ & \hskip 1in
\times 
\widetilde{\cal T}_{[Q\bar{Q}] \to H(P)}(z,u,v; P)
\bigg]
dPS(P),
\label{eq:QQxsection_lp}
\end{eqnarray}
where 
\begin{eqnarray}
\hat{P}^\mu_Q &=& \left(\frac{p_c^+}{2}+q_1^+\right)\bar{n}^\mu
=\frac{1}{2}\left(1+\frac{2q_1^+}{p_c^+}\right)p_c^+ \bar{n}^\mu
\equiv \frac{1}{2}\left(1+\zeta_1\right)\frac{p^\mu}{z}
=\frac{u}{z}\, p^\mu\, , 
\nonumber\\
\hat{P}'^\mu_Q &=& 
\left(\frac{p_c^+}{2}+q_2^+\right)\bar{n}^\mu
=\frac{1}{2}\left(1+\frac{2q_2^+}{p_c^+}\right)p_c^+ \bar{n}^\mu
\equiv \frac{1}{2}\left(1+\zeta_2\right)\frac{p^\mu}{z}
=\frac{v}{z}\, p^\mu\, , 
\nonumber\\
\hat{P}^\mu_{\bar{Q}} 
&=&  
\left(\frac{p_c^+}{2}-q_1^+\right)\bar{n}^\mu
=\frac{1}{2}\left(1-\zeta_1\right)\frac{p^\mu}{z}
=(1-u)\frac{p^\mu}{z} = \frac{\bar{u}}{z}\, p^\mu\, ,
\nonumber\\
\hat{P}'^\mu_{\bar{Q}} 
&=&
\left(\frac{p_c^+}{2}-q_2^+\right)\bar{n}^\mu
=\frac{1}{2}\left(1-\zeta_2\right)\frac{p^\mu}{z}
=(1-v)\frac{p^\mu}{z} = \frac{\bar{v}}{z}\, p^\mu\, ,
\label{eq:hq-hat}
\end{eqnarray}
with momentum fractions: $\zeta_1\equiv 2q_1^+/p_c^+$, $\zeta_2\equiv 2q_2^+/p_c^+$,
and $z$, $u$ and $v$, which  are defined in Eq.~(\ref{eq:zzetadef}).  
In Eq.~(\ref{eq:QQxsection_lp}), the collinear heavy quark pair correlation function is
\begin{eqnarray}
\widetilde{\cal T}_{[Q\bar{Q}] \to H}(z,u,v; P)
&=&
\int \frac{d^4p_c}{(2\pi)^4}\, \frac{d^4q_1}{(2\pi)^4}\, \frac{d^4q_2}{(2\pi)^4}\,
\delta\left(z-\frac{p^+}{p_c^+}\right)
\delta\left(u-\frac{1}{2}\left(1+\frac{2q_1^+}{p_c^+}\right)\right)
\nonumber\\
& &
\times\
\delta\left(v-\frac{1}{2}\left(1+\frac{2q_2^+}{p_c^+}\right)\right)
\hat{\cal T}_{[Q\bar{Q}] \to H}(P_Q, P_{\bar{Q}}, P'_Q, P'_{\bar{Q}}; P)\, .
\label{eq:QQfrag0}
\end{eqnarray}
To complete the derivation of exact operator definitions of heavy quark pair fragmentation functions, 
we need to factorize the color and spinor traces between $\hat{\cal H}$ and $\widetilde{T}$ 
in Eq.~(\ref{eq:QQxsection_lp}).

For a pair of produced heavy quarks, the color of the pair can be in either a color singlet ``[1]" or a color octet ``[8]" state with the projection operators proportional to $\delta_{ab}$ and $(t^B)_{ab}$, respectively, where $a,b=1,2 \dots N_c$ are color indices for the heavy quark pair as shown in Fig.~\ref{fig:fac1}, and $t^B$ is the generator in the fundamental representation of the group SU(N$_c$) color.  
We can assume that the function $\tilde {\cal T}$ is a linear combination of singlet and color-averaged octet contributions.
The projection operators that project on these $[Q\bar{Q}]$-fragmentation functions in definite color representation can be taken as
\begin{eqnarray}
{\cal C}_{ab,cd}^{[1]}  &= &
\left[ \frac{\delta_{ab}}{\sqrt{N_c}} \right] \left[ \frac{\delta_{cd}}{\sqrt{N_c}} \right] \, ,
\nonumber\\
{\cal C}_{ab,cd}^{[8]} &= &
\frac{1}{N_c^2-1} \sum_{B}  \left[\sqrt{2} \left(t^B\right)_{ab}\right]
                                             \left[\sqrt{2} \left(t^B\right)_{cd}\right] ,
\label{eq:color-cv}
\end{eqnarray}
and the corresponding color projection operators for the partonic hard part as
\begin{eqnarray}
\widetilde{\cal C}_{ba,dc}^{[1]} &= &
\left[\frac{\delta_{ba}}{\sqrt{N_c}} \right] \left[ \frac{\delta_{dc}}{\sqrt{N_c}} \right] \, ,
\nonumber\\
\widetilde{\cal C}_{ba,dc}^{[8]} &= &
\sum_{A} \left[\sqrt{2} \left(t^A\right)_{ba}\right]\, 
               \left[\sqrt{2} \left(t^A\right)_{dc}\right]\, .
\label{eq:color-pj}
\end{eqnarray}
As required, the color projection operators satisfy the normalization condition,
\begin{equation}
\sum_{abcd} {\cal C}_{ab,cd}^{[I]} \,
\widetilde{\cal C}_{ba,dc}^{[J]} =\delta^{IJ}
\label{eq:color-normal}
\end{equation}
with $I,J=1,8$.  However, the exact coefficients of ${\cal C}$ and $\widetilde{\cal C}$ are not unique 
and any constant factor can be moved between them, as long as they satisfy the normalization 
condition in Eq.~(\ref{eq:color-normal}).  Our choice of normalization here matches the convention often adopted in NRQCD factorization.

The separation of the spinor traces of heavy quarks between the short-distance function $\hat{\cal H}$ 
and the long-distance part $\widetilde{\cal T}$ in Eq.~(\ref{eq:QQxsection_lp}),
is implemented by a Fierz reshuffling of spinor indices.
In the limit $m_Q/p_T \rightarrow 0$, there are only three leading power 
spin projection operators for the produced heavy quark pair in Fig.~\ref{fig:fac1},  
\begin{eqnarray}
(\gamma\cdot p)_{ji}, \quad\quad
(\gamma\cdot p\, \gamma_5)_{ji}, \quad\quad
(\gamma\cdot p\, \gamma_\perp^\alpha)_{ji}, 
\end{eqnarray}
where the superscript ``$\alpha$'' has two independent values.  These three projection operators, up to a choice of normalization factors to be discussed below, cover the total four spin degrees of freedom of the produced heavy quark pair.  They are the vector ($v$), axial vector ($a$), and tensor ($t$) forms of the $\gamma$ matrices, respectively.   The same projection operators also apply to the produced heavy quark pair in the complex conjugate of the scattering amplitude.  Note that, because the heavy quark mass $m_Q$ of the partonic hard parts is set to zero in the hard scattering part ${\cal H}$, the tensor projection gives nonvanishing contributions only if the trace of heavy quark spin in ${\cal H}$ includes both projections $\gamma\cdot p\gamma^\alpha_\perp$, which have even numbers of gamma matrices.  

The choice of coefficients for these spin projection operators is not unique, as long as the projection operators for the production of the heavy quark pair and the corresponding spin projection for the cut-vertices defining the fragmentation functions of the pair are normalized to unity.  
We adopt the following spin projection operators for the partonic hard part, that is,  the part below the dashed line in Fig.~\ref{fig:fac1}, 
\begin{eqnarray}
\widetilde{\cal P}^{(v)}(p)_{ji,kl}
&=& 
\left(\gamma\cdot p\right)_{ji}\,
\left(\gamma\cdot p \right)_{kl}
\, ,
\nonumber\\
\widetilde{\cal P}^{(a)}(p)_{ji,kl}
&=& 
\left(\gamma\cdot p\, \gamma_5 \right)_{ji}\,
\left(\gamma\cdot p\, \gamma_5 \right)_{kl}
\, ,
\nonumber\\
\widetilde{\cal P}^{(t)}(p)_{ji,kl}
&=& 
\sum_{\alpha=1,2}\left(\gamma\cdot p\gamma_\perp^\alpha \right)_{ji}\,
\left(\gamma\cdot p\gamma_\perp^\alpha \right)_{kl}
\, ,
\label{eq:spin-pj}
\end{eqnarray}
which are independent of the momentum fractions of fragmenting quarks and antiquarks. 
Correspondingly, we have spin projections for the cut-vertices that
define the heavy quark pair fragmentation functions,
the part above the dashed line in Fig.~\ref{fig:fac1}, 
\begin{eqnarray}
{\cal P}^{(v)}(p)_{ij,lk}
&=& 
\frac{1}{4p\cdot n}\left(\gamma\cdot n\right)_{ij}\,
\frac{1}{4p\cdot n}\left(\gamma\cdot n\right)_{lk}
\, ,
\nonumber\\
{\cal P}^{(a)}(p)_{ij,lk}
&=& 
\frac{1}{4p\cdot n}\left(\gamma\cdot n\, \gamma_5  \right)_{ij}\,
\frac{1}{4p\cdot n}\left(\gamma\cdot n\, \gamma_5  \right)_{lk}
\, ,
\nonumber\\
{\cal P}^{(t)}(p)_{ij,lk}
&=& 
\frac{1}{2}\sum_{\beta=1,2}
\frac{1}{4p\cdot n}\left(\gamma\cdot n\gamma_\perp^\beta \right)_{ij}\,
\frac{1}{4p\cdot n}\left(\gamma\cdot n\gamma_\perp^\beta \right)_{lk}
\, ,
\label{eq:spin-cv}
\end{eqnarray}
where the light-cone vector $n$ was introduced in the last section to pick up 
the ``+" light-cone momentum component.    
Similar to the color decomposition, the projection operators for spin decomposition satisfy 
a orthonormality condition, 
\begin{equation}
\sum_{ijlk} {\cal P}_{ij,lk}^{(s)}(p)\,
\widetilde{\cal P}_{ji,kl}^{(s')}(p) =\delta^{ss'}
\label{eq:spin-normal}
\end{equation}
with $s,s'=v,a,t$.

Using these projection operators, we can verify our claim that the gluon in the quark pair-single gluon interference diagram, Fig.\ \ref{fig:absent-mix} has no pole in $p_c^2$, so that this contribution can be absorbed into the normal diagonal quark pair fragmentation.    The only projection operators that can contribute to the interference term in Fig.\ \ref{fig:absent-mix} are the vector and axial vector, because the trace of the heavy quark loop in the hard scattering would vanish otherwise (there is no matching tensor projection in the hard part on the right of the cut where the gluon of momentum $p_c$ emerges.)  After integration of the internal loop momenta of the diagram above the dashed line in the figure (at fixed values of the variable $u$ on the left), the only vectors that are left to couple to the gluon are $p_c^\mu$ and $n^\mu$.    Choosing the light-cone gauge, $n\cdot A=0$, the gluon propagator $G^{\mu\nu}(p_c,n)$ is orthogonal to $n^\mu$, while $p_c^\mu G_{\mu}{}^\nu(p_c,n)=n^\nu/p_c\cdot n$, which depends only on the total longitudinal momentum of the quark pair.   If we choose a covariant gauge, the same cancellation follows by applying Ward identities to the hard scattering below the dashed line and to the heavy quark loop above the line.    Note that this argument assumes that the spin of the heavy quark state is not fixed.   

\bef
\includegraphics[width=0.25\textwidth]{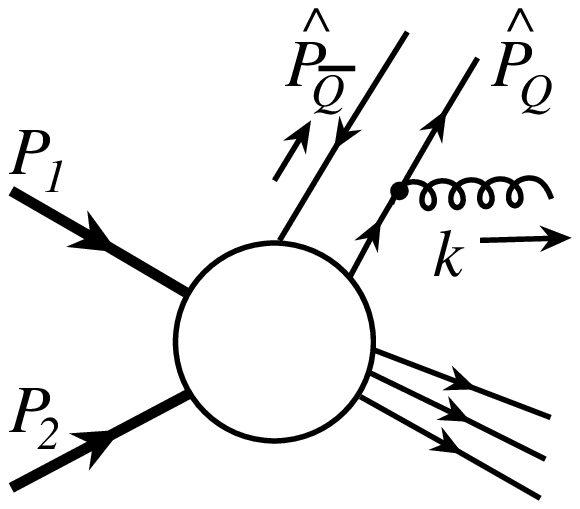}
\caption{Feynman diagram that has an endpoint singularity when the heavy quark momentum $\hat{P}_{Q}$ vanishes. } 
\label{fig:endpoint_pole}
\eef

Before discussing the matrix element realization of the pair fragmentation functions, we will point out a technical issue characteristic of multi-parton factorization expansions, like the one here.    Consider first, the single parton fragmentation channel of heavy quarkonium production, 
on the left of Fig.~\ref{fig:fac1}.   The light-cone momentum of the fragmenting parton (e.g., $P_g^+$) 
is always larger than final state quarkonium momentum, $p^+$.  The same is the case for the total momentum of the fragmenting heavy
quark pair on the right of the figure.   In the heavy quark pair fragmentation channel, however, the light-cone momentum of one of the  
heavy quarks or antiquarks in the process vanishes when $u$, $\bar{u}$, $v$, or $\bar{v}$ vanishes,
even though the total light-cone momentum of the pair $p_c^+$ is always larger than 
the observed quarkonium momentum, $p^+$ in both the amplitude and complex conjugate.
This is illustrated by the correction to the hard function, ${\cal H}$ in Fig.~\ref{fig:endpoint_pole}, in which
the heavy quark, of momentum $(\hat{P}_{Q}+k)^\mu$,
radiates a gluon of momentum $k$ before going on-shell with  momentum
 $\hat{P}_{Q}^\mu=(P_{Q}^+,0,{\bf 0}_T)=((u/z)\, p^+,0,{\bf 0}_T)$.   As appropriate for a line emerging from the hard part, $\hat P^2_Q=0$, and
 \begin{equation}
(\hat{P}_{Q}+k)^2 = 2P_{Q}^+ k^- 
= \frac{u}{z}\, p^+ k^- \, ,
\end{equation}
which vanishes as $u\to 0$.   This produces a $1/{u}$ ``endpoint" singularity in the corresponding contribution to ${\cal H}$ at ${u}\to 0$.  
Similarly, a diagram with the gluon radiated from the heavy antiquark would give an endpoint singularity 
proportional to $1/\bar{u}$.  
In general, taking into account the complex conjugate of the scattering amplitude,
the partonic hard part, when 
calculated by using the projection operators in Eq.~(\ref{eq:spin-pj}), 
will have terms with endpoint singularities proportional to 
$1/u\bar{u}v\bar{v}=1/u(1-u)v(1-v)$, where as many as two of these factors may vanish simultaneously.

Such endpoint singularities are by no means unique to heavy quarkonium production, and appear, for example, in the calculation of exclusive processes \cite{Brodsky:1989pv}.   They will not result in divergences in the production cross sections as long as the heavy quark pair fragmentation functions vanish when the quark and antiquark have very different collinear momenta.  With this in mind, we could choose to move systematically such end point singularities from the partonic hard parts to the corresponding fragmentation functions by using a different combination of spin projection operators 
and cut vertices.    This option is discussed in Appendix~\ref{sec:appendix-spin}, but here we will retain the projections described above.

Using the color and spinor projection operators in Eqs.~(\ref{eq:color-cv}), (\ref{eq:color-pj}), (\ref{eq:spin-pj}), and (\ref{eq:spin-cv}), we separate the color and spinor traces in Eq.~(\ref{eq:QQxsection_lp}), and derive the factorized contribution to heavy quarkonium production from the heavy quark pair fragmentation channel, 
as presented in Eq.~(\ref{eq:pqcd_fac0}),
\begin{eqnarray}
d\sigma_{AB\to [Q\bar{Q}]\to H(P)}
&\approx &
\sum_{[Q\bar{Q}(\kappa)]} \int dz \, du\, dv\
{\cal D}_{[Q\bar{Q}(\kappa)]\to H}(z,u,v;m_Q)
\label{eq:QQ-fac}\\
&\ & 
\times 
d\hat{\sigma}_{A+B\to [Q\bar{Q}(\kappa)](p_c)+X}
\left(
P_Q=up_c^+,P_{\bar{Q}}=\bar{u}p_c^+,
P'_Q=vp_c^+, P'_{\bar{Q}}=\bar{v}p_c^+\right)
\, ,
\nonumber
\end{eqnarray}
where $p_c^+ = p^+/z$, 
and where $[Q\bar{Q}(\kappa)]$ labels the color/spin state of the heavy quark pair
produced at short distances.  We adopt the notation
$\kappa=sI$, $s=v,a,t$ for spin and $I=1,8$ for color. The sum over $\kappa$
runs over all spin and color states of the pair, 
and the factorization scale dependence is suppressed.
The perturbative hard-scattering functions  $d\hat\sigma$ in Eq.~(\ref{eq:QQ-fac}) are constructed from the short-distance functions ${\cal H}$ of Eq.~(\ref{eq:QQxsection_lp}), by
\begin{eqnarray}
&&
d\hat{\sigma}_{A+B\to [Q\bar{Q}(\kappa)](p_c)+X}\left(
P_Q=up_c^+,P_{\bar{Q}}=\bar{u}p_c^+,
P'_Q=vp_c^+,P'_{\bar{Q}}=\bar{v}p_c^+\right)
\ \nonumber\\
&\ & \hskip 0.5in
=\left[
\hat{\cal H}\left(
P_Q=up_c^+,P_{\bar{Q}}=\bar{u}p_c^+,
P'_Q=vp_c^+,P'_{\bar{Q}}=\bar{v}p_c^+\right)
\widetilde{\cal P}^{(s)}(p_c^+))\, \widetilde{\cal C}^{[I]} 
\right]
dPS(p_c) \, , \nonumber\\
\label{eq:QQ-hard}
\end{eqnarray}
where $dPS(p_c)$ is the differential phase space of the produced 
heavy quark pair of momentum $p_c$.
The corresponding heavy quark pair fragmentation functions are given by
\begin{eqnarray}
&\ & {\cal D}_{[Q\bar{Q}(\kappa)]\to H}(z,u,v;m_Q)
=
\left[
z^2\, {\cal P}^{(s)}(p_c)\, {\cal C}^{[I]}\, 
\widetilde{\cal T}_{Q\bar{Q} \to H}(z,u,v;P)
\right]
\label{eq:QQfrag-lg} \\
\ \nonumber\\
&\ & \hskip 0.3in
=
\int \frac{d^4p_c}{(2\pi)^4}\, \frac{d^4q_1}{(2\pi)^4}\, \frac{d^4q_2}{(2\pi)^4}\,
z^2\, \delta\left(z-\frac{p^+}{p_c^+}\right)
\delta\left(u-\frac{1}{2}\left(1+\frac{2q_1^+}{p_c^+}\right)\right)
\nonumber \\
&\ & \hskip 0.5in
\times\
\delta\left(v-\frac{1}{2}\left(1+\frac{2q_2^+}{p_c^+}\right)\right)
\left[{\cal P}^{(s)}(p_c)\, {\cal C}^{[I]}
\hat{\cal T}_{Q\bar{Q} \to H}(P_Q, P_{\bar{Q}}, P'_Q, P'_{\bar{Q}}; P)
\right]
\nonumber\\
\ \nonumber\\
&\ & \hskip 0.3in
=
\int \frac{p^+ dy^-}{2\pi}\, {\rm e}^{-i(p^+/z)y^-}
\int \frac{p^+ dy_1^-}{2\pi}\, {\rm e}^{i(p^+/z)(1-v)y_1^-}
\int \frac{p^+ dy_2^-}{2\pi}\, {\rm e}^{-i(p^+/z)(1-u)y_2^-}
\nonumber \\
&\ & \hskip 0.5in
\times 
{\cal P}^{(s)}_{ij,lk}(p)\, {\cal C}^{[I]}_{ab,cd}\ \sum_X
\langle 0|\overline{\psi}_{c,l}(y_1^-)\psi_{d,k}(0)|H(p)X\rangle
\langle H(p)X| \overline{\psi}_{a,i}(y^-)\psi_{b,j}(y^- + y_2^-)|0\rangle \, ,
\nonumber
\end{eqnarray} 
where the factor $z^2$ in the second equality is a result of changing the final-state phase space 
for producing a heavy quarkonium of momentum $P$ to a heavy quark pair of momentum $p_c$, 
so that $d^3P/E_P\approx z^2 d^3p_c/E_c$.
Like parton distribution functions (PDFs) and multi-parton correlation functions (MPCFs) 
of colliding hadrons, fragmentation functions (FFs) of a single parton,  a heavy quark pair,
or other combination of  partons, are fundamental in QCD.  
Unlike PDFs and MPCFs, FFs are defined in terms of a product of two matrix elements 
(one from scattering amplitude and the other from its complex conjugate).
The moments of FFs are not necessary local, because only final states with a heavy quarkonium are included in the  sum over final states,
$H+X$, in Eq.\ (\ref{eq:QQfrag-lg}).  (In the following, we will suppress the explicit sum over states.)
Fragmentation functions carry the fundamental information of how color neutral hadrons emerge from the colored   parton(s) produced in high energy collisions.

\bef
\includegraphics[width=0.35\textwidth]{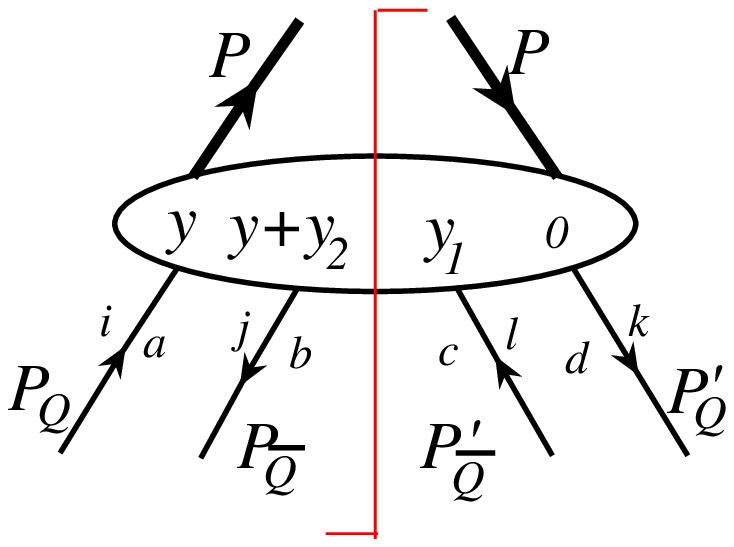}
\caption{Feynman diagram representation of heavy quark pair fragmentation functions.} 
\label{fig:QQfrag}
\eef

The heavy quark pair fragmentation functions ${\cal D}_{[Q\bar{Q}(\kappa)]\to H}(z,u,v;m_Q)$ 
can also be generated by Feynman diagrams in Fig.~\ref{fig:QQfrag} 
in terms of the momentum space cut-vertices,
\begin{eqnarray}
{\cal V}_{[Q\bar{Q}(\kappa)]}(z,u,v) 
&= &
\int \frac{d^4p_c}{(2\pi)^4}\, \frac{d^4q_1}{(2\pi)^4}\, \frac{d^4q_2}{(2\pi)^4}\
{\cal P}^{(s)}_{ij,lk}(p_c)\ {\cal C}^{[I]}_{ab,cd}\
\nonumber\\
&\ & \times
z^2\, \delta\left(z-\frac{p^+}{p_c^+}\right)
\delta\left(u-\frac{1}{2}\left(1+\frac{2q_1^+}{p_c^+}\right)\right)
\delta\left(v-\frac{1}{2}\left(1+\frac{2q_2^+}{p_c^+}\right)\right)\, ,
\label{eq:cv}
\end{eqnarray}
where heavy quark momenta are specified as in Eq.~(\ref{eq:QQmomentum}).  

The heavy quark pair fragmentation functions in Eq.~(\ref{eq:QQfrag-lg}) are appropriate to the light-cone gauge,
but because the quark fields are at different points in space-time, they are not in a gauge invariant form.  
With insertion of gauge links, we obtain an explicitly  gauge invariant definition of our
heavy quark fragmentation functions,
\begin{eqnarray}
&\ & {\cal D}_{[Q\bar{Q}(\kappa)]\rightarrow H}(z,u,v;m_Q)
=
\int \frac{p^+ dy^-}{2\pi}  \frac{p^+ dy_1^-}{2\pi} \frac{p^+ dy_2^-}{2\pi}\, 
\nonumber\\
&\ & \hskip 0.7in
\times 
{\rm e}^{-i(p^+/z)y^-} {\rm e}^{i(p^+/z)(1-v)y_1^-} {\rm e}^{-i(p^+/z)(1-u) y_2^-}
\nonumber\\
&\ & \hskip 0.7in
\times 
{\cal P}^{(s)}_{ij,lk}(p)\, {\cal C}^{[I]}_{ab,cd}\
\langle 0| 
\overline{\psi}_{c',l}(y_1^-)[\Phi^{(F)}_{{n}}(y_1^-)]^{\dagger}_{c'c}
[\Phi^{(F)}_{{n}}(0)]_{dd'}\psi_{d',k}(0)
|H(p)X\rangle
\nonumber\\
&\ & \hskip 0.7in
\times 
\langle H(p)X|
\overline  {\psi}_{a',i}(y^-) [\Phi^{(F)}_{{n}}(y^-)]^\dagger_{a'a}
[\Phi^{(F)}_{{n}}(y^- + y_2^-)]_{bb'} \psi_{b',j}(y^- + y_2^-)
|0\rangle \, ,
\label{eq:QQfrag}
\end{eqnarray}
where all fields are located on the light-cone in the $n$-direction, with zero ``+" and ``$\perp$'' components, 
and where repeated indices are summed.
In Eq.~(\ref{eq:QQfrag}), the gauge link in the matrix  color representation $j=F,A$ (for fundamental and adjoint, respectively) is given by
\begin{equation}
\Phi^{(j)}_{{n}}(y^-) = {\cal P} \exp \left[ -ig\int_{y^-}^\infty d\lambda\, {n}\cdot A^{(j)}({n}\lambda)\right]\, ,
\label{eq:gaugelink}
\end{equation}
where ${\cal P}$ denotes path ordering and $A^{(j)}$ is the gauge field in the representation $j$.   
In Eq.~(\ref{eq:QQfrag}), the heavy quark pair fragmentation functions are defined with gauge links 
in fundamental representation.  Such detailed gauge links are a universal feature of fragmentation functions involving colored partons \cite{Collins:1981uw}, and are required for gauge invariance.  We shall not review arguments for their presence here, except to note that in the terminology of Sec.\ \ref{sec:fac} above they are necessary to match the fragmentation function to the leading pinch surfaces of the diagrams in covariant gauges.  
Eq.~(\ref{eq:QQfrag}) is our operator definition for heavy quark pair fragmentation functions.
 
For the fragmentation of a color singlet heavy quark pair, ${\cal C}^{[1]}$ in Eq.\ (\ref{eq:color-cv}),
the two gauge links for each matrix element 
in Eq.~(\ref{eq:QQfrag}) reduce to a single gauge link between the positions of two quark fields, 
by the identity
\begin{eqnarray}
\left[\Phi_n^{(j)}(y^-)\right]^\dagger_{a'a}\, 
\left[\Phi_n^{(j)}(y^-+y_2^-)\right]_{ab'} = \left[U^{(j)}(y^-,y^-+y_2^-)\right]_{a'b'}\, ,
\end{eqnarray}
where we define a path ordered exponential with arbitrary beginning and endpoints by
\begin{eqnarray}
U^{(j)}(x^-_2,x^-_1) = {\cal P} \exp \, \left[ -ig \int_{x^-_1}^{x^-_2} n\cdot A^{(j)}(n\lambda)\right]\, ,
\end{eqnarray} 
noting that 
\begin{eqnarray}
U^{(j)}(x_2,x_1)^\dagger = U^{(j)}(x_1,x_2)\, .
\label{eq:Udagger}
\end{eqnarray}
For a color-singlet projection, the two path-ordered exponentials thus overlap and cancel each other outside the region between the heavy quark pair.  

The octet projection in Eq.\ (\ref{eq:color-cv}), can be introduced in the final state at $x^-=\infty$, 
or at a finite distance, which we may think of as a three-gauge link ``junction".
In this case, only an adjoint gauge link will connect
the two matrix elements, and should be chosen to
extend to infinity in the $n^\mu$ direction within the matrix elements, to preserve gauge invariance.   
A similar construction is necessary for octet matrix elements in nonelativistic QCD \cite{Nayak:2005rw}.

A general form consistent with gauge invariance
is, for the right-hand matrix element in Eq.\ (\ref{eq:QQfrag}),
\begin{eqnarray}
&\ & 
\langle H(p)X|
\overline  {\psi}_{a',i}(y^-) [\Phi^{(F)}_{{n}}(y^-)]^\dagger_{a'a}
\left( t^C\right)_{ab}\, 
[\Phi^{(F)}_{{n}}(y^- + y_2^-)]_{bb'} \psi_{b',j}(y^- + y_2^-)
|0\rangle
\nonumber\\
&= &
\langle H(p)X|\overline \psi_{a',i}(y^- )
[U^{(F)}_{{n}}(y^- ,x^-)]_{a'a} [U^{(A)}_{{n}}(\infty,x^-)]_{CC'} 
\left( t^{C'}\right)_{ab}
\nonumber\\
&\ &  \hspace{2in} 
\times  \left[ U^{(F)}(x^-,y^-+ y_2^-)\right]_{bb'}
{\psi}_{b',j}(y^-+ y_2^-)|0\rangle
\, ,
\label{eq:links}
\end{eqnarray}
where now adjoint index $C'$ will be summed against the conjugate (left-hand) matrix element.
The junction has been chosen at an arbitrary point $x^-$, and we show below that
the product does not depend on $x^-$.  The identity in Eq.~(\ref{eq:links}) is illustrated 
 in Fig.~\ref{fig:links}.
  
\bef
\includegraphics[width=0.5\textwidth]{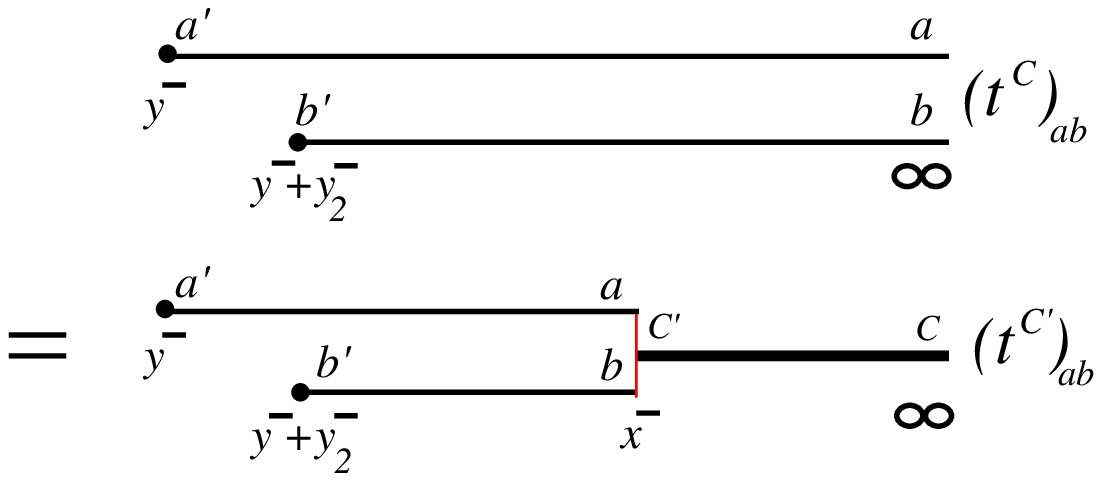}
\caption{A schematic illustration of Eq.~(\ref{eq:links}), where thin lines are for gauge links in the fundamental representation, while the thick line is for the gauge link in the adjoint representation.} 
\label{fig:links}
\eef

Independence of $x^-$ follows from the defining differential equations for an ordered exponential
in the minus direction,
  \begin{eqnarray}
  \frac{\partial}{\partial w^-}\, U^{(j)}(w^-,z^-) &=& -ig t_e^{(j)} A_e^+(w^-)\,  U^{(j)}(w^-,z^-) \, ,
  \nonumber\\
  \frac{\partial}{\partial z^-}\, U^{(j)}(w^-,z^-) &=&   U^{(j)}(w^-,z^-)\, \left[ig t_e^{(j)} A_e^+(z^-)\right]\, ,
  \label{eq:link_deriv}
  \end{eqnarray}
where the product is in group indices of representation $j$.  
    
Equations (\ref{eq:link_deriv})  allow us to evaluate the change of the 
junction product with $x^-$,
\begin{eqnarray}
 \frac{\partial}{\partial x^-}\, 
[U^{(F)}_{{n}}(y^- ,x^-)]_{a'a} [U^{(A)}_{{n}}(\infty,x^-)]_{CC'}
\left( t^{C'}\right)_{ab}
\left[ U^{(F)}(x^-,y^-+ y_2^-)\right]_{bb'}
 &\ &
 \nonumber\\
 &\ & 
\hspace{-120mm}
 = ig  A_E^+(x^-)\, [U^{(F)}_{{n}}(y^- ,x^-)]_{a'a}\, 
 \nonumber\\
 &\ & \hspace{-110mm} \times\ [U^{(A)}_{{n}}(\infty,x^-)]_{CD}\;
 \left( \left[  t^{E}\, ,\, t^{D}\right]_{ab} 
 + \left(T^{E}\right)_{DC'} (t^{C'})_{ab} \right)\;  
 \nonumber\\
 &\ & \hspace{-100mm}\times  U^{(F)}_{bb'}(x^-,y^-)\, 
 \, ,
 \end{eqnarray}
where we have relabeled index $C'$ to $D$ in the first term, and $(T^E)_{DC'}$ in the second term is the generator in the adjoint representation.  We now recall the Lie algebra basics,
 \begin{eqnarray}
 \left[  t^E\, ,\, t^D\right]_{ij} = if^{EDC'} \left(t^{C'}\right)_{ij}
\, ,\quad 
 \left(T^E\right)_{DC'} = -\, if^{EDC'} \, ,
 \end{eqnarray}
to confirm that the derivative vanishes.  As a result, the junction between the gauge links
in fundamental and adjoint representation may be placed anywhere on the light cone.

In summary, for color singlet fragmentation functions, no gauge link is
necessary between the two matrix elements of the heavy quark pair, while for
fragmentation of an octet pair, the matrix elements can be connected by a gauge link
in adjoint representation, as in NRQCD \cite{Nayak:2005rt,Nayak:2005rw}.
In the latter case, the adjoint link connects at each end to a junction of two finite gauge links 
in fundamental representation at an arbitrary location on the light cone.

%% file: sec4ab.tex
Ä
\section{The evolution of fragmentation functions}
\label{sec:evolution}

In this section, we develop a closed set of evolution equations for both single-parton and 
heavy quark pair fragmentation functions, which are necessary to evaluate heavy quarkonium production at collider energies at the accuracy of $1/p_T^2$ power corrections.  We calculate the evolution kernels that appear in these equations to lowest order.

\subsection{The evolution equations}
\label{subsec:evolution}

A QCD factorization formalism for a physical observable, like the one in Eq.~(\ref{eq:pqcd_fac0}), necessarily leads to evolution equations for the variation of the factorization scale. This is because a physical cross section for heavy quarkonium production should not depend on the choice of the factorization scale $\mu$, \begin{eqnarray}
0\ &=&\ \frac{d}{d\ln\mu^2} d\sigma_{A+B\to H+X}(P)
\nonumber\\
&=& \frac{d}{d\ln\mu^2} \left\{ D_{f\rightarrow H}(\mu,m_Q) \otimes d\hat \sigma_{A+B\rightarrow f+X}(\mu,p_T) + \right.
\nonumber\\
&\ & \left. \hspace{20mm}+\ D_{[Q\bar Q(\kappa)]}(\mu,m_Q)\ \otimes\ d\hat \sigma_{A+B\rightarrow {[Q\bar Q(\kappa)]}+X}(\mu,p_T) \right\}\, ,
\label{eq:scale_indep}
\end{eqnarray}
where the symbol $\otimes$ represents  the convolutions of parton momentum fraction shown in Eq.~(\ref{eq:pqcd_fac0}).  
As indicated by the choice of arguments, the fragmentations functions, $D$, are independent of $p_T$ and other kinematic invariants of the hard scattering, and the hard-scattering functions, $d\hat \sigma$, are independent of the heavy quark mass.   They share dependence only on the convolution variables and $\alpha_s(\mu)$.   
By applying the physical condition (\ref{eq:scale_indep}) to the factorization formula in Eq.~(\ref{eq:pqcd_fac0}) perturbatively, 
and demanding that the LP and NLP terms in $1/p_T^2$ vanish, we derive a closed set of evolution equations for the single-parton and heavy quark pair fragmentation functions to this accuracy, as we now show.
\bef
\includegraphics[width=0.4\textwidth]{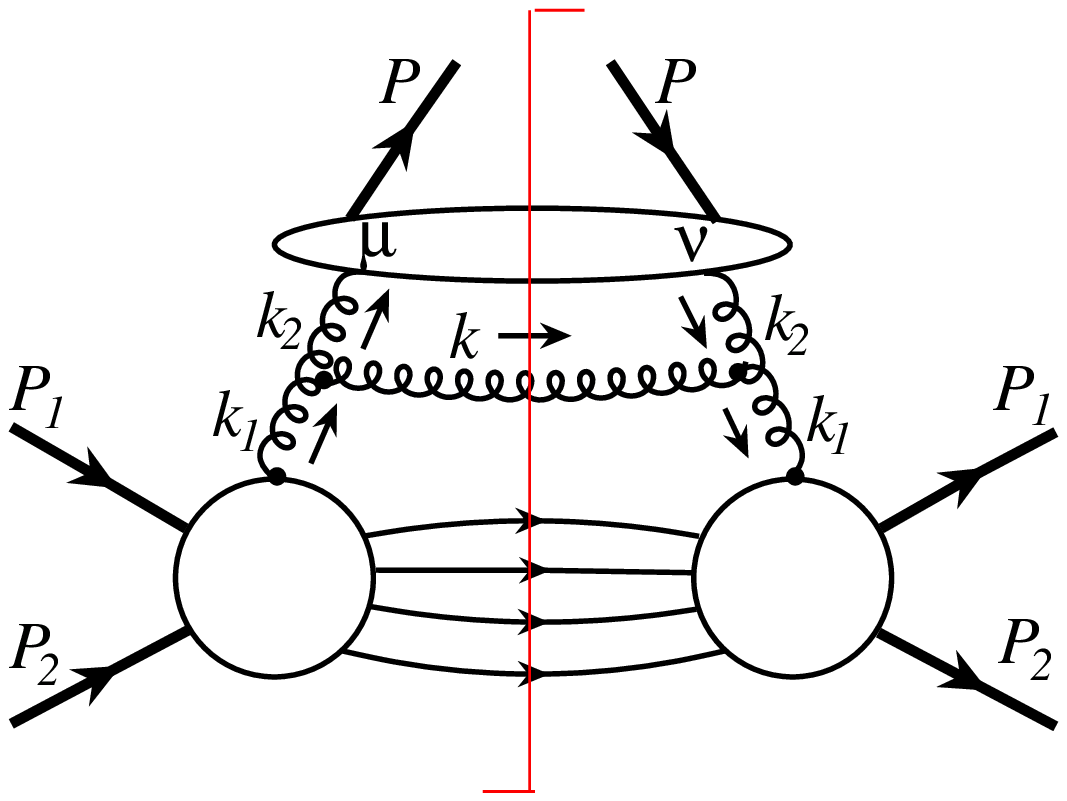}
\hskip 0.8in
\includegraphics[width=0.4\textwidth]{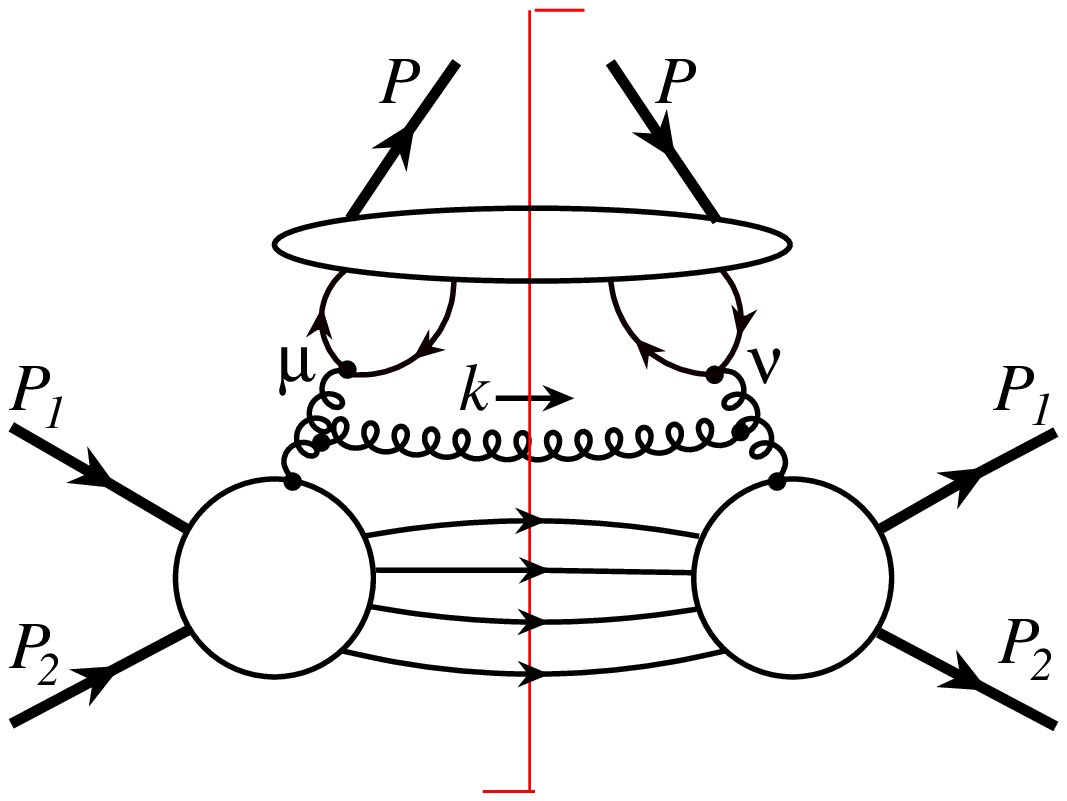}
\caption{Left: A sample Feynman diagram  responsible for the leading logarithmic fragmentation contribution to heavy quarkonium production at the leading power in $1/p_T$.  Right: A sample diagram  that generates power-like collinear divergence, responsible for the creation of heavy quark pair between the distance scale from $1/p_T$ to $1/\mu_0\sim 1/(2m_Q)$, and for the non-linear mixing evolution from a gluon to a fragmenting heavy quark pair, as well as power-suppressed short-distance contributions to the production of heavy quark pair.}
\label{fig:evo-g}
\eef

Consider first the leading power in the $1/p_T^2$ expansion, keeping only the first term 
on the right-hand-side of Eq.~(\ref{eq:pqcd_fac0}).  The physical condition 
in Eq.~(\ref{eq:scale_indep}) requires
\begin{eqnarray}
&& \left[
\frac{\partial}{\partial\ln\mu^2} D_{f\to H}(z,\mu^2;m_Q) \right]
\otimes d\hat{\sigma}_{A+B\to f(p_c)+X}(p_{c}=p/z,\mu^2)
\nonumber\\
&& \hspace{20mm} +\
D_{f\to H}(z,\mu^2;m_Q)
\otimes
\left[
\frac{\partial}{\partial\ln\mu^2} d\hat{\sigma}_{A+B\to f(p_c)+X}(p_{c}=p/z,\mu^2)\right]
= 0\, ,
\label{eq:scale_indep_lp}
\end{eqnarray}
order-by-order in powers of $\alpha_s$. 
In Eq.~(\ref{eq:scale_indep_lp}), the factorization scale, $\mu$-dependence of the short distance function,
$d\hat{\sigma}_{A+B\to f(p_c)+X}$, is a result of subtracting the collinear logarithmic divergences of  the partonic cross section, which are generated by radiation from the fragmenting (single) parton.  For example, the diagram in Fig.~\ref{fig:evo-g}(left) contributes to the production cross section, and has the following logarithmic collinear divergence from the radiation of the final-state gluon,
\begin{equation}
\int^{{\cal O}(p_T^2)} \frac{dk_T^2}{k_T^2} 
\equiv \int_{\mu^2}^{{\cal O}(p_T^2)} \frac{dk_T^2}{k_T^2} 
+ \int^{\mu^2} \frac{dk_T^2}{k_T^2}\, .
\label{eq:frag-log}
\end{equation}
The first term on the right is short-distance in nature and gives the factorization scale 
$\mu$ dependence to $\hat{\sigma}_{A+B\to f(p_c)+X}$ in Eq.~(\ref{eq:scale_indep_lp}), 
while the second term is absorbed into the gluon fragmentation function to a heavy quarkonium.
That is, $\partial d\hat{\sigma}/\partial\ln\mu^2$ is one power of $\alpha_s$ higher than 
 $d\hat{\sigma}$ in Eq.~(\ref{eq:scale_indep_lp}), and perturbatively, we have
\begin{eqnarray}
\frac{\partial}{\partial\ln\mu^2} D_{f\to H}(z,\mu^2;m_Q)
&=&
\sum_{f'} 
\int_z^1 \frac{dz'}{z'}\, D_{f'\to H}(z',\mu^2;m_Q) \
\gamma_{f\to f'}(z/z',\alpha_s)\, ,
\label{eq:evo_dglap}
\end{eqnarray}
where $\sum_{f'}$ runs over all parton flavors and 
$\gamma_{f\to f'}(z/z',\alpha_s)$ are the perturbatively calculable evolution kernels for a parton of flavor $f$ and momentum fraction $z$ to evolve into another parton of flavor $f'$ carrying the momentum fraction $z'$. 
Equation (\ref{eq:evo_dglap}) is the well-known leading power 
DGLAP evolution equation for the fragmentation functions.  Notice that to calculate the kernel at this order we do not need an explicit regularization for the $k^2_T\to 0$ limit, which appears in these calculations only in intermediate steps.
We now show how to extend this reasoning to the next power in $p_T$.

Applying the physical condition in Eq.~(\ref{eq:scale_indep}) to the factorization formalism 
in Eq.~(\ref{eq:pqcd_fac0}), including the NLP term,  we obtain
\begin{eqnarray}
0\ &= & 
D'_{f\to H} \otimes \hat{\sigma}_{A+B\to f(p_c)+X} 
+  {\cal D}'_{[Q\bar{Q}(\kappa)]\to H}
\otimes d\hat{\sigma}_{A+B\to [Q\bar{Q}(\kappa)](p_c)+X}
\nonumber\\
&\ & \hspace{10mm} +\
D_{f\to H} \otimes \hat{\sigma}'_{A+B\to f(p_c)+X}
+ {\cal D}_{[Q\bar{Q}(\kappa)]\to H}
\otimes d\hat{\sigma}'_{A+B\to [Q\bar{Q}(\kappa)](p_c)+X}\, ,
\label{eq:scale_indep_nlp}
\end{eqnarray}
where the prime represents $\partial/\partial\ln\mu^2$, and repeated partonic indices are summed over.   In this expression, the leading-power hard scattering function
$d\hat \sigma_{A+B\rightarrow f+X}(p_T,\mu)$ is already fully determined at leading power, including its $\mu$-dependence, which is specified by Eq.\ (\ref{eq:evo_dglap}), and which we can represent as
\begin{eqnarray}
d\hat{\sigma}'_{A+B\to f+X}\ =\ -\ \gamma_{f'\rightarrow f}\ \otimes\ d\hat{\sigma}_{A+B\to f'+X}\, .
\label{eq:lp-gamma-fprimef}
\end{eqnarray} 
In contrast, the nonleading-power  short distance function in Eq.~(\ref{eq:scale_indep_nlp}), $d\hat{\sigma}_{A+B\to [Q\bar{Q}(\kappa)](p_c)+X}$, which describes the production of a heavy quark pair,  has two types of factorization scale dependence.    

The first source of $\mu$-dependence in $d\hat{\sigma}_{A+B\to [Q\bar{Q}(\kappa)](p_c)+X}$ comes from absorbing collinear logarithmic divergences 
in the evolution of the heavy quark pairs themselves.    This is analogous to the $\mu$-dependence in Eq.\ (\ref{eq:evo_dglap}).   It is proportional to $d\hat \sigma_{A+B\rightarrow [Q\bar Q(\kappa)]}$ and hence its kernel is dimensionless.    The second source of factorization scale dependence is from the production of a heavy quark pair from a light parton, or a single heavy quark or antiquark, and is hence proportional to $d\hat{\sigma}_{A+B\to f+X}$.   As a result, this kernel has dimensions of inverse mass squared.  The only available scale is $\mu$ itself, because the heavy quark distribution shares only this scale with the hard scatterings.    We thus have for the $\mu$-derivatives of quark pair hard scattering functions,
\begin{eqnarray}
d\hat{\sigma}'_{A+B\to [Q\bar{Q}(\kappa)](p_c)+X}\ &=&\  
-\ \Gamma_{[Q\bar{Q}(\kappa')]\to [Q\bar{Q}(\kappa)]}\ \otimes\ d\hat{\sigma}_{A+B\to [Q\bar{Q}(\kappa')]+X}
\nonumber\\
&\ & \hspace{10mm}
-\ \frac{1}{\mu^2} \, \gamma_{f'\rightarrow [Q\bar Q(\kappa)]}\ \otimes\ d\hat{\sigma}_{A+B\to f'+X}\, ,
\label{eq:sigma-var-qqbar}
\end{eqnarray}
where $\gamma_{f'\to [Q\bar Q[\kappa]}$  dimensionless.
To cancel the resulting dependence in Eq.\ (\ref{eq:scale_indep_nlp}), 
we must supplement the leading power evolution equation in Eq.~(\ref{eq:evo_dglap}) by adding a power correction in $\mu^2$, as (first in schematic, then in detailed form),
\ben
\frac{\partial}{\partial\ln\mu^2} D_{f\to H}(z,\mu^2;m_Q)\
&=&\ {D}_{f'\to H}\ \otimes\ \gamma_{f\to f'}\ +\ \frac{1}{\mu^2} \, {\mathcal D}_{[Q\bar{Q}(\kappa')]\to H}\ \otimes\ \gamma_{f\to [Q\bar{Q}(\kappa')]}
\nonumber\\
&=&\
\sum_{f'}     
\int_{z}^1 \frac{dz'}{z'} \,
{D}_{f'\to H}(z', \mu^2;m_Q)\  \gamma_{f\to f'} (z/z',\alpha_s) 
\nnu
& + & 
\frac{1}{\mu^2}\  
\sum_{[Q\bar{Q}(\kappa')]}
\int_{z}^1 \frac{dz'}{z'} \int_{0}^1 du'  \int_{0}^1 dv' \, 
{\mathcal D}_{[Q\bar{Q}(\kappa')]\to H}(z', u',  v', \mu^2; m_Q)
\nnu
&\ &\hskip 1.6in
\times 
\gamma_{f\to [Q\bar{Q}(\kappa')]}(z/z', u', v',\alpha_s) \, ,
\label{eq:evo_1p}
\een
where the mass dimension of $\mu^2$ compensates the dimension of the heavy quark pair fragmentation function ${\cal D}_{[Q\bar{Q}(\kappa')]\to H}$, and where the evolution kernels are process-independent 
and can be calculated perturbatively, as will be done in next subsection.

The pattern just illustrated, in which a pair of partons (in this case heavy quarks) is produced in the course of evolution of single partons (whether quark, antiquark or gluon) is analogous to a similar effect in the evolution of gluon distributions, as analyzed in the context of nuclear shadowing in Refs.~\cite{Mueller:1985wy,Qiu:1986wh,Kang:2007nz}.    By analogy to that case, for very large values of $p_T$, leading-power evolution should dominate the relevant cross section.   At the same time, measured fragmentation functions may be expected to show substantial contributions from lower, but still perturbative, values of factorization scales, where they mix strongly with quark pair fragmentation functions.

\bef
\includegraphics[width=0.4\textwidth]{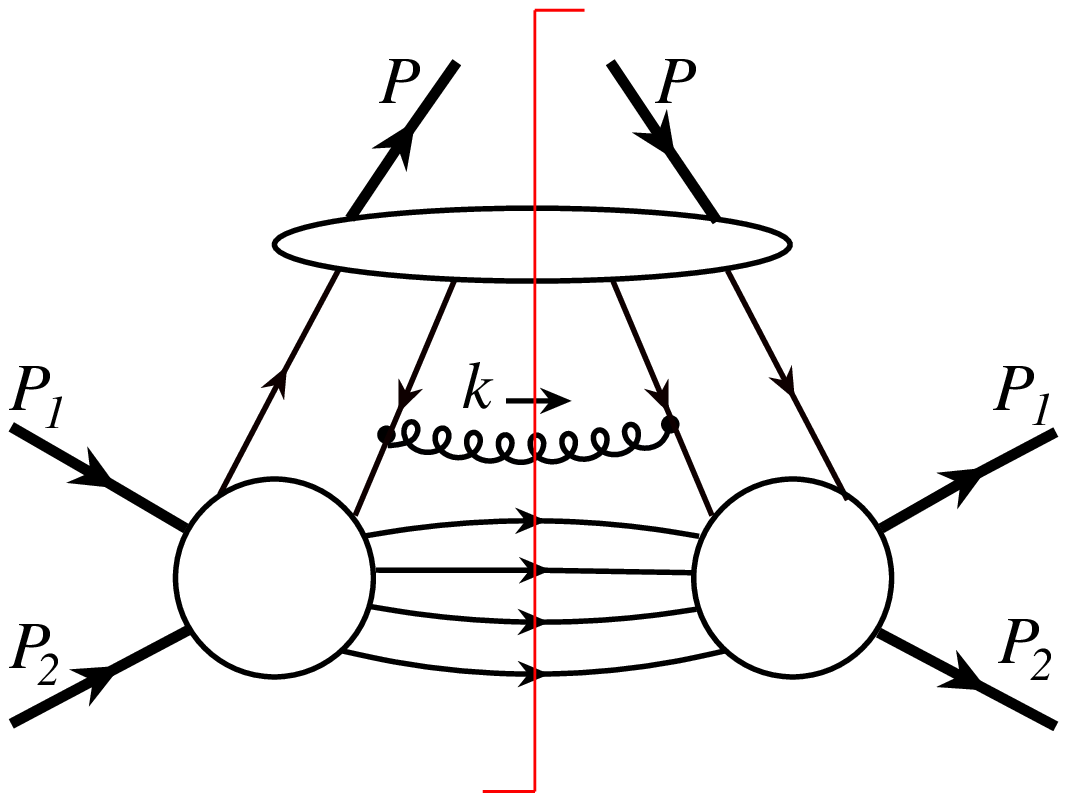}
\hskip 0.8in
\includegraphics[width=0.4\textwidth]{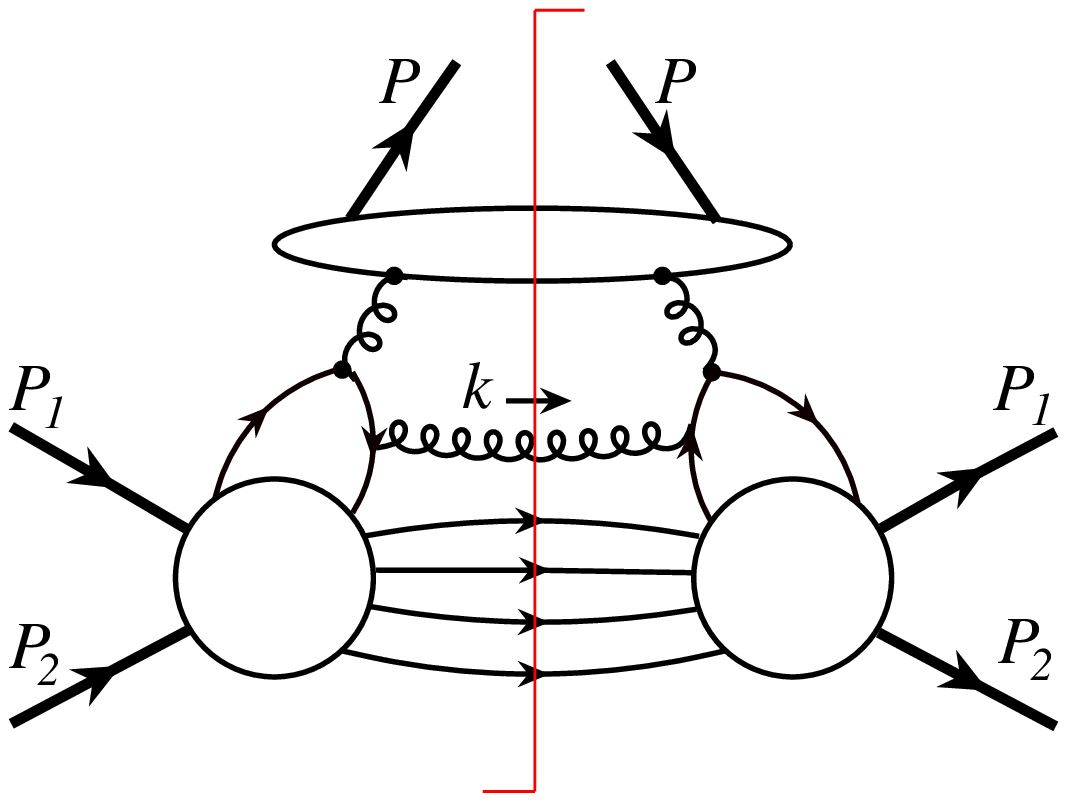}
\caption{Left: A sample Feynman diagram responsible for the leading logarithmic fragmentation contribution to heavy quarkonium production at the next-to-leading power in the $1/p_T$ expansion.  Right: A diagram that gives a correction to the short distance function for producing a single fragmenting gluon.  Because it is free of collinear divergence, this diagram does not generate mixed evolution from a heavy quark pair to a single fragmenting parton. }
\label{fig:evoQQbar}
\eef

In addition to diagrams similar to the one in Fig.~\ref{fig:evo-g}(right), which generate the power-like 
collinear divergence, there are diagrams, like the one in Fig.~\ref{fig:evoQQbar}(left), that contribute to the partonic hard part $d\hat{\sigma}_{A+B\to [Q\bar{Q}(\kappa)](p_c)+X}$ in Eq.~(\ref{eq:scale_indep_nlp}) with logarithmic factorization scale, $\mu$-dependence from subtracting logarithmic collinear divergences generated by radiation from the pair.  Just as in Eq.~(\ref{eq:frag-log}), the logarithmic divergences are absorbed into the nonperturbative fragmentation functions, in this case, into the heavy quark pair fragmentation functions, while the finite first term contributes to the partonic hard part $d\hat{\sigma}_{A+B\to [Q\bar{Q}(\kappa)](p_c)+X}$ in Eq.~(\ref{eq:scale_indep_nlp}).  The derivative $\partial/\partial\ln\mu^2$ on $d\hat{\sigma}_{A+B\to [Q\bar{Q}(\kappa)](p_c)+X}$, when combined perturbatively with the term proportional to the derivative of ${\cal D}_{[Q\bar{Q}(\kappa)]\to H}$ in Eq.~(\ref{eq:scale_indep_nlp}), because of their common overall $1/p_T^2$ dependence, leads to a linear evolution equation for the heavy quark pair fragmentation functions,
\ben
\frac{\partial}{\partial{\ln\mu^2}}{\mathcal D}_{[Q\bar{Q}(\kappa)]\to H}(z, u, v, \mu^2;m_Q)
&=&
\sum_{[Q\bar{Q}(\kappa')]}
\int_{z}^1 \frac{dz'}{z'} 
\int_{0}^1 du'  \int_{0}^1 dv'
\nnu
&\ & \hskip 0.3in
\times \
{\mathcal D}_{[Q\bar{Q}(\kappa')]\to H}(z', u', v', \mu^2;m_Q)
\nnu
&\ & \hskip 0.3in
\times \
\Gamma_{[Q\bar{Q}(\kappa)]\to [Q\bar{Q}(\kappa')]}(z/z', u, v; u', v', \alpha_s) \, ,
\label{eq:evo_2p}
\een
where the evolution kernels, $\Gamma$, are process-independent.   They will be calculated 
to first non-trivial order later in this section.
In Eqs.~(\ref{eq:evo_1p}) and (\ref{eq:evo_2p}), we keep only the first subleading power corrections involving fragmentation of heavy quark pairs, and neglect other power correction terms, as discussed in Sec.~\ref{sec:fac}.  Combining Eqs.~(\ref{eq:evo_1p}) and (\ref{eq:evo_2p}), we have a closed set of evolution equations for single-parton and heavy quark pair fragmentation functions to heavy quarkonium.  

As we shall see, the power dependence in Eq.\ (\ref{eq:sigma-var-qqbar}) arises diagrammatically from subtracting power-like collinear divergences in the partonic cross section, which result from a single parton evolving into a heavy quark pair by radiating a light parton, for example, as illustrated by the diagram in Fig.~\ref{fig:evo-g}(right).   
For the heavy quark pair produced in either a vector, an axial vector, or a tensor state, as discussed in Sec.~\ref{sec:frag} below Eq.~(\ref{eq:spin-normal}), or from explicit calculations below in this section, only the contact term of the gluon propagator in Fig.~\ref{fig:evo-g}(right) contributes to the partonic cross section to produce a pair of heavy quarks.  In the limit $p_T\gg m_Q$, or equivalently neglecting the heavy quark mass, all partonic diagrams similar to Fig.~\ref{fig:evo-g}(right), including those in which the final-state gluon is radiated from either the gluon or the heavy (anti-)quark, have a power-like collinear divergence,
\begin{equation}
\int^{{\cal O}(p_T^2)} \frac{dk_T^2}{(k_T^2)^2} 
\equiv \int_{\mu^2}^{{\cal O}(p_T^2)} \frac{dk_T^2}{(k_T^2)^2} 
+ \int^{\mu^2} \frac{dk_T^2}{(k_T^2)^2}\, .
\label{eq:frag-power}
\end{equation}
Similar to Eq.~(\ref{eq:frag-log}), the first term on the right in Eq.~(\ref{eq:frag-power}) contributes to the partonic hard parts, while the second term is absorbed into the nonperturbative fragmentation function, in this case, into the single gluon fragmentation function.  As in Eq.\ (\ref{eq:frag-power}), we need not regulate the $k_T^2\to 0$ limit, because we need only the dependence on $\mu$ in the short-distance functions.

Taken together, Eqs.\ (\ref{eq:evo_1p}) and (\ref{eq:evo_2p}) control the evolution and mixing of single-parton and quark pair 
fragmentation functions.   As we have seen, they are a direct result of factorization at leading and next-to-leading power, Eq.\ (\ref{eq:pqcd_fac0}).   Mixing proceeds through the evolution of the single-parton fragmentation functions,
which feeds into the heavy pair fragmentation functions at order $1/\mu^2$, while the evolution of the heavy
pair fragmentation function is diagonal.   As noted above, this is the same pattern that is encountered in the evolution equations linking single- to two-gluon
distributions in nuclear shadowing \cite{Mueller:1985wy,Qiu:1986wh,Kang:2007nz}.   
The power correction to single-parton evolution is necessary to organize the production of heavy quark pairs
at time scales   between the short-distance scale $1/p_T$ and the scale associated with the heavy quark mass, $1/m_Q$.
In this evolution, the heavy quark mass serves as a collinear infrared cutoff, so that the single-power correction 
dominates over the entire range of evolution to which perturbation theory can be applied.  \footnote{We also note in passing that these evolution equations, in which the logarithmic derivatives of matrix elements of lower-dimension operators give terms proportional to (``mix with") with matrix elements of higher dimension operators, but not vice-versa (or not from the process like the one in Fig.~\ref{fig:evoQQbar}(right)), is the opposite of the mixing found from the renormalization of composite operators with dimension greater than four.   The matrix elements of such operators will involve positive powers of a UV cutoff $\mu$ in general, and derivatives of these matrix elements will generate matrix elements of operators with lower, rather than higher, dimension.   The essential difference is that the latter are ultraviolet divergences, and the former collinear singularities.}

The combination of the factorization formula in Eq.~(\ref{eq:pqcd_fac0}) and 
the evolution equations in Eq.~(\ref{eq:evo_1p})  organizes  contributions to
the production of heavy quark pairs according to distance scales (or times) where (or when) the
pair was produced.   The first term in Eq.~(\ref{eq:pqcd_fac0}) describes the production of the heavy quark pairs 
at any fixed time after the initial hard collision.   This term behaves as $1/p_T^4$.
The second term describes pair production right at the hard collision, and behaves as $1/p_T^6$.  
It is important to emphasize, however, that the $1/p_T^4$ term includes contributions 
from pairs produced at {\it any fixed} time scale smaller than $1/m_Q$.   
The full evolution equation of the single parton fragmentation function in Eq.~(\ref{eq:evo_1p})
describes how the heavy quark pair is produced at any intermediate, but still perturbative, scale in the fragmentation 
process.      The DGLAP, leading power contribution,
 the first term on the right of Eq.~(\ref{eq:evo_1p}), describes the evolution of the single active parton 
before the creation of the heavy quark pair. The power suppressed second term in Eq.~(\ref{eq:evo_1p}) then
organizes the production of the heavy quark pair at any stage during the evolution.  
Pairs produced at intermediate times then evolve according to Eq.\ (\ref{eq:evo_2p}), in general changing their spin and color through radiation.
If  the evolution equation (\ref{eq:evo_1p}) has only the DGLAP term on the right, 
the evolved single parton fragmentation function is restricted to the situation 
when the heavy quark pair is produced at the latest times, $1/m_Q$ or beyond.  
The presence of the second term in the evolution equation Eq.~(\ref{eq:evo_1p}),
 the mixing term,  completes the picture by allowing the production of heavy quark pairs  between the time scale of the initial hard collision, $1/p_T$ 
and the time scale of $1/m_Q$.

\subsection{Kernels of mixed evolution}
\label{sec:frag_mixing}

In this subsection, we present our calculation of the new evolution kernels,
$\gamma_{f\to [Q\bar{Q}(\kappa')]}(z/z',u',v',\alpha_s)$, 
which are responsible for the evolution of a single parton of flavor $f$ to a heavy quark pair, 
at the first non-trivial order in $\alpha_s$.  

We will extract evolution kernels below by evaluating the factorization scale dependence of 
parton fragmentation functions.  Because evolution kernels are perturbative, 
we can derive them by studying the scale dependence 
of parton fragmentation functions to  partonic states.
More specifically, for extracting the kernel, $\gamma_{f\to [Q\bar{Q}(\kappa')]}(z/z',u',v',\alpha_s)$, 
we apply the factorized evolution equations in Eq.~(\ref{eq:evo_1p}) to  states of 
a perturbative heavy quark pair,  $H\ =\ [Q\bar{Q}(\kappa')]$
where $\kappa'$ represents a spin and color state of the pair.  
In this way, both single quark and gluon fragmentation functions to a heavy quark pair
can be represented by Feynman diagrams, as shown in Fig.~\ref{fig:frag-1p2QQ}, 
with the following cut vertex, 
\begin{eqnarray}
{\cal V}_q(z) = \int \frac{d^4p_c}{(2\pi)^4}\ z^2 \delta\left(z-\frac{p\cdot n}{p_c\cdot n}\right) 
\left[\frac{1}{N_c}\sum_{i=1}^{N_c}\ \delta_{i'i}\ \frac{\gamma\cdot n}{4p_c\cdot n}\right] 
\label{eq:cv_q}
\end{eqnarray}
for a quark fragmentation function, and cut vertex,
\begin{eqnarray}
{\cal V}_g(z) = \int \frac{d^4p_c}{(2\pi)^4}\ z^2 \delta\left(z-\frac{p\cdot n}{p_c\cdot n}\right) 
\left[\frac{1}{N_c^2-1}\sum_{a=1}^{N_c^2-1}\ \delta_{a'a}\
\left( \frac{1}{2} \widetilde{d}^{\mu\nu}(p_c) \right)\right]
\label{eq:cv_g}
\end{eqnarray}
with
\begin{equation}
\widetilde{d}^{\mu\nu}(p_c)
=-g^{\mu\nu} + \frac{p_c^\mu n^\nu + n^\mu p_c^\nu}{p_c\cdot n} 
-\frac{p_c^2}{(p_c\cdot n)^2}\, n^\mu n^\nu\, ,
\label{eq:cv_g0}
\end{equation}
for a gluon fragmentation function.  In Eqs.~(\ref{eq:cv_q}) and (\ref{eq:cv_g}), 
``$i$" and ``$a$'' are color indices of fragmenting quark and gluon, respectively. 

\bef
\psfig{file=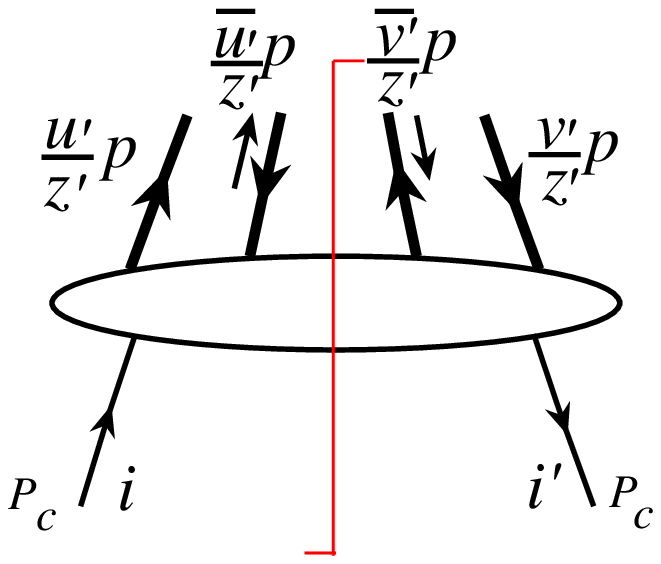, width=1.7in}
\hskip 0.5in
\psfig{file=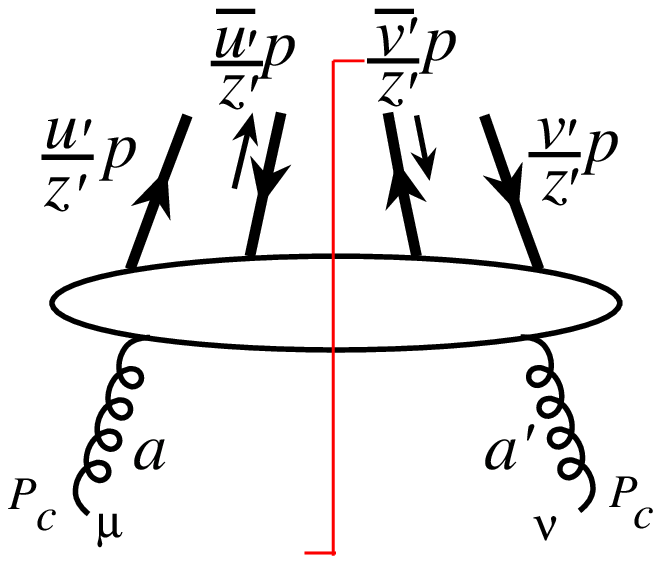, width=1.7in}
\caption{The generic Feynman diagrams for an off-shell single parton (quark or gluon) to 
fragment into a heavy quark pair.}
\label{fig:frag-1p2QQ}
\eef

To calculate the evolution kernels  
$\gamma_{f\to [Q\bar{Q}(\kappa')]}^{(2)}(z/z',u',v')$ with $f=g,q,\bar{q},Q$ and $\bar{Q}$, we first apply 
the evolution equation in Eq.~(\ref{eq:evo_1p}) to the production of a heavy quark pair  
$[Q\bar{Q}(\kappa')](z',u',v')$ with  total momentum $p/z'$, 
where $z'$, $u'$ and $v'$ specify the ``+'' components of quark and antiquark momenta, 
as shown in Fig.~\ref{fig:frag-1p2QQ}.  We then expand both sides of 
Eq.\ (\ref{eq:evo_1p}) to order $\alpha_s^2$,
\ben
\frac{\partial}{\partial\ln\mu^2} D^{(2)}_{f\to [Q\bar{Q}(\kappa')]}(z/z',\mu^2;u',v')
&=&
\int_{z}^{z'} \frac{dz_1}{z_1} 
{D}^{(1)}_{g\to [Q\bar{Q}(\kappa')]}(z_1/z',\mu^2;u',v')\ \gamma^{(1)}_{f\to g} (z/z_1)
\nnu
&\ & \hskip -0.6in
+\ \frac{1}{\mu^2}\ 
\int_{z}^{z'} \frac{dz_1}{z_1} \int_{0}^1 du_1  \int_{0}^1 dv_1 \, 
\label{eq:evo_1pf2}\\
&\ & \hskip - 1 in 
\times \
{\cal D}^{(0)}_{[Q\bar{Q}(\kappa_1)]\to [Q\bar{Q}(\kappa')]}(z_1/z',u_1,v_1; u',  v')\,
\gamma^{(2)}_{f\to [Q\bar{Q}(\kappa_1)]}(z/z_1, u_1, v_1) \, ,
\nonumber
\een
where the superscript ``$(i)$'' with $i=0,1,2$ indicates the power in $\alpha_s$.  
With the zeroth order fragmentation function of the heavy quark pair,
\begin{equation}
{\cal D}^{(0)}_{[Q\bar{Q}(\kappa_1)]\to [Q\bar{Q}(\kappa')]}(z_1/z',u_1,v_1; u',  v')
=\delta^{\kappa_1\,\kappa'}\, 
\delta(1-z_1/z')\, \delta(u_1-u')\, \delta(v_1-v')\, ,
\label{eq:DQQ0}
\end{equation}
we can rewrite Eq.~(\ref{eq:evo_1pf2}) as
\begin{eqnarray}
\frac{1}{\mu^2}\ \gamma^{(2)}_{f\to [Q\bar{Q}(\kappa')]}(z/z', u', v')
&=& 
\frac{\partial}{\partial\ln\mu^2} D^{(2)}_{f\to [Q\bar{Q}(\kappa')]}(z/z',\mu^2;u',v')
\nonumber\\
&-& 
\int_{z}^{z'} \frac{dz_1}{z_1} 
{D}^{(1)}_{g\to [Q\bar{Q}(\kappa')]}(z_1/z',\mu^2;u',v')\ \gamma^{(1)}_{f\to g} (z/z_1)\, ,
\label{eq:f2QQ}
\end{eqnarray}
where $D^{(2)}_{f\to [Q\bar{Q}(\kappa')]}(z/z',\mu^2;u',v')$ is the order $\alpha_s^2$ 
fragmentation function for a single parton of flavor $f$ and momentum fraction $z$ 
to fragment into a heavy quark pair $[Q\bar{Q}(\kappa')]$ with quark and antiquark
momentum fractions $(z',u',v')$, as shown in Fig.~\ref{fig:frag-1p2QQ}. 
For example, the left diagram in Fig.~\ref{fig:frag_q2nd} contributes 
to the light-quark fragmentation function at this order.  
Equation~(\ref{eq:f2QQ}) exhibits how evolution kernels 
are proportional to the variation of the fragmentation functions.  
The second term on the right in Eq.~(\ref{eq:f2QQ})  automatically removes contributions to the variation 
that have been included in the normal DGLAP evolution.  
Higher order corrections to the evolution kernels can be derived systematically 
in the same way by expanding the evolution equation for fragmentation to a heavy quark pair 
to higher order in $\alpha_s$. 

To derive the mixing evolution kernel, 
$\gamma^{(2)}_{f\to [Q\bar{Q}(\kappa')]}(z/z', u', v')$, 
for a single fragmenting parton of flavor $f$ to evolve to a heavy quark pair, 
we need, according to Eq.~(\ref{eq:f2QQ}), to calculate the order $\alpha_s^2$ 
single parton fragmentation function, $D^{(2)}_{f\to [Q\bar{Q}(\kappa')]}(z/z',\mu^2;u',v')$, 
and the order $\alpha_s$ heavy quark pair fragmentation function, 
$D^{(1)}_{g\to [Q\bar{Q}(\kappa')]}(z_1/z',\mu^2;u',v')$ from a gluon of intermediate momentum fraction $z_1$, 
since $\gamma^{(1)}_{f\to g} (z/z_1)$, the first order DGLAP evolution, kernel is known.  

\bef
\psfig{file=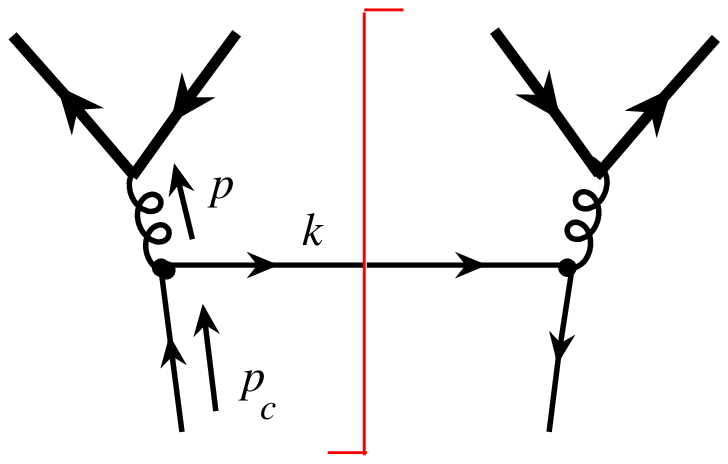, width=1.8in}
\hskip 0.3in
\psfig{file=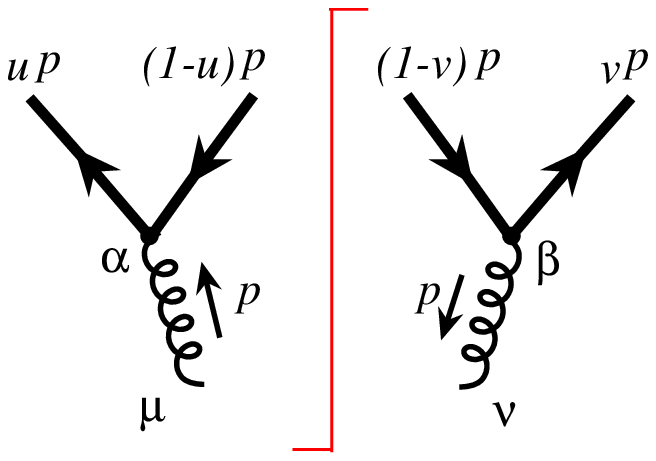, width=1.8in}
\hskip 0.3in
\psfig{file=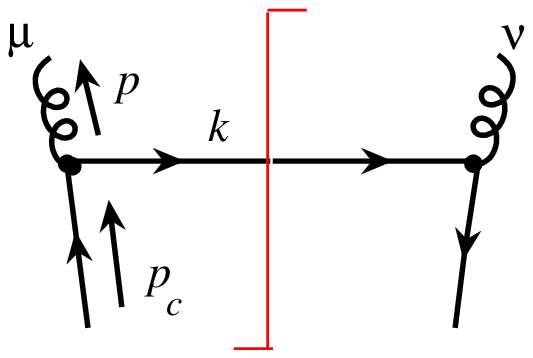, width=1.4in}
\caption{Left:\ Lowest order Feynman diagram ($\alpha_s^2$) for a light quark to fragment into a heavy quark pair.\  Center:\ Lowest order Feynman diagram ($\alpha_s$) for a gluon to fragment into a heavy quark pair.\
Right:\ Lowest order Feynman diagram ($\alpha_s$) for a quark to split into a gluon and a quark.
}
\label{fig:frag_q2nd}
\eef

For the fragmentation of a light quark, we need to evaluate the Feynman diagrams 
in Fig.~\ref{fig:frag_q2nd}, where, without losing generality, we can take $z'=1$, $u'=u$, $v'=v$ and 
$\kappa'=\kappa$ for simplicity of  notation.
The diagram on the left contributes to 
$D^{(2)}_{q\to [Q\bar{Q}(\kappa)]}(z,\mu^2;u,v)$, 
while the diagram in the middle and the one on the right contribute to 
$D^{(1)}_{g\to [Q\bar{Q}(\kappa)]}(z_1,\mu^2;u,v)$ and 
$\gamma^{(1)}_{q\to g} (z/z_1)$, respectively.  
From the decay of a gluon, as shown in Fig.~\ref{fig:frag_q2nd},
we can only have a color octet heavy quark pair with a vector spin projection, $\kappa=v8$, 
which we impose by the operator $\widetilde{\cal P}^{(v)}(p)_{ji,kl}$, Eq.\ (\ref{eq:spin-pj}).
From the diagram on the left in Fig.~\ref{fig:frag_q2nd}, we have in a light-cone gauge,
\begin{eqnarray}
D^{(2)}_{q\to [Q\bar{Q}(v8)]}(z,\mu^2;u,v)
&=& g_s^4\, {\cal C}_g \int \frac{d^4 p_c}{(2\pi)^4}\,
\theta\left(\mu^2-p_c^2\right)
z^2\, \delta\left( z-\frac{p^+}{p_c^+} \right)
{\rm Tr}[\gamma\cdot p\, \gamma^\alpha]\, 
{\rm Tr}[\gamma\cdot p\, \gamma^\beta]\,
\nonumber \\
&\ & \hskip -1.0in \times
\frac{{\cal P}_{\alpha\mu}(p)}{p^2}\,
\frac{{\cal P}_{\beta\nu}(p)}{p^2}\,
{\rm Tr}\left[\frac{\gamma\cdot n}{4p_c^+}\, \frac{\gamma\cdot p_c}{p_c^2}\,
\gamma^\nu\, \gamma\cdot (p_c-p)\, \gamma^\mu\, 
\frac{\gamma\cdot p_c}{p_c^2}\right]
(2\pi)\delta((p_c-p)^2)\, ,
\label{eq:dq2} 
\end{eqnarray}
where the $\mu^2$ indicates the factorization scale dependence of the partonic 
fragmentation function (see more discussion below), $g_s$ is the strong coupling constant,
${\cal C}_g=(N_c^2-1)/(4N_c)$ is the color factor, and 
${\cal P}_{\alpha\mu}(p)$ (or ${\cal P}_{\beta\nu}(p)$) is the gluon's polarization tensor
in $n\cdot A=0$ light-cone gauge,
\begin{equation}
{\cal P}_{\alpha\mu}(p) = - g_{\alpha\mu} + \frac{p_\alpha n_\mu + n_\mu p_\alpha}{p\cdot n}\, .
\label{eq:gluepol}
\end{equation}
Since we take the limit $p^2\to 0$ for all evolution kernels and perturbative hard parts, 
the parton-level fragmentation function $D^{(2)}_{q\to [Q\bar{Q}(v8)]}(z,\mu^2;u,v)$
in Eq.~(\ref{eq:dq2}) has a potential power singularity in  $1/p^2$, which we shall see is absent.  
The same $1/p^2$ singularity also appears in 
$D^{(1)}_{g\to [Q\bar{Q}(v8)]}(z_1,\mu^2;u,v)$ from the middle diagram 
in Fig.~\ref{fig:frag_q2nd}.  The factorization formalism ensures the cancelation
of power singularities in $p^2$ between the two terms in Eq.~(\ref{eq:f2QQ}). 
At this order, such cancelation can be handled analytically, by reorganizing $D^{(2)}_{q\to [Q\bar{Q}(v8)]}(z,\mu^2;u,v)$.  
From Eq.~(\ref{eq:gluepol}), we have
\begin{eqnarray}
{\cal P}_{\alpha\mu}(p) 
&=& \left[ - g_{\alpha\mu} + \frac{p_\alpha n_\mu + n_\mu p_\alpha}{p\cdot n}
				- \frac{p^2}{(p\cdot n)^2}\, n_\alpha\, n_\mu \right]
+  \frac{p^2}{(p\cdot n)^2}\, n_\alpha\, n_\mu\, ,
\label{eq:gluepolt}
\\
&\equiv & 
\widetilde{d}_{\alpha\mu}(p) +  \frac{p^2}{(p\cdot n)^2}\, n_\alpha\, n_\mu
\nonumber
\end{eqnarray}
where $\widetilde{d}_{\alpha\mu}(p)$ is defined as in the cut vertex of Eq.~(\ref{eq:cv_g0}), and represents 
the sum over the gluon's physical polarizations.   It is transverse to both $n^\mu$ and $p^\mu$,
\begin{equation}
   p^\alpha\, \widetilde{d}_{\alpha\mu}(p) = 
   n^\alpha\, \widetilde{d}_{\alpha\mu}(p) = 0\, .
\end{equation}
Using Eq.~(\ref{eq:gluepolt}), we rewrite the gluon propagator as
\begin{equation}
G^{\alpha\mu}(p) 
=\frac{i{\cal P}^{\alpha\mu}(p) }{p^2}
=\frac{i\widetilde{d}^{\alpha\mu}(p)}{p^2} + \frac{i\, n^\alpha\, n^\mu}{(p\cdot n)^2}\, ,
\label{eq:glueprop}
\end{equation}
where the first term is the pole term, proportional to the gluon's physical polarization tensor, and 
the second term is a contact term, or the ``special propagator" 
\cite{Qiu:1988dn},
\begin{equation}
G^{\alpha\mu}_s(p) 
\equiv \frac{i\, n^\alpha\, n^\mu}{(p\cdot n)^2}\, .
\label{eq:glues}
\end{equation}
Also, by construction, the two terms of the gluon propagator in Eq.~(\ref{eq:glueprop})
are orthogonal, $\widetilde{d}_{\alpha\mu}(p) G^{\mu\nu}_s(p)=0$.  

In general, in Eq.~(\ref{eq:dq2}), the term with the apparent $1/p^2$ mass singularity 
is to be exactly canceled by the subtraction term in Eq.~(\ref{eq:f2QQ}), 
and the term with the special gluon propagator, which does not have the mass singularity, 
is the only one that contributes to the short-distance evolution kernel.  
In this case, however,
since ${\rm Tr}[\gamma\cdot p\gamma^\alpha]=4p^\alpha$ 
and $p^\alpha \widetilde{d}_{\alpha\mu}(p)=0$
the terms with the apparent $1/p^2$ mass singularities in Eq.~(\ref{eq:dq2})
vanish,
\begin{eqnarray}
D^{(1)}_{g\to [Q\bar{Q}(v8)]}(z,\mu^2;u,v)
&=& g_s^2\, {\cal C}_g^{(1)} \int \frac{d^4 p_c}{(2\pi)^4}\, 
\theta\left(\mu^2-p_c^2\right)\,
z^2\, \delta\left( z-\frac{p^+}{p_c^+} \right)
\nnu
&\ & \hskip -1.0in  \times 
{\rm Tr}[\gamma\cdot p\, \gamma^\alpha]\, 
{\rm Tr}[\gamma\cdot p\, \gamma^\beta]\,
\frac{{\cal P}_{\alpha\mu}(p_c)}{p_c^2}\,
\frac{{\cal P}_{\beta\nu}(p_c)}{p_c^2}\,
\frac{1}{2}\, \widetilde{d}^{\mu\nu}(p_c)\,
(2\pi)^4 \delta^4(p-p_c)
\nonumber\\
&\ &  \hskip - 1.0in  
= 0 \, ,
\label{eq:Dg2QQ1}
\end{eqnarray}
and similarly for the other terms involving the physical propagator.

By applying Eq.~(\ref{eq:glueprop}), we reexpress
$D^{(2)}_{q\to [Q\bar{Q}(v8)]}(z,\mu^2;u,v)$ in Eq.~(\ref{eq:dq2}) as 
\begin{eqnarray}
D^{(2)}_{q\to [Q\bar{Q}(v8)]}(z,\mu^2;u,v)
&=& g_s^4\, {\cal C}_g \int \frac{d^4 p_c}{(2\pi)^4}\, 
\theta\left(\mu^2-p_c^2\right)\,
z^2\, \delta\left( z-\frac{p^+}{p_c^+} \right)
\nonumber \\
&\ &  \times 
{\rm Tr}[\gamma\cdot p\, \gamma^\alpha]\, 
{\rm Tr}[\gamma\cdot p\, \gamma^\beta]
\left[
\left(  \frac{n_\alpha\, n_\mu}{(p\cdot n)^2}\right)
\left(  \frac{n_\beta\, n_\nu}{(p\cdot n)^2}\right)
\right]
\label{eq:dq2new}\\
&\ & \times
{\rm Tr}\left[\frac{\gamma\cdot n}{4p_c^+}\, \frac{\gamma\cdot p_c}{p_c^2}\,
\gamma^\nu\, \gamma\cdot (p_c-p)\, \gamma^\mu\, 
\frac{\gamma\cdot p_c}{p_c^2}\right]
(2\pi)\delta((p_c-p)^2)\, .
\nonumber\
\end{eqnarray}
Therefore, from Eq.~(\ref{eq:f2QQ}), 
we find the mixing evolution kernel for a light quark to a heavy quark pair,
\begin{eqnarray}
\frac{1}{\mu^2}\ \gamma^{(2)}_{q\to [Q\bar{Q}(v8)]}(z, u, v)
&=& 
\frac{\partial}{\partial\ln\mu^2} D^{(2)}_{q\to [Q\bar{Q}(v8)]}(z,\mu^2;u,v)_c\, ,
\label{eq:q2QQs}
\end{eqnarray}
where the subscript ``$c$'' indicates the  contact term, or special propagator.  
The function $D^{(2)}_{q\to [Q\bar{Q}(v8)]}(z,\mu^2;u,v)_c$ is  the fragmentation function 
for a light quark to produce a heavy quark pair with the mass singularity removed.
It can be represented by the diagram in Fig.~\ref{fig:evo_q2QQ}, 
where the gluon line with a short bar represents the special gluon propagator 
defined in Eq.~(\ref{eq:glues}) \cite{Qiu:1988dn}.
\bef
\psfig{file=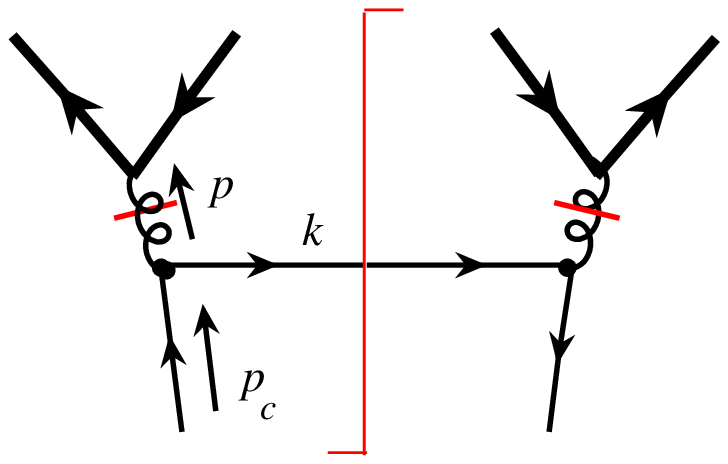, width=1.8in}
\caption{The lowest order ($\alpha_s^2$) contribution to the mixing evolution kernel for a light quark to fragment into a heavy quark pair.  The gluon line with a short bar is given by the special propagator 
in Eq.~(\ref{eq:glues}).
}
\label{fig:evo_q2QQ}
\eef

From Eq.~(\ref{eq:dq2new}), or directly from the diagram in Fig.~\ref{fig:evo_q2QQ}, we have
\begin{eqnarray}
D^{(2)}_{q\to [Q\bar{Q}(v8)]}(z,\mu^2;u,v)_c\
&=& \
\int^{\mu^2} {\hskip -0.08in} 
\frac{dp_c^2}{(p_c^2)^2}\, 
\alpha_s^2 \left[\frac{N_c^2-1}{4N_c}\right]
\left(\frac{64(1-z)}{z^2}\right)\, 
\label{eq:q2QQ2}\, .
\end{eqnarray}
As above, the mixing kernel depends only on the integrand at $p_c^2=\mu^2$, and is independent of the unphysical power singularity at $p_c^2=0$.
We note that we could have used an alternate choice for the factorization scale, as an upper limit in an integration over  $p_{c\perp}^2 = p_c^2 \, (1-z)/z$,
\begin{eqnarray}
D^{(2)}_{q\to [Q\bar{Q}(v8)]}(z,\mu^2;u,v)_c^{(\perp)}\ &=&\
\int^{\mu_{\perp}^2} {\hskip -0.08in}
\frac{dp_{c\perp}^2}{(p_{c\perp}^2)^2}\, 
\alpha_s^2 \left[\frac{N_c^2-1}{4N_c}\right]
\left(\frac{64(1-z)^2}{z^3}\right)\, .
\label{eq:q2QQ2p}
\end{eqnarray}
In either case, it is clear from Eq.~(\ref{eq:q2QQ2}) that 
the fragmentation function for a single light quark to a heavy quark pair  
has a power-like collinear divergence, in contrast to the logarithmic collinear divergence of 
leading twist single parton fragmentation functions.  As noted above,
this is because the single parton and heavy quark pair fragmentation functions have different mass dimension.   The mixing evolution kernels thus have  dimension  $1/{\rm mass}^2$.  Because of the mass dimension, the normalization of this evolution kernel is 
sensitive to the choice of the factorization scale $\mu^2$.  
In Eq.~(\ref{eq:q2QQ2}), we identify the factorization scale as a cutoff
on the invariant mass  of fragmenting quark, $p_c^2$ \cite{Qiu:2001nr,Berger:2001wr}. 
Instead of the invariant mass, we could have used another variable to regularize 
the power collinear divergence, such as the transverse momentum
of the fragmenting quark, $p_{c\perp}^2$, as in Eq.~(\ref{eq:q2QQ2p}).  
Different functional choices of the factorization scale leads to a different $z$-dependence of the 
quark fragmentation to heavy pair fragmentation function, and hence the corresponding mixing evolution kernel given below. 
Since the fragmentation process is kinematically similar to a decay process for the active fragmenting parton, however,
the cut-off on the invariant mass of the fragmenting parton not only regularizes 
the collinear divergence, but also controls the available phase space 
for the fragmentation process and gives the correct threshold behavior if we produce
a massive particle, such as heavy quarkonium \cite{Qiu:2001nr,Berger:2001wr}.
This is the choice we shall make.

In summary, then, from Eq.~(\ref{eq:q2QQ2}), we obtain the mixing evolution kernel for a quark to 
fragment into a vector heavy quark pair with  octet color, 
\begin{equation}
\gamma^{(2)}_{q\to [Q\bar{Q}(v8)]}(z, u, v)
=\alpha_s^2 \left[\frac{N_c^2-1}{4N_c}\right]
\left(\frac{64(1-z)}{z^2}\right)\, ,
\label{eq:evo_q2QQ2}
\end{equation}
when the factorization scale is chosen to be a cutoff on the invariant mass of the fragmenting quark.
Similarly, we find that the  evolution kernel for the fragmentation of a light antiquark 
to a color octet vector heavy quark pair, is equal to that of the corresponding quark,
\begin{eqnarray}
\gamma^{(2)}_{\bar{q}\to [Q\bar{Q}(v8)]}(z, u, v) = \gamma^{(2)}_{q\to [Q\bar{Q}(v8)]}(z, u, v)\, ,
\end{eqnarray}
while the mixing evolution kernels for a light quark to other channels of heavy quark pairs vanish.

\bef
\psfig{file=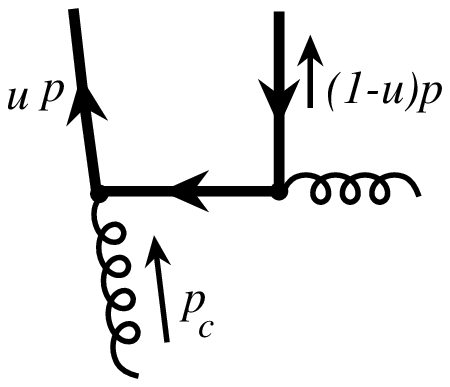, width=1.2in}
\hskip 0.3in
\psfig{file=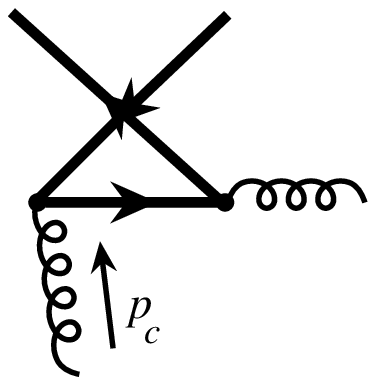, width=1.1in}
\hskip 0.3in
\psfig{file=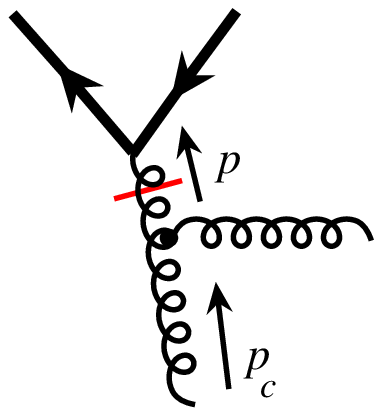, width=1.0in}
\caption{ Diagrams that contribute to the splitting function 
for a gluon to fragment into a heavy quark pair .}
\label{fig:frag_g2qqbar1}
\eef

Similar to the analysis of quark fragmentation after Eq.~(\ref{eq:q2QQs}), following the same reasoning for gluon fragmentation we find,
\begin{equation}
\frac{1}{\mu^2}\, \gamma^{(2)}_{g\to [Q\bar{Q}(\kappa)]}(z, u, v)
=
\frac{\partial}{\partial\ln\mu^2} D^{(2)}_{g\to [Q\bar{Q}(\kappa)]}(z,\mu^2;u,v)_c\, ,
\label{eq:g2QQb}
\end{equation}
where the subscript ``$c$'' again indicates the use of contact terms or special propagators. 
In this case, the order $\alpha_s^2$ fragmentation function for a gluon to a heavy quark pair is 
given by the square of the three diagrams in Fig.~\ref{fig:frag_g2qqbar1}.  
The diagram with a gluon special propagator is necessary for gauge invariance of 
the mixing kernels for a gluon to fragment into a color octet heavy quark pair.  
As an example, we present here the detailed derivation of the kernel 
for a gluon to a color singlet heavy quark pair with the vector spin projection $[Q\bar Q(v1)]$.

The diagram on the right  in Fig.~\ref{fig:frag_g2qqbar1} does not contribute to the production
of a color singlet pair.  We only need to evaluate 
contributions from the other two diagrams.  
From the square of the diagram on the left in Fig.~\ref{fig:frag_g2qqbar1}, we have
\begin{eqnarray}
D^{(2-ll)}_{g\to [Q\bar{Q}(v1)]}(z,\mu^2;u,v)
&=& g_s^4\, {\cal C}_g^{(s)} \int \frac{d^4 p_c}{(2\pi)^4}\, 
\theta\left(\mu^2-p_c^2\right)\,
z^2\, \delta\left( z-\frac{p^+}{p_c^+} \right)\,
\left( \frac{1}{2}\, \widetilde{d}^{\mu\nu}(p_c)\right)
\nonumber\\
&\ & \hspace{-15mm} \times
{\rm Tr}[\gamma\cdot p\, \gamma^\rho\, \gamma\cdot(up-p_c)\gamma^\alpha]\, 
{\rm Tr}[\gamma\cdot p\, \gamma^\beta\, \gamma\cdot(vp-p_c)\gamma^\sigma]\,
\frac{{\cal P}_{\mu\rho}(p_c)}{p_c^2}\,
\frac{{\cal P}_{\sigma\nu}(p_c)}{p_c^2}
\nonumber\\
&\ & \hspace{-15mm} \times
\frac{1}{(up-p_c)^2}\, \frac{1}{(vp-p_c)^2}\,
{\cal P}_{\alpha\beta}(p_c-p)
(2\pi)\delta((p_c-p)^2) 
\nnu
&\ & \hspace{-15mm} =\
\int^{\mu^2} {\hskip -0.08in} \frac{dp_c^2}{(p_c^2)^2}\,
\alpha_s^2 \left[ \frac{1}{4N_c}\right] 
\left[ \frac{8z^2(2u-1)(2v-1)-8z(u+v-1)+4}{(1-u)(1-v)} \right]\, ,
\label{eq:Dg2v1-ll}
\end{eqnarray}
where the superscript ``$(2-ll)$'' indicates the square of the diagram on the left at the order of $\alpha_s^2$.
The color factor for the singlet channel is ${\cal C}_g^{(s)}=1/(4N_c)$.  
After adding contributions that include the crossed diagrams, we derive the partonic fragmentation function 
from a gluon to a color singlet heavy quark pair,
\begin{equation}
D^{(2)}_{g\to [Q\bar{Q}(v1)]}(z,\mu^2;u,v)
=
\int^{\mu^2} {\hskip -0.08in} \frac{dp_c^2}{(p_c^2)^2}
\left(4\alpha_s^2\right) \left[\frac{1}{4N_c}\right]  
\left[z^2+(1-z)^2\right]
\left[ \frac{(u-\bar{u})(v-\bar{v})}{u\,\bar{u}\,v\,\bar{v}} \right]\, ,
\label{eq:Dg2v1}
\end{equation}
where the factor $[1/4N_c]$ represents the color factor, and where as above $\bar{u}=1-u$ and $\bar{v}=1-v$.  
As discussed above, if we choose the factorization scale to be a cutoff on the transverse momentum of fragmenting gluon, $p_{c\perp}^2$, we will have an extra factor $(1-z)/z$ on the right of this expression.
 
From Eqs.~(\ref{eq:g2QQb}) and (\ref{eq:Dg2v1}), we obtain
the mixing evolution kernel for a gluon to fragment into a heavy quark pair of the state ``$\kappa=v1$'',
\begin{equation}
\gamma^{(2)}_{g\to [Q\bar{Q}(v1)]}(z, u, v)
=
\left(4\alpha_s^2\right) \left[\frac{1}{4N_c}\right] 
\left[z^2+(1-z)^2\right]
\left[ \frac{(u-\bar{u})(v-\bar{v})}{u\,\bar{u}\,v\,\bar{v}} \right]\, .
\label{eq:Gg2v1}
\end{equation}
A different choice of the factorization scale would result in a different expression for the evolution kernel in Eq.~(\ref{eq:Gg2v1}), but, the difference would be finite and perturbative, and absorbed into the corresponding fragmentation function.
Evolution kernels for a gluon to fragment into a heavy quark pair 
in other color and spin states are derived similarly, and the results are presented 
in Appendix~\ref{sec:appendix-mix}.

\bef
\psfig{file=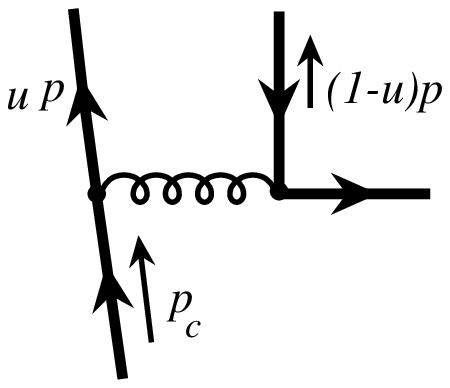, width=1.2in}
\hskip 0.5in
\psfig{file=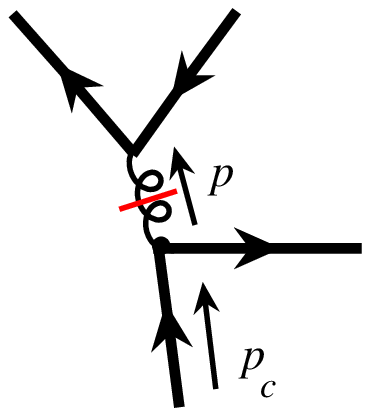, width=1.0in}
\caption{Diagrams that contribute to the mixing evolution kernel for a heavy quark to fragment into a heavy quark pair .}
\label{fig:frag_q2qqbar}
\eef

In addition to the evolution kernels from a light quark and a gluon to a heavy quark pair, 
we also need evolution kernels for a heavy quark (and antiquark) to 
fragment into a heavy quark pair, which could be important when $p_T\gg m_Q$. 
The heavy quark (or antiquark) evolution kernels 
can be derived in the same way,
\begin{equation}
\frac{1}{\mu^2}\, \gamma^{(2)}_{Q\to [Q\bar{Q}(\kappa)]}(z, u, v)
=
\frac{\partial}{\partial\ln\mu^2} D^{(2)}_{Q\to [Q\bar{Q}(\kappa)]}(z,\mu^2;u,v)_c\, ,
\label{eq:HQ2QQb}
\end{equation}
where the subscript ``$c$'' indicates the use of contact terms
(with a bar as shown in Fig.~\ref{fig:frag_q2qqbar}).  
The fragmentation function for a heavy quark to a pair can be derived from 
the square of diagrams in Fig.~\ref{fig:frag_q2qqbar}.  
Differently from the fragmentation of a light quark or a gluon at this order, 
a heavy quark can fragment into a heavy quark pair with a transverse spin 
from square of the diagram on the left in Fig.~\ref{fig:frag_q2qqbar}.
In this case, we find
\begin{eqnarray}
D^{(2)}_{Q\to [Q\bar{Q}(t1)]}(z,\mu^2;u,v)_c
&=& 
g_s^4\, {\cal C}_Q^{(s)} 
\int \frac{d^4 p_c}{(2\pi)^4}\,
\theta\left(\mu^2-p_c^2\right)\,
z^2\, \delta\left( z-\frac{p^+}{p_c^+} \right)\,
(2\pi)\delta((p_c-p)^2)
\nonumber\\
&\ &\times
\frac{1}{4p\cdot n}
{\rm Tr}[\gamma\cdot n\, \gamma\cdot p_c\, \gamma^\nu\, \gamma_\perp^\rho\, \gamma\cdot p\, 
\gamma^\mu\, \gamma\cdot(p_c-p)\gamma^\beta\, \gamma\cdot p\, \gamma_\perp^\sigma\, 
\gamma^\alpha\, \gamma\cdot p_c]
\nnu
&\ &\times
\left(\frac{1}{p_c^2}\right)^2
\frac{{\cal P}_{\alpha\beta}(p_c-up)}{(p_c-up)^2}\,
\frac{{\cal P}_{\mu\nu}(p_c-vp)}{(p_c-vp)^2}\, 
{\cal P}_{\rho\sigma}(p)
\nnu
&=&
\int^{\mu^2} {\hskip -0.08in} \frac{dp_c^2}{(p_c^2)^2}
\alpha_s^2 \left[\frac{C_F^2}{N_c}\right]  
\frac{8(1-z)z^2}{\bar{u}\,\bar{v} (1-zu)(1-zv)}\, ,
\label{eq:DQ2t1}
\end{eqnarray}
where $C_F=(N_c^2-1)/2N_c$ and the color factor ${\cal C}_Q^{(s)} = C_F^2/N_c$.
From Eq.~(\ref{eq:HQ2QQb}), we obtain the mixing evolution kernel for a heavy quark
to fragment into a heavy quark pair with a transverse spin,
\begin{equation}
\gamma^{(2)}_{Q\to [Q\bar{Q}(t1)]}(z, u, v)
=
\alpha_s^2 \left[\frac{C_F^2}{N_c}\right] 
\left[ \frac{8(1-z)z^2}{\bar{u}\bar{v}(1-uz)(1-vz)} \right]\, .
\label{eq:GQ2t1}
\end{equation}
Other mixing evolution kernels from a heavy quark or an antiquark to 
various color and spin states of a heavy quark pair are given in 
Appendix~\ref{sec:appendix-mix}.

%% file: sec4c.tex
\subsection{Kernels for quark pair evolution}

In this subsection, we present the calculation of evolution kernels 
for a heavy quark pair to fragment into another heavy quark pair: 
$\Gamma_{[Q\bar{Q}(\kappa)]\to [Q\bar{Q}(\kappa')]}(z/z',u,v;u',v',\alpha_s)$ 
at order $\alpha_s$.  

Similarly to the calculation of evolution kernels for a single parton
to fragment into a heavy quark pair, 
we apply the evolution equation in Eq.~\eqref{eq:evo_2p} 
to the production of a heavy quark pair $\left[Q\bar{Q}(\kappa')\right]$
of total momentum $p/z'$, 
using the pair as state $H$, 
as shown in Fig.~\ref{fig:frag-QQ2QQ}. 
We then expand the both sides of Eq.~\eqref{eq:evo_2p} to order $\alpha_s$,
\begin{align}
\frac{\partial}{\partial{\ln\mu^2}}
{\mathcal D}^{(1)}_{[Q\bar{Q}(\kappa)]\to [Q\bar{Q}(\kappa')]}(z/z', u, v; u', v';\mu^2)
&= \sum_{[Q\bar{Q}(\kappa_1)]} \int_{z}^{z'} \frac{dz_1}{z_1}
\int_{0}^1 du_1 \int_{0}^1 dv_1 
\nnu 
& \hskip 0.2in 
\times 
{\mathcal D}^{(0)}_{[Q\bar{Q}(\kappa_1)]\to [Q\bar{Q}(\kappa')]}(z_1/z', u_1, v_1;u', v') 
\nnu 
& \hskip 0.2in \times
\Gamma^{(1)}_{[Q\bar{Q}(\kappa)]\to [Q\bar{Q}(\kappa_1)]}(z/z_1, u,v; u_1, v_1)\, .
\end{align} 
Using the zeroth order fragmentation in Eq.~\eqref{eq:DQQ0}, we find
\begin{align}
\Gamma^{(1)}_{[Q\bar{Q}(\kappa)]\to [Q\bar{Q}(\kappa')]}(z/z', u, v; u',v')
=\frac{\partial}{\partial{\ln\mu^2}}
{\mathcal D}^{(1)}_{[Q\bar{Q}(\kappa)]\to [Q\bar{Q}(\kappa')]}(z/z', u, v; u',v';\mu^2)\,.
\label{eq:evo_sol}
\end{align}
That is, the evolution kernels of heavy quark pair fragmentation functions are given by
the variation of the heavy quark pair fragmentation functions with
respect to the factorization scale.  

\bef
\psfig{file=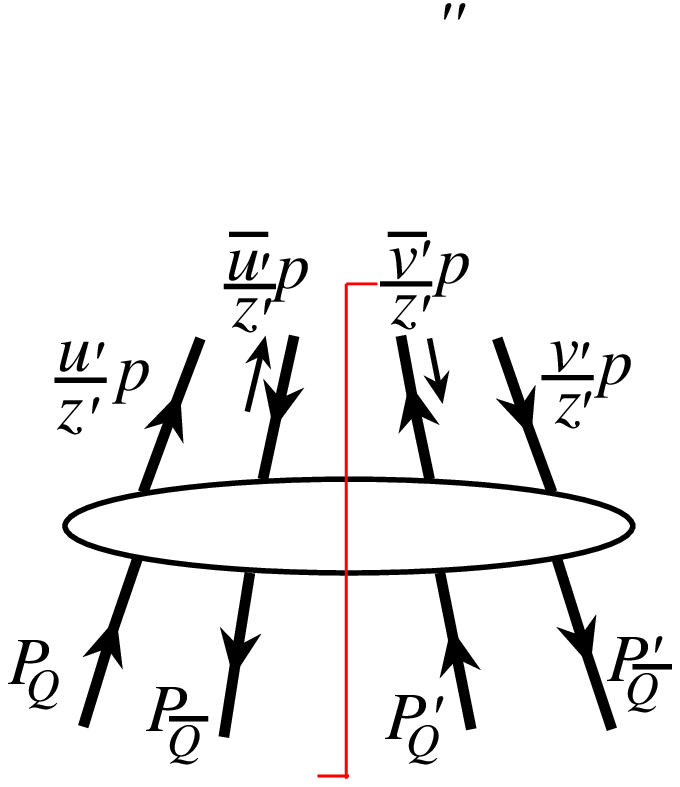, width=1.8in}
\caption{Generic Feynman diagrams for an off-shell pair of heavy quarks to 
fragment into an on-shell heavy quark pair.}
\label{fig:frag-QQ2QQ}
\eef

The heavy quark pair fragmentation functions for a pair of total momentum $p_c=P_Q + P_{\bar{Q}}$ 
with spin-color state $\left[Q\bar{Q}(\kappa)\right]$ to evolve into another pair of total momentum 
$p/z'$ with spin-color state $\left[Q\bar{Q}(\kappa')\right]$ are given by calculating 
the cut diagrams represented in Fig.~\ref{fig:frag-QQ2QQ}.  The bottom of the diagram is contracted 
with the momentum space cut vertex given in Eq.~(\ref{eq:cv}), while the top of the diagram
is contracted with the spin and color projection operators, $\widetilde{\cal P}^{(s)}(p/z')$ and 
$\widetilde{C}^{(I)}$, in Eqs.\ (\ref{eq:spin-pj}) and (\ref{eq:color-pj}), respectively.   Without losing any generality, we can set $z'=1$ in Eq.~(\ref{eq:evo_sol}).

\bef
\psfig{file=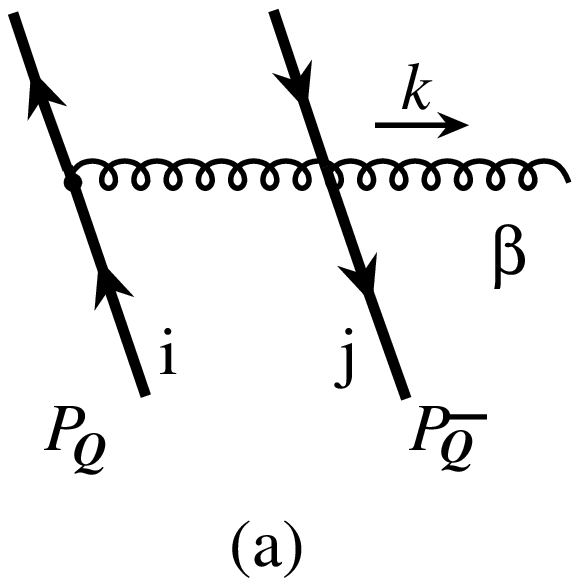, height=1.1in}
\hskip 0.3in
\psfig{file=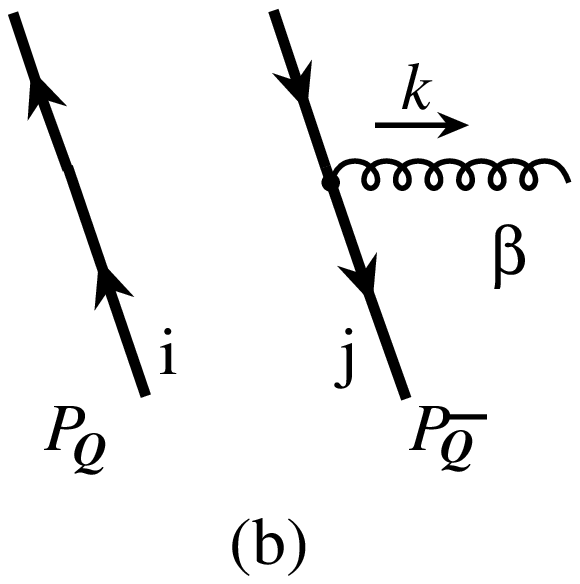, height=1.1in}
\hskip 0.3in
\psfig{file=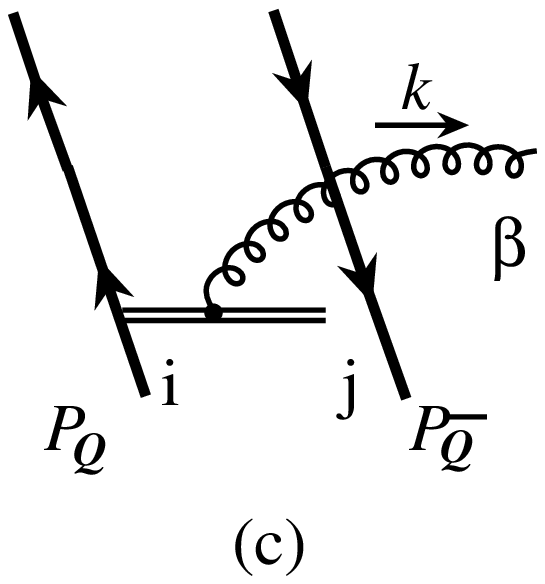, height=1.1in}
\hskip 0.3in
\psfig{file=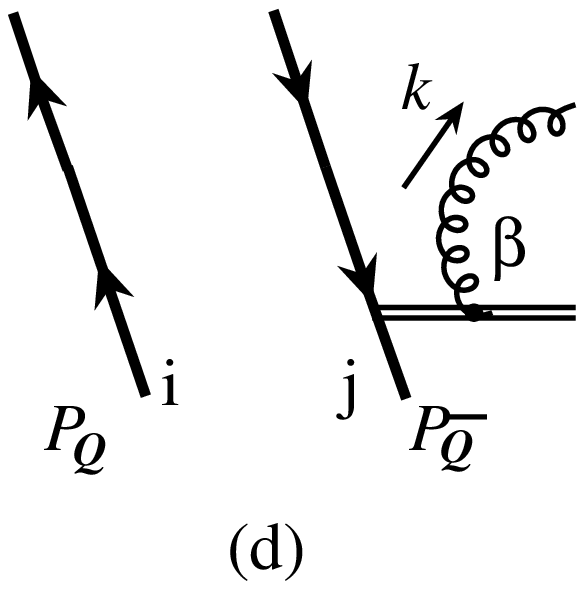, height=1.1in}
\caption{Feynman diagrams with real gluon radiation that contribute to the first order evolution kernels of heavy quark pair fragmentation functions..  The double line represents the eikonal propagator from the ordered exponentials (gauge links).}
\label{fig:evo1-real}
\eef

\bef
\psfig{file=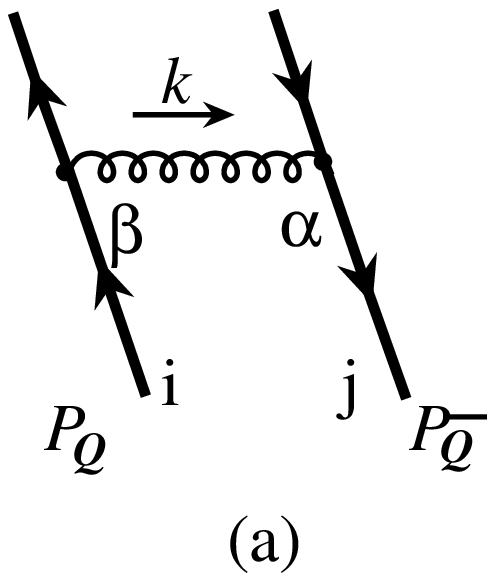, height=1.0in}
\hskip 0.3in
\psfig{file=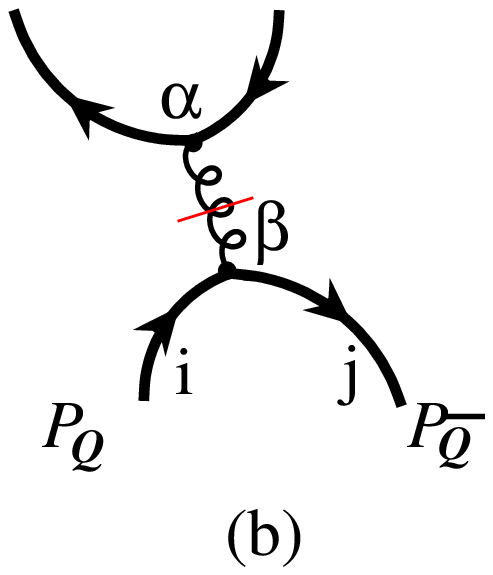, height=1.0in}
\hskip 0.25in
\psfig{file=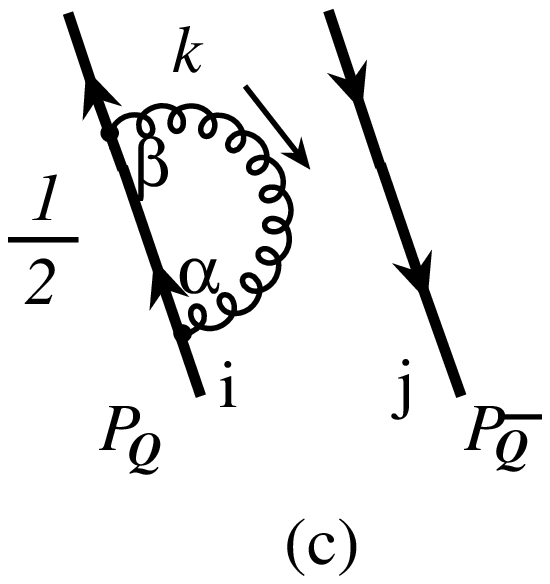, height=1.0in}
\hskip 0.2in
\psfig{file=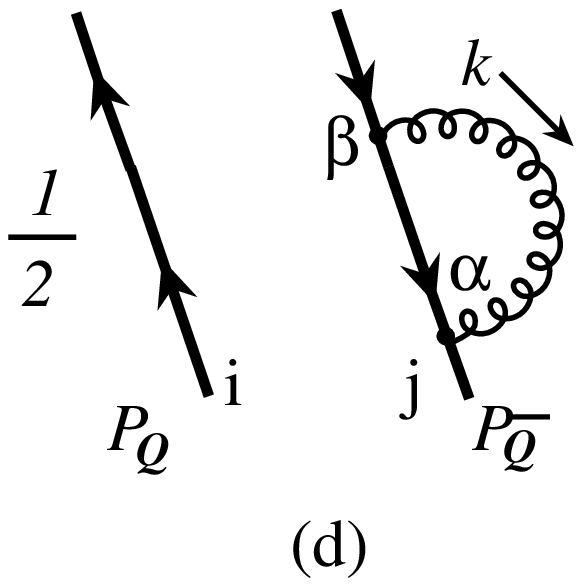, height=1.0in}
\hskip 0.3in
\psfig{file=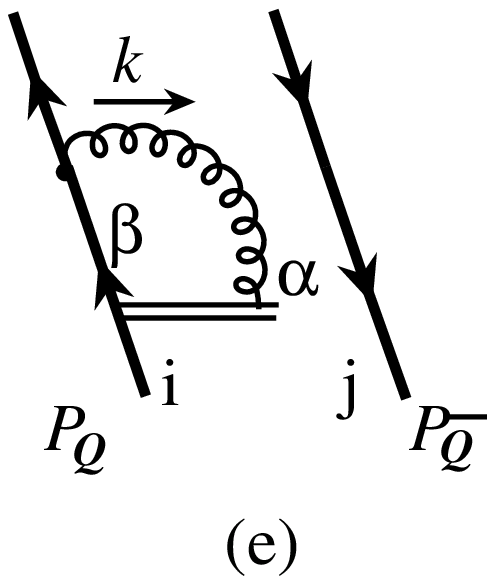, height=1.0in}
\\
\vspace{0.1in}
\hskip 0.1in
\psfig{file=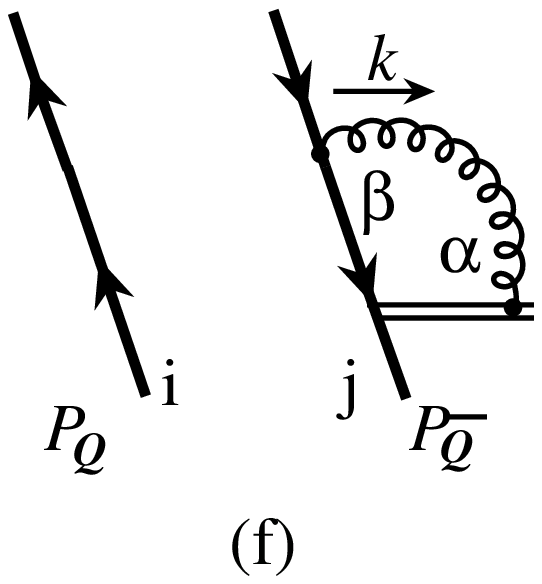, height=1.0in}
\hskip 0.3in
\psfig{file=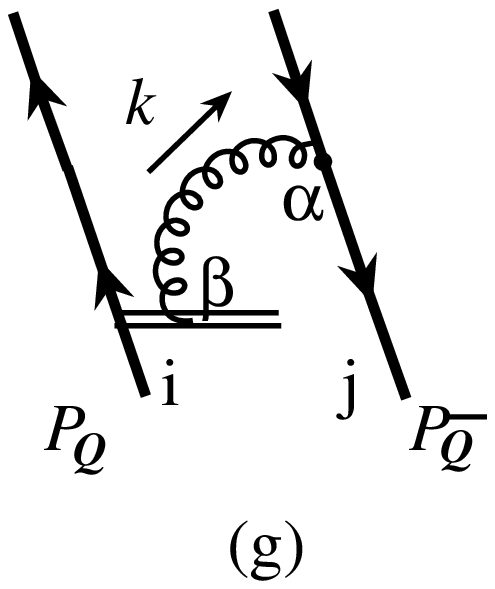, height=1.0in}
\hskip 0.3in
\psfig{file=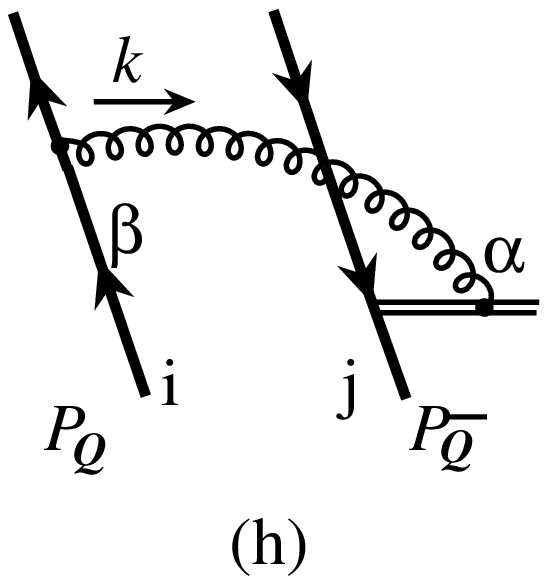, height=1.0in}
\hskip 0.25in
\psfig{file=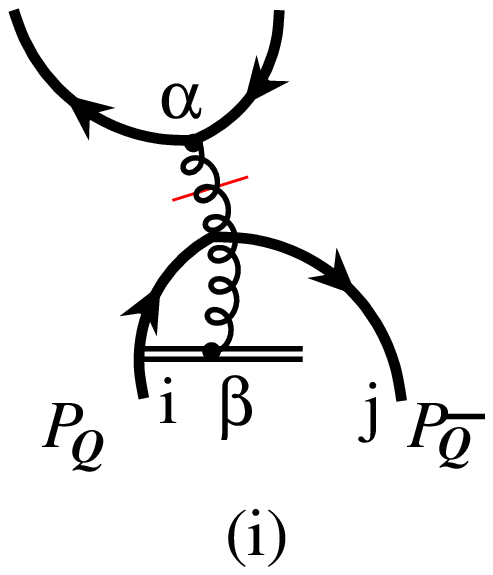, height=1.0in}
\hskip 0.3in
\psfig{file=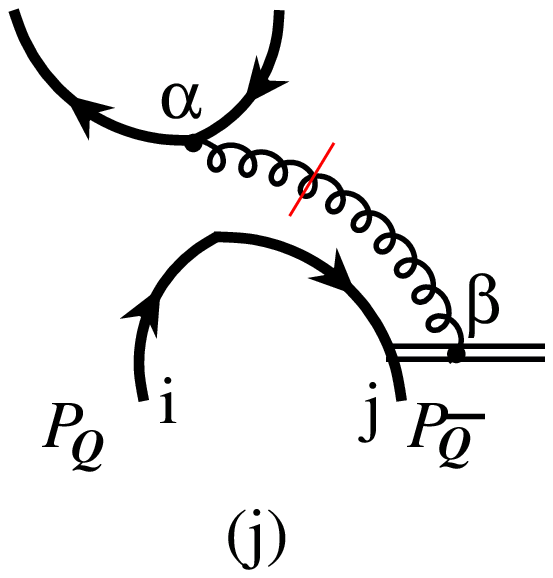, height=1.0in}
\caption{Feynman diagrams 
with a virtual gluon that contribute to the first order evolution kernels 
of heavy quark pair fragmentation functions.}
\label{fig:evo1-virtual}
\eef

At order $\alpha_s$, the heavy quark pair fragmentation function 
can receive contributions from the squares of diagrams with real gluon radiation 
in Fig.~\ref{fig:evo1-real}, as well as from interference between the diagrams with 
a virtual gluon in Fig.~\ref{fig:evo1-virtual} and the lowest order diagram 
(same as the diagram (a) without the gluon).  
Both real and virtual contributions to the parton level fragmentation functions 
have logarithmic ultraviolet and collinear singularities that share 
the same coefficients at this order, while the infrared divergences of these 
diagrams cancel among themselves, as they must, from the factorization.
From Eq.~(\ref{eq:evo_sol}), the evolution kernels 
$\Gamma^{(1)}_{[Q\bar{Q}(\kappa)]\to [Q\bar{Q}(\kappa')]}$ 
can be read off as the coefficients of the logarithmic divergences, and 
do not depend on the regularization and factorization scheme.  
In the following, we describe our calculation for the evolution kernels
by deriving the coefficients of the logarithmic divergences 
from both the real and virtual diagrams 
in Figs.~\ref{fig:evo1-real} and \ref{fig:evo1-virtual}.
As an example, we provide in the remainder of this subsection 
the detailed derivation of the evolution kernel for a heavy quark
pair of quantum numbers $v8$ to another pair with the same quantum
numbers, $v8$.   Calculations for evolution kernels between other quark-antiquark states 
are very similar, and complete results are given in Appendix~\ref{sec:appendix-qqb}.


\bef
\psfig{file=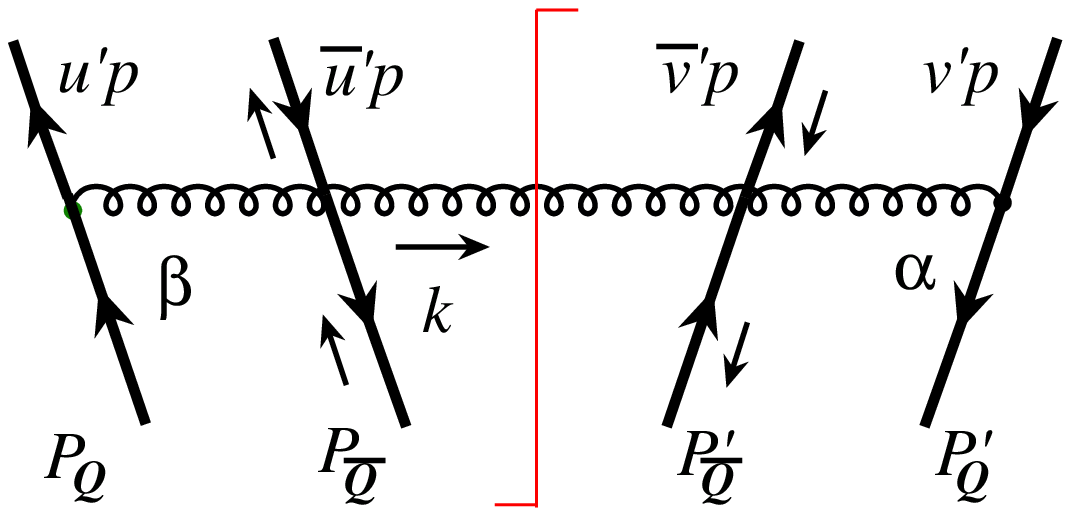, height=1.0in}
\caption{Sample cut diagram with a real gluon that contributes to the 
heavy quark pair fragmentation function at order of $\alpha_s$.}
\label{fig:qqfrag-real1}
\eef
  
In $n\cdot A=0$ light-cone gauge, the diagrams (c) and (d)  in Fig.~\ref{fig:evo1-real} 
do not contribute.   As an example, the square of diagram (a), as sketched in Fig.~\ref{fig:qqfrag-real1}, 
has the expression,
\begin{eqnarray}
{\mathcal D}^{(1,R-aa^\dagger)}_{[Q\bar{Q}(v8)]\to [Q\bar{Q}(v8)]}(z, u, v; u',v';\mu^2)
&=& g_s^2\, C_a 
\int \frac{d^4p_c}{(2\pi)^4}
\frac{d^4q_1}{(2\pi)^4} \frac{d^4q_2}{(2\pi)^4}
\frac{d^4k}{(2\pi)^4}\,
\theta\left(\mu^2-p_{c\perp}^2\right)\,
\nonumber\\
&\ & \hskip -1.6in \times
(2\pi)^4\,\delta^4(p_c-p^+-k)\, 
z^2 \,\delta\left(z-\frac{p^+}{p_c^+}\right)
\delta\left(u-\frac{1}{2}-\frac{q_1^+}{p_c^+}\right)
\delta\left(v-\frac{1}{2}-\frac{q_2^+}{p_c^+}\right)
\nonumber\\
&\ & \hskip -1.6in \times 
(2\pi)^4\delta^{4}\left(\frac{p_c}{2}-q_1-\bar{u}'p^+\right)
(2\pi)^4\delta^{4}\left(\frac{p_c}{2}-q_2-\bar{v}'p^+\right)
(2\pi)\,\delta(k^2) {\cal P}_{\alpha\beta}(k)\, 
\nonumber\\
&\ & \hskip -1.6in \times 
\frac{1}{4p_c^+}{\rm Tr}\left[
\gamma\cdot n\gamma\cdot p \gamma^\beta \gamma\cdot(p_c/2+q_1)\right]
\frac{1}{(p_c/2+q_1)^2 +i\varepsilon}
\nonumber\\
&\ & \hskip -1.6in \times 
\frac{1}{4p_c^+}{\rm Tr}\left[
\gamma\cdot n \gamma\cdot(p_c/2+q_2)\gamma^\alpha\gamma\cdot p \right]
\frac{1}{(p_c/2+q_2)^2 -i\varepsilon}\, ,
\label{eq:evo_r}
\end{eqnarray}
where $\mu^2$ dependence can be a cutoff on either $p_c^2$ or $p_{c\perp}^2$ 
since we are only interested in the coefficient of the logarithmic divergence.   In this expression,
$C_a=(N_c^2-2)/(2N_c)$ is color factor and ${\cal P}_{\alpha\beta}(k) $ 
is the gluon polarization tensor given in Eq.~(\ref{eq:gluepol}).  
Using the $\delta$-functions to fix phase space integration 
over $p_c,q_1$, and $q_2$, and 
\begin{eqnarray}
\int \frac{d^4k}{(2\pi)^4}\, 
\theta\left(\mu^2-k_\perp^2\right)\,
(2\pi)\,\delta(k^2)=
\frac{1}{16\pi^2}
\int^{\mu^2}dk_\perp^2
\int_0^\infty \frac{dk^+}{k^+}\,
\int dk^- \delta\left(k^- - \frac{k_\perp^2}{2k^+}\right)\, ,
\label{eq:evo_rint}
\end{eqnarray}
we obtain,
\begin{eqnarray}
{\mathcal D}^{(1,R-aa^\dagger)}_{[Q\bar{Q}(v8)]\to [Q\bar{Q}(v8)]}(z, u, v;u', v';\mu^2)
&=&
\int^{\mu^2} \frac{dk_\perp^2}{k_\perp^2} 
\left(\frac{\alpha_s}{2\pi}\right)\frac{N_c^2-2}{2N_c}
\left[\frac{u}{u'}+z\right]\left[\frac{v}{v'}+z\right]
\nonumber\\
&\ & \hskip -1.2in \times
\delta(\bar{u}-z\bar{u}')\delta(\bar{v}-z\bar{v}')\,
\frac{z}{2}\int_0^{p^+/z} \frac{dk^+}{k^+}
\delta\left(1-z+\frac{k^+}{p^+/z}\right)\, .
\label{eq:evo_real_aa}
\end{eqnarray}
where the upper limit of the $k^+$-integration is constrained by ${p}^+/{z}$ due to
the delta function. 
The square of the diagram (b) in Fig.~\ref{fig:evo1-real},
and the interference contribution from diagrams (a) and (b), have similar expressions and 
we obtain total real contribution at this order,
\begin{eqnarray}
{\mathcal D}^{(1,R)}_{[Q\bar{Q}(v8)]\to [Q\bar{Q}(v8)]}(z, u, v;u', v';\mu^2)
&=&
\int^{\mu^2} \frac{dk_\perp^2}{k_\perp^2} 
\left(\frac{\alpha_s}{2\pi}\right)
\frac{z}{2}\int_0^{p^+/z} \frac{dk^+}{k^+}
\delta\left(1-z+\frac{k^+}{p^+/z}\right)
\nonumber\\
&\ & \hskip -1.2in \times \bigg\{
\frac{N_c^2-2}{2N_c}
\left[\frac{u}{u'}+z\right]\left[\frac{v}{v'}+z\right]
\delta(\bar{u}-z\bar{u}')\delta(\bar{v}-z\bar{v}')
\nonumber\\
&\ & \hskip -1.1in +
\frac{1}{N_c}
\left[\frac{u}{u'}+z\right]\left[\frac{\bar{v}}{\bar{v}'}+z\right]
\delta(\bar{u}-z\bar{u}')\delta({v}-z{v}')
\nonumber\\
&\ & \hskip -1.1in +
\frac{1}{N_c}
\left[\frac{\bar{u}}{\bar{u}'}+z\right]\left[\frac{v}{v'}+z\right]
\delta({u}-z{u}')\delta(\bar{v}-z\bar{v}')
\nonumber\\
&\ & \hskip -1.1in +
\frac{N_c^2-2}{2N_c}
\left[\frac{\bar{u}}{\bar{u}'}+z\right]\left[\frac{\bar{v}}{\bar{v}'}+z\right]
\delta({u}-z{u}')\delta({v}-z{v}')
\bigg\}
\label{eq:evo_r2}\\
&\ & \hskip -1.2in \equiv
\int^{\mu^2} \frac{dk_\perp^2}{k_\perp^2} 
\left(\frac{\alpha_s}{2\pi}\right)
\left(\frac{1}{2N_c}\right)
S_+ \Delta_{-}^{[8]}\,
\frac{z}{2}\int_0^{p^+/z} \frac{dk^+}{k^+}
\delta\left(1-z+\frac{k^+}{p^+/z}\right),
\nonumber
\end{eqnarray}
where we define
\begin{align}
S_\pm=&
\left[\frac{u}{u'}\pm\frac{\bar{u}}{\bar{u}'}\right]
\left[\frac{v}{v'}\pm\frac{\bar{v}}{\bar{v}'}\right],
\nonumber\\
\Delta_\pm^{[8]}=&
\left\{(N_c^2-2)\left[\delta(u-{z}u^\prime)\delta(v-{z}v^\prime)+
\delta(\bar{u}-{z}\bar{u}^\prime)\delta(\bar{v}-{z}\bar{v}^\prime)\right]\right.
\nonumber\\
&\left.\mp\, 2\left[\delta(u-{z}u^\prime)\delta(\bar{v}-{z}\bar{v}^\prime)+
\delta(\bar{u}-{z}\bar{u}^\prime)\delta(v-{z}v^\prime)\right]\right\}.
\label{eq:def-sdelta}
\end{align}
In Eq.~(\ref{eq:evo_r2}), the $k^+$-integration has a pole at ${z}=1$, 
which corresponds to the infrared divergence of the real contribution 
when the momentum fraction of the radiating gluon vanishes. 
(This is sometimes refered as a rapidity divergence.)  
To make manifest the infrared cancelation between the real and virtual contributions, we 
regularize this $z\to 1$ divergence by separating out a plus distribution,
\begin{eqnarray}
&\ & 
\hspace{-20mm} \int_0^{{p}^+/{z}} \frac{d k^+}{k^+}
\delta\left(1-{z}-\frac{k^+}{{p}^+/{z}}
\right)\nonumber\\
&=&
\int_0^{{p}^+/{z}} \frac{d k^+}{k^+}
\left[\delta\left(1-{z}-\frac{k^+}{{p}^+/{z}}
\right)-\delta\left(1-{z} \right)\right] 
+ \int_0^{{p}^+/{z}}
\frac{d k^+}{k^+}\, \delta\left(1-{z} \right)
\label{eq:regularize}
\\
&\equiv&
\frac{1}{(1-{z})_+} + \delta\left(1-{z} \right) \int_0^{{p}^+}
\frac{d k^+}{k^+}\, .
\label{eq:decom_r}
\end{eqnarray}
To identify the distribution in Eq.~(\ref{eq:decom_r}), we have changed variables  in the first term 
on the right in Eq.~(\ref{eq:regularize}), which is not singular, using $k^+=(1-x)p^+/z$, so that
\begin{eqnarray}
\int_0^{{p}^+/{z}} \frac{d k^+}{k^+}
\left[\delta\left(1-{z}-\frac{k^+}{{p}^+/{z}}
\right)-\delta\left(1-{z} \right)\right] 
&=&
\int_0^1 \frac{dx}{1-x}\left[\delta(x-z) - \delta(1-z)\right]
\nonumber\\
\equiv
\int_0^1 dx\ \frac{\delta(x-z)}{(1-x)_+}
&=&
\frac{1}{(1-{z})_+}\, ,  
\label{eq:def-plus}
\end{eqnarray}
which is the standard plus distribution of $(1-z)$,
with the property
\begin{align}
\int_a^1\frac{dz}{\left(1-z\right)_+} f(z)\equiv - f(1)
\ln\frac{1}{1-a} + \int_a^1\frac{dz}{1-z} \left[f(z)-f(1)\right]\, ,
\end{align}
for a smooth test function $f(z)$.
For the second term in Eq.~\eqref{eq:decom_r}, which is
divergent, we will combine its integrand directly with corresponding terms 
from virtual corrections.

Substituting Eq.~(\ref{eq:decom_r}) into Eq.~(\ref{eq:evo_r2}), 
we obtain a compact expression for the full real contribution,
\begin{eqnarray}
&\ &
{\mathcal D}^{(1,R)}_{[Q\bar{Q}(v8)]\to [Q\bar{Q}(v8)]}(z, u, v;u', v';\mu^2)
\nonumber\\
&\ & \hskip 0.5in 
=
\int^{\mu^2} \frac{dk_\perp^2}{k_\perp^2} 
\left(\frac{\alpha_s}{2\pi} \right)
\left[ \frac{1}{2N_c}\,  
S_+ \Delta^{[8]}_{-} \, \frac{1}{2} \frac{{z}}{(1-{z})_+} 
+ 2C_A \Delta_{0} \int_0^{{p}^+}
\frac{d k^+}{k^+} \right] \, , 
\label{eq:evo_r3}
\end{eqnarray}
where $C_A=N_c$, $S_+$ and $\Delta^{[8]}_{-}$ are given in Eq.~(\ref{eq:def-sdelta}), and 
\begin{eqnarray}
\Delta_{0}= \delta\left(1-{z} \right)\delta\left(u-u'
\right)\delta\left(v-v' \right) \, .
\end{eqnarray}


\bef
\psfig{file=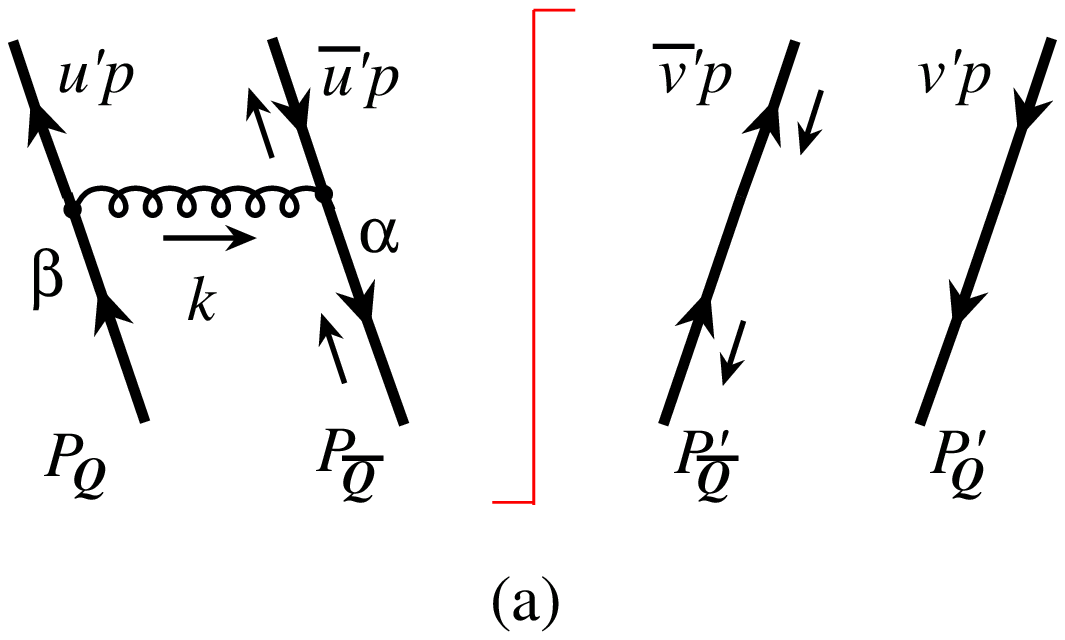, height=1.2in}
\hskip 0.8in
\psfig{file=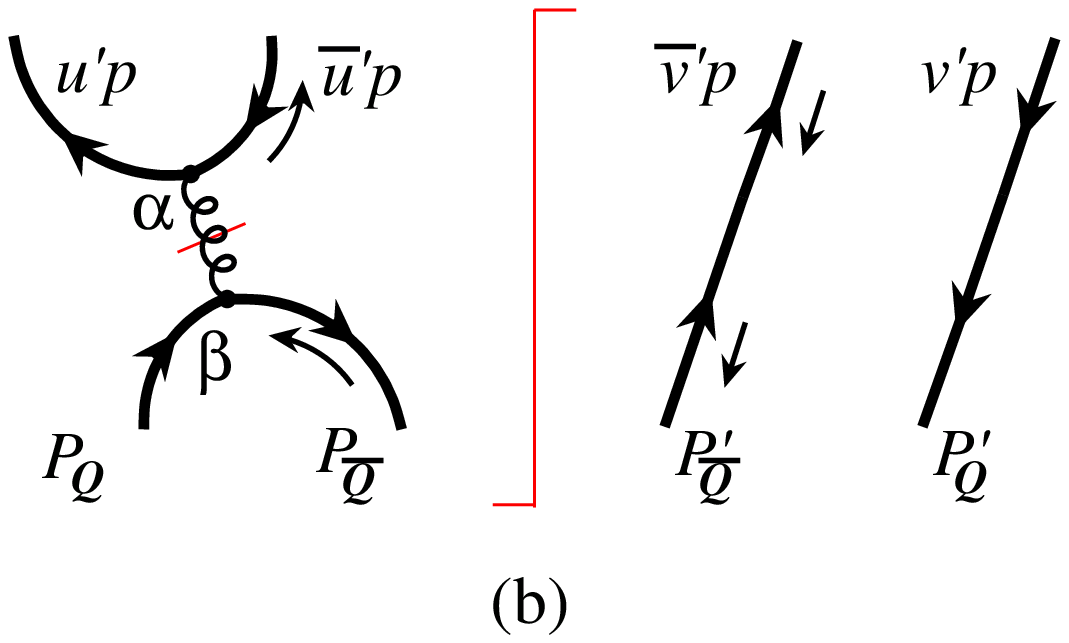, height=1.2in}
\\
\vspace{0.1in}
\psfig{file=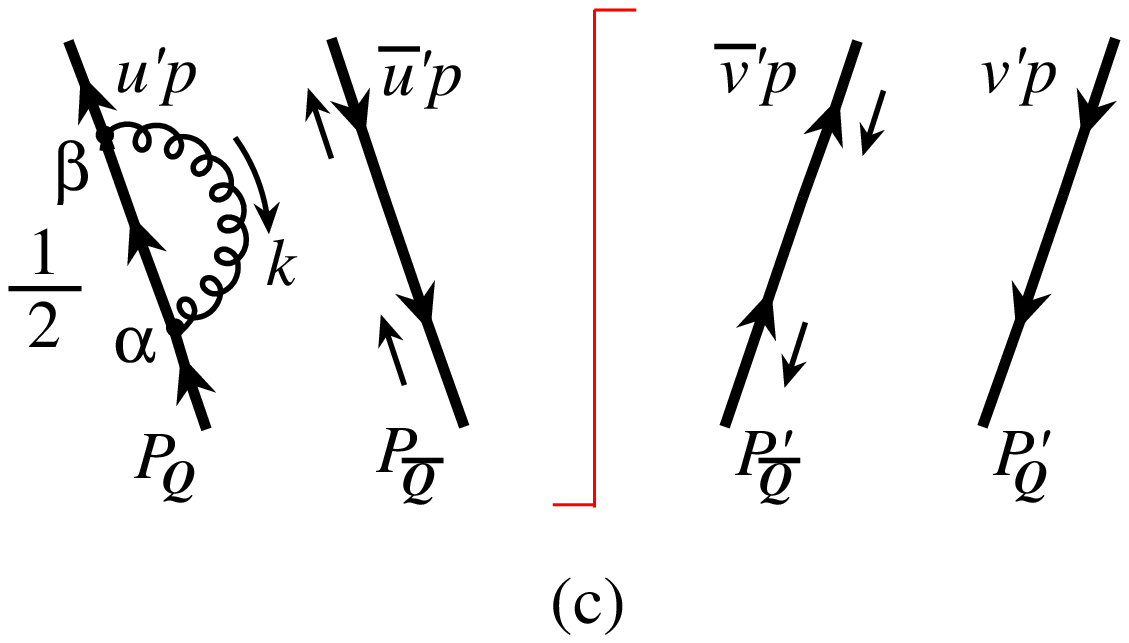, height=1.2in}
\hskip 0.8in
\psfig{file=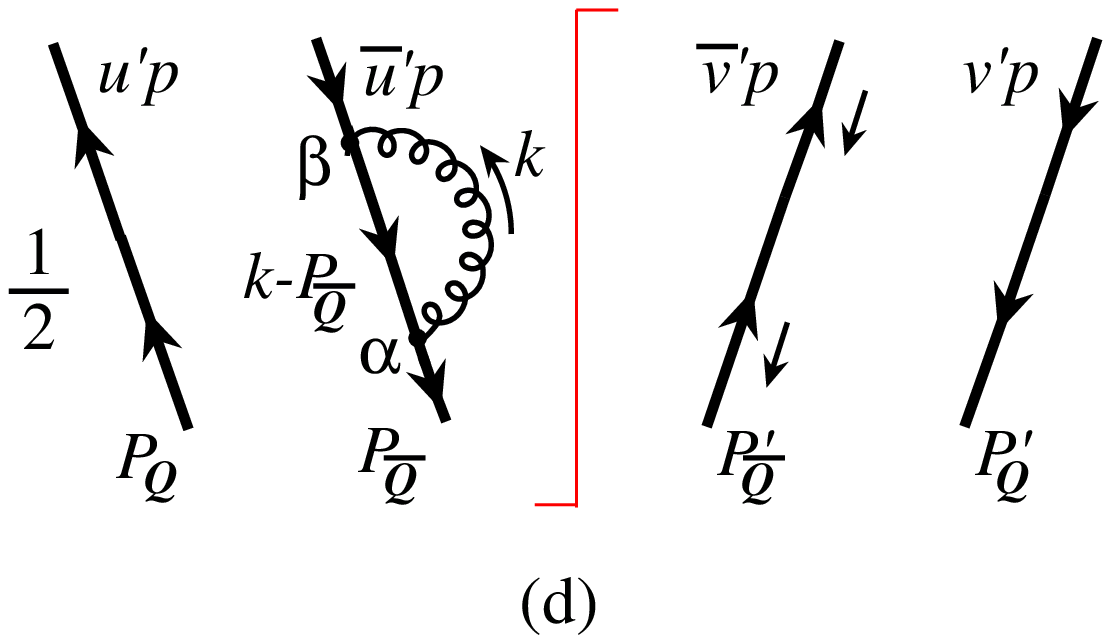, height=1.2in}
\caption{Cut diagrams that give the virtual contribution to heavy quark pair fragmentation function at order of $\alpha_s$.}
\label{fig:qqfrag-virtual1}
\eef

The virtual contribution to the first-order fragmentation functions in
Eq.~(\ref{eq:evo_sol}), which we will combine with Eq.\ (\ref{eq:evo_r3}) to derive the evolution kernels, 
is given by the interference between the lowest order diagram 
and the diagrams with a virtual gluon loop in Fig.~\ref{fig:evo1-virtual}.
In the light-cone gauge, diagrams (e)-(j) in Fig.~\ref{fig:evo1-virtual}
do not contribute.  The total virtual contribution
is given by the cut diagrams in Fig.~\ref{fig:qqfrag-virtual1} and their complex conjugates. 

In detail, the cut diagram (a) in Fig.~\ref{fig:qqfrag-virtual1} gives 
\begin{eqnarray}
{\mathcal D}^{(1,V-a)}_{[Q\bar{Q}(v8)]\to [Q\bar{Q}(v8)]}(z, u, v; u',v';\mu^2)
&=& 
\int \frac{d^4p_c}{(2\pi)^4}
\frac{d^4q_1}{(2\pi)^4} \frac{d^4q_2}{(2\pi)^4}
\frac{d^4k}{(2\pi)^4}\,
\theta\left(\mu^2-p_{c\perp}^2\right)\,
\nonumber\\
&\ & \hskip -1.6in \times
g_s^2\, C_{V_a}\,
(2\pi)^4\,\delta^4(k+u'p^+-p_c/2-q_1)
\nonumber\\
&\ & \hskip -1.6in \times
z^2 \,\delta\left(z-\frac{p^+}{p_c^+}\right)
\delta\left(u-\frac{1}{2}-\frac{q_1^+}{p_c^+}\right)
\delta\left(v-\frac{1}{2}-\frac{q_2^+}{p_c^+}\right)
\nonumber\\
&\ & \hskip -1.6in \times 
(2\pi)^4\delta^{4}\left(\frac{p_c}{2}+q_2-v'p^+\right)
(2\pi)^4\delta^{4}\left(\frac{p_c}{2}-q_2-\bar{v}'p^+\right)
{\cal T}_a
\label{eq:evo_va0}
\end{eqnarray}
where $C_{V_a}=-1/(2N_c)$ is color factor.  The trace term ${\cal T}_{a}$ 
in Eq.~(\ref{eq:evo_va0}) is given by
\begin{eqnarray}
{\cal T}_a 
&=&
\frac{1}{4p_c^+}{\rm Tr}\left[\gamma\cdot n\gamma\cdot p\right]\,
\frac{1}{4p_c^+}{\rm Tr}\left[
\gamma\cdot n \gamma\cdot(q_1-p_c/2) \gamma^\alpha
\gamma\cdot p \gamma^\beta \gamma\cdot(p_c/2+q_1) \right]
\nonumber\\
&\ & \times
\frac{i\, {\cal P}_{\alpha\beta}(k)}
{[(p_c/2-q_1)^2+i\varepsilon] [(p_c/2+q_1)^2+i\varepsilon][k^2+i\varepsilon]}\, ,
\label{eq:va-trace}
\end{eqnarray}
where as above, ${\cal P}_{\alpha\beta}(k) $ is the gluon polarization tensor given in Eq.~(\ref{eq:gluepol}).  
Using the $\delta$-functions to fix phase space integration over $p_c,q_1$, and $q_2$, 
we obtain,
\begin{eqnarray}
{\mathcal D}^{(1,V-a)}_{[Q\bar{Q}(v8)]\to [Q\bar{Q}(v8)]}(z, u, v; u',v';\mu^2)
&=& 
\int^{\mu^2} dk_\perp^2 \left(\frac{\alpha_s}{2\pi}\right)\left(-\frac{1}{2N_c}\right)
\delta(z-1)\,\delta(v-v')
\nonumber\\
&\times &
\int dk^+ \delta(u-u' - k^+/p^+) 
\left[\frac{1}{2\pi} \int_{-\infty}^\infty dk^-\, {\cal T}_a \right]\, .
\label{eq:evo_va1}
\end{eqnarray}
The $\int_{-\infty}^\infty dk^-$ integration above can be carried out 
by examining the  pole structure of ${\cal T}_a$ in $k^-$ as follows.  
From the denominator of the trace term in Eq.~(\ref{eq:va-trace}), we 
have three poles in $k^-$,
\begin{eqnarray}
(1) & k^2+i\varepsilon = 0 
& \Rightarrow \
k^- = \frac{k_\perp^2}{2k^+} - i\varepsilon\, \mbox{sgn}(k^+)\, ,
\label{eq:va-pole1} \\
(2) & (k+u'p^+)^2+i\varepsilon = 0 
& \Rightarrow \
k^- = \frac{k_\perp^2}{2(k^++u'p^+)} - i\varepsilon\, \mbox{sgn}(k^++u'p^+)\, ,
\label{eq:va-pole2} \\
(3) & (k-\bar{u}'p^+)^2+i\varepsilon = 0 
& \Rightarrow \
k^- = \frac{k_\perp^2}{2(k^+-\bar{u}'p^+)} - i\varepsilon\, \mbox{sgn}(k^+-\bar{u}'p^+)\, .
\label{eq:va-pole3} 
\end{eqnarray}
In Fig.~\ref{fig:pole-va}, we show the positions of these poles on the complex $k^-$-plane 
as a function of $k^+/p^+$.  
When $k^+ < -u'p^+$ and $k^+ > \bar{u}'p^+$, the $k^-$-integration in Eq.~(\ref{eq:evo_va1}) 
vanishes, because all poles are in the same half-plane.  
After carrying out $k^-$-integration, we obtain,
\begin{eqnarray}
{\mathcal D}^{(1,V-a)}_{[Q\bar{Q}(v8)]\to [Q\bar{Q}(v8)]}(z, u, v; u',v';\mu^2)
&=& 
\int^{\mu^2} \frac{dk_\perp^2}{k_\perp^2} 
\left(\frac{\alpha_s}{2\pi}\right)\left[-\frac{1}{2N_c}\right]
\delta(z-1)\,\delta(v-v')
\nonumber\\
&\ & \hskip -1.1in \times
\bigg[
\left(\frac{\bar{u}}{\bar{u}'}\right)\left(\bar{u}'+ u \right)
\int_0^{\bar{u}'p^+} \frac{dk^+}{k^+} \delta(u-u' - k^+/p^+)
\nonumber\\
&\ & \hskip -1in
- 
\left(\frac{{u}}{{u}'}\right)\left(u' + \bar{u} \right)
\int_{-{u}'p^+}^0 \frac{dk^+}{k^+} \delta(u-u' - k^+/p^+)
\bigg]\, ,
\label{eq:evo_va2}
\end{eqnarray}
where the $k^+$-integration is singular when $u=u'$.  
This singularity will be cancelled by the corresponding singularities of diagrams (b) and (c)
in Fig.~\ref{fig:qqfrag-virtual1}, as required by the factorization.  

\bef
\psfig{file=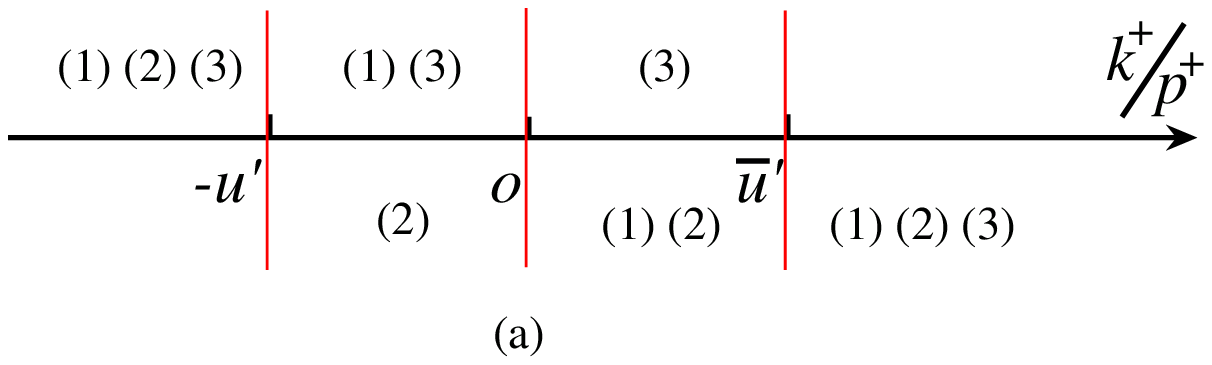, height=1.0in}
\caption{Positions of $k^-$-poles of cut diagram (a) in Fig.~\ref{fig:qqfrag-virtual1} 
as a function of $k^+/p^+$.}
\label{fig:pole-va}
\eef

The cut diagram (b) in Fig.~\ref{fig:qqfrag-virtual1} gives the following virtual contribution 
to the heavy quark pair fragmentation function, 
\begin{eqnarray}
&\ &
{\mathcal D}^{(1,V-b)}_{[Q\bar{Q}(v8)]\to [Q\bar{Q}(v8)]}(z, u, v; u',v';\mu^2)
\nonumber\\
&\ & \hskip 1.0in 
= g_s^2\, C_{V_b}
\int \frac{d^4p_c}{(2\pi)^4}
\frac{d^4q_1}{(2\pi)^4} \frac{d^4q_2}{(2\pi)^4}\,
\theta\left(\mu^2-p_{c\perp}^2\right)\,
\nonumber\\
&\ & \hskip 1.1in \times
z^2 \,\delta\left(z-\frac{p^+}{p_c^+}\right)
\delta\left(u-\frac{1}{2}-\frac{q_1^+}{p_c^+}\right)
\delta\left(v-\frac{1}{2}-\frac{q_2^+}{p_c^+}\right)
\nonumber\\
&\ & \hskip 1.1in \times 
(2\pi)^4\delta^{4}\left(\frac{p_c}{2}+q_2-v'p^+\right)
(2\pi)^4\delta^{4}\left(\frac{p_c}{2}-q_2-\bar{v}'p^+\right)
{\cal T}_b\, .
\label{eq:evo_cont0}
\end{eqnarray}
In Eq.~(\ref{eq:evo_cont0}), $C_{V_b}=1/2$ is the color factor, and 
\begin{eqnarray}
{\cal T}_b
&=& -
\frac{1}{4p_c^+}{\rm Tr}\left[\gamma\cdot n\gamma\cdot p\right]\,
\frac{1}{4p_c^+}{\rm Tr}\left[
\gamma\cdot n \gamma\cdot(q_1-p_c/2) \gamma^\beta
\gamma\cdot(p_c/2+q_1) \right]
\nonumber\\
&\ & \times
{\rm Tr}\left[\gamma\cdot p \gamma^\alpha\right]
\left[\frac{i\, n_\alpha n_\beta}{(p_c\cdot n)^2}\right]
\frac{1}{[(p_c/2-q_1)^2+i\varepsilon] [(p_c/2+q_1)^2+i\varepsilon]}\, ,
\label{eq:cont-trace}
\end{eqnarray}
where $i n_\alpha n_\beta/(p_c\cdot n)^2$ is the contact term of the gluon propagator.  
Using the $\delta$-functions to fix phase space integration over $p_c$ and $q_2$,
and replacing the loop momentum $q_1$ by $k$, we obtain
\begin{eqnarray}
{\mathcal D}^{(1,V-b)}_{[Q\bar{Q}(v8)]\to [Q\bar{Q}(v8)]}(z, u, v; u',v';\mu^2)
&=& -
\int^{\mu^2} dk_\perp^2 \left(\frac{\alpha_s}{2\pi}\right) \left[\frac{1}{2}\right]
\delta(z-1)\,\delta(v-v') \left[4u(1-u)\right]
\nonumber\\
&\ & \hskip -1.5in \times 
\left(\frac{-ip_c^+}{\pi}\right) \int dk^- 
\left[\frac{1}{2up_c^+ k^- - k_\perp^2 + i\varepsilon}\right]
\left[\frac{1}{2(1-u)p_c^+ (-k^-) - k_\perp^2 + i\varepsilon}\right]
\nonumber\\
&\ & \hskip -1.5in 
= - \int^{\mu^2} \frac{dk_\perp^2}{k_\perp^2} 
\left(\frac{\alpha_s}{2\pi}\right) \left[\frac{1}{2}\right]
\delta(z-1)\,\delta(v-v') \left[4u(1-u)\right]\, .
\label{eq:evo_cont}
\end{eqnarray}


The cut diagram (c) in Fig.~\ref{fig:qqfrag-virtual1} gives the following virtual contribution 
to the heavy quark pair fragmentation function, 
\begin{eqnarray}
&\ &
{\mathcal D}^{(1,V-c)}_{[Q\bar{Q}(v8)]\to [Q\bar{Q}(v8)]}(z, u, v; u',v';\mu^2)
\nonumber\\
&\ & \hskip 0.5in 
= \frac{1}{2}\, g_s^2\, C_{V_c}
\int \frac{d^4p_c}{(2\pi)^4}
\frac{d^4q_1}{(2\pi)^4} \frac{d^4q_2}{(2\pi)^4}\,
z^2 \,\delta\left(z-\frac{p^+}{p_c^+}\right)
\delta\left(u-\frac{1}{2}-\frac{q_1^+}{p_c^+}\right)
\delta\left(v-\frac{1}{2}-\frac{q_2^+}{p_c^+}\right)
\nonumber\\
&\ & \hskip 0.6in \times 
(2\pi)^4\delta^{4}\left(\frac{p_c}{2}-q_1-\bar{u}'p^+\right)
(2\pi)^4\delta^{4}\left(\frac{p_c}{2}+q_2-v'p^+\right)
(2\pi)^4\delta^{4}\left(\frac{p_c}{2}-q_2-\bar{v}'p^+\right)
\nonumber\\
&\ & \hskip 0.6in \times 
\frac{1}{4p_c^+}{\rm Tr}\left[\gamma\cdot n\gamma\cdot p\right]\,
\frac{1}{4p_c^+}{\rm Tr}\left[
\gamma\cdot n \gamma\cdot p
\left(i\hat{\Sigma}(P_Q,n;\mu^2)\right) 
\left(\frac{i\gamma\cdot P_Q}{P_Q^2+i\varepsilon} \right)
\right]
\nonumber\\
&\ & \hskip 0.5in 
=
\frac{1}{2}\, g_s^2 \, C_{V_c}\, 
\delta(z-1)\, \delta(u-u')\, \delta(v-v')
\nonumber\\
&\ & \hskip 0.6in \times
\frac{1}{4p^+}{\rm Tr}\left[
\gamma\cdot n \gamma\cdot p
\left(i\hat{\Sigma}(P_Q,n;\mu^2)\right) 
\left(i\gamma\cdot P_Q\right)\right]
\left(\frac{1}{P_Q^2+i\varepsilon} \right)\, .
\label{eq:evo_vb0}
\end{eqnarray}
In Eq.~(\ref{eq:evo_vb0}), $C_{V_c}=C_F=N_c/2-1/(2N_c)$ is the color factor, and 
\begin{equation}
i\hat{\Sigma}(P_Q,n;\mu^2)
=
\int \frac{d^4k}{(2\pi)^4}\, 
\theta\left(\mu^2-k_{\perp}^2\right)\,
\left[(-i\gamma^\beta)\frac{i\gamma\cdot(k+P_Q)}{(k+P_Q)^2+i\varepsilon}
(-i\gamma^\alpha)\right]
\left(\frac{i{\cal P}_{\alpha\beta}(k)}{k^2+i\varepsilon}\right)\, ,
\label{eq:sigma-hat}
\end{equation}
where $\mu^2$ is the renormalization scale dependence from the wave function renormalization
at this order, ${\cal P}_{\alpha\beta}(k) $ is the gluon polarization tensor given in Eq.~(\ref{eq:gluepol}).
Since $\hat{\Sigma}(P_Q,n;\mu^2)$ is a function of vectors $P_Q$ and $n$,
we define \cite{Collins:1988wj},
\begin{equation}
i\hat{\Sigma}(P_Q,n;\mu^2)
\equiv
\gamma\cdot P_Q\ f_1 + \frac{P_Q^2}{2P_Q\cdot n}\, \gamma\cdot n\ f_2\, ,
\label{eq:vb-trace}
\end{equation}
where $f_1$ and $f_2$ are scalar functions depending only on Lorentz invariants of four-vectors 
$P_Q$ and $n$, and the factorization scale $\mu^2$.  From Eq.~(\ref{eq:vb-trace}), we have
\begin{eqnarray}
\frac{\partial}{\partial P_Q^-}\left(i\hat{\Sigma}(P_Q,n;\mu^2)\right)
=\gamma\cdot n\left(f_1+f_2\right)\, ,
\label{eq:match-d}
\end{eqnarray}
plus terms that vanish as $P_Q^2\to 0$.  
Substituting Eq.~(\ref{eq:vb-trace}) into Eq.~(\ref{eq:evo_vb0}), 
and using the identity in Eq.~(\ref{eq:match-d}), we obtain
\begin{eqnarray}
{\mathcal D}^{(1,V-c)}_{[Q\bar{Q}(v8)]\to [Q\bar{Q}(v8)]}(z, u, v; u',v';\mu^2)
=
\frac{1}{2}\, g_s^2 \left[\frac{N_c}{2}-\frac{1}{2N_c}\right]
\delta(z-1)\, \delta(u-u')\, \delta(v-v')\, 
{\cal T}_c\, ,
\nonumber\\
\label{eq:evo_vb1}
\end{eqnarray}
with
\begin{eqnarray}
{\cal T}_c 
&=& 
\frac{i}{4p^+}{\rm Tr}\left[
\gamma\cdot p
\left(\frac{\partial}{\partial P_Q^-}\left(i\hat{\Sigma}(P_Q,n;\mu^2)\right)
\right)\right]
\label{eq:vb-trace1}\\
&=&
\frac{-i}{4p^+}
\int\frac{d^4k}{(2\pi)^4}\,
\theta\left(\mu^2-k_{\perp}^2\right)\,
\frac{{\cal P}_{\alpha\beta}(k)}
       {[k^2+i\varepsilon][(k+P_Q)^2+i\varepsilon]^2}
\nonumber\\
&\ & \times
{\rm Tr}\left[
\gamma\cdot p\, \gamma^\beta
\gamma\cdot(k+P_Q) \gamma\cdot n\, \gamma\cdot(k+P_Q) 
\gamma^\alpha\right]\,        .
\nonumber
\end{eqnarray}
From the denominator above, we  have two  poles in the $k^-$-integration of ${\cal T}_c$,
\begin{eqnarray}
(1) & k^2+i\varepsilon = 0 
& \Rightarrow \
k^- = \frac{k_\perp^2}{2k^+} - i\varepsilon\, \mbox{sgn}(k^+)\, ,
\label{eq:vb-pole1} \\
(2) & (k+P_Q)^2+i\varepsilon = 0 
& \Rightarrow \
k^- = -P_Q^- + \frac{k_\perp^2}{2(k^++P_Q^+)} 
- i\varepsilon\, \mbox{sgn}(k^++P_Q^+)\, .
\label{eq:vb-pole2} 
\end{eqnarray}
We show the positions of these poles in the complex $k^-$-plane as a function of $k^+/p^+$
in Fig.~\ref{fig:pole-vbc}(c), where we use $P_Q^+ = up^+$.  
As in Fig.~\ref{fig:pole-va}, the  $k^-$-integration in Eq.~(\ref{eq:vb-trace1}) vanishes
when $k^+$ is outside of the range $[-up^+,0]$.  
Integrating over $k^-$, we obtain
\begin{eqnarray}
{\cal T}_c
=
\frac{1}{8\pi^2}
\int^{\mu^2} \frac{dk_\perp^2}{k_\perp^2} \left[
\frac{3}{2}+2\int_{-up^+}^0\frac{dk^+}{k^+} \right]
\label{eq:vb-trace2}
\end{eqnarray}
and the virtual contribution from diagram (b) in Fig.~\ref{fig:qqfrag-virtual1}, is
\begin{eqnarray}
{\mathcal D}^{(1,V-c)}_{[Q\bar{Q}(v8)]\to [Q\bar{Q}(v8)]}(z, u, v; u',v';\mu^2)
&=&
\int^{\mu^2} \frac{dk_\perp^2}{k_\perp^2} 
\left(\frac{\alpha_s}{2\pi}\right)\left[\frac{N_c}{2}-\frac{1}{2N_c}\right]
\nonumber\\
&\ & \hskip -1.0in  \times
\frac{1}{2}\, \delta(z-1)\, \delta(u-u')\, \delta(v-v')
\left[
\frac{3}{2}+2\int_{-up^+}^0\frac{dk^+}{k^+} \right]\, .
\label{eq:evo_vb2}
\end{eqnarray}
Similarly, with the pole structure of the $k^-$-integration in Fig.~\ref{fig:pole-vbc}(d), 
we obtain the virtual contribution from diagram (c) in Fig.~\ref{fig:qqfrag-virtual1} as
 \begin{eqnarray}
{\mathcal D}^{(1,V-d)}_{[Q\bar{Q}(v8)]\to [Q\bar{Q}(v8)]}(z, u, v; u',v';\mu^2)
&=&
\int^{\mu^2} \frac{dk_\perp^2}{k_\perp^2} 
\left(\frac{\alpha_s}{2\pi}\right)\left[\frac{N_c}{2}-\frac{1}{2N_c}\right]
\nonumber\\
&\ & \hskip -1.0in  \times
\frac{1}{2}\, \delta(z-1)\, \delta(u-u')\, \delta(v-v')
\left[
\frac{3}{2}-2\int_0^{\bar{u}p^+}\frac{dk^+}{k^+} \right]\, .
\label{eq:evo_vc2}
\end{eqnarray}
\bef
\psfig{file=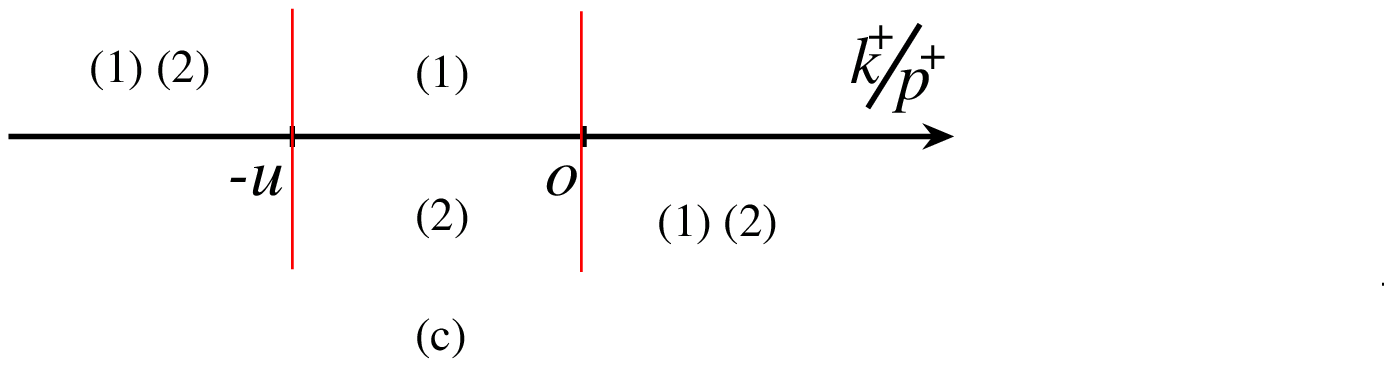, height=1.0in}
\hskip 0.3in
\psfig{file=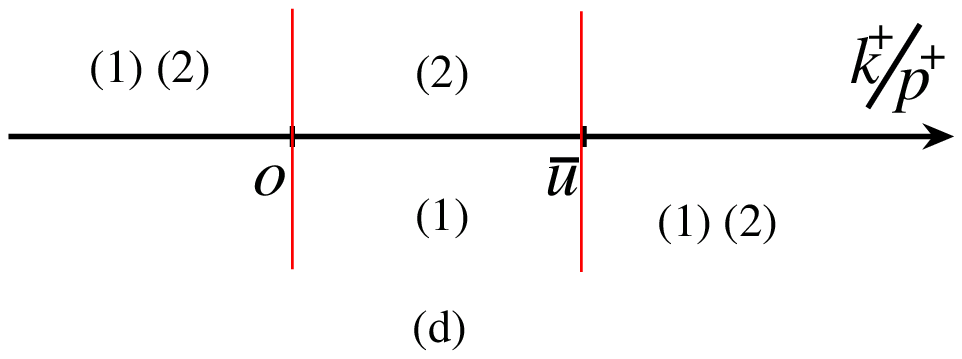, height=1.0in}
\caption{Positions of $k^-$-poles of cut diagrams (c) and (d) in Fig.~\ref{fig:qqfrag-virtual1} 
as a function of $k^+/p^+$.}
\label{fig:pole-vbc}
\eef

\noindent
Combining Eqs.~(\ref{eq:evo_va2}), (\ref{eq:evo_cont}), (\ref{eq:evo_vb2}), and (\ref{eq:evo_vc2}), we have
 \begin{eqnarray}
&\ & \hskip -0.1in
{\mathcal D}^{(1,Vabcd)}_{[Q\bar{Q}(v8)]\to [Q\bar{Q}(v8)]}(z, u, v; u',v';\mu^2)
\nonumber\\
&\ & \hskip 0.1in 
=
\int^{\mu^2} \frac{dk_\perp^2}{k_\perp^2} 
\left(\frac{\alpha_s}{2\pi}\right) \delta(z-1)\, \delta(v-v')
\nonumber\\
&\ & \hskip 0.1in  \times
\Bigg\{
\left[\frac{N_c}{2}\right] 
\bigg\{
\delta(u-u')\left[\frac{3}{4}+\int_{-up^+}^0\frac{dk^+}{k^+}\right]
+
\delta(\bar{u}-\bar{u}')\left[\frac{3}{4}-\int_0^{\bar{u}p^+}\frac{dk^+}{k^+} \right]
\bigg\}
\nonumber\\
&\ & \hskip 0.2in 
- \left[\frac{1}{2}\right] 
\big\{4u(1-u)\big\}
\label{eq:evo_vabc}\\
&\ & \hskip 0.2in 
+\left[\frac{1}{2N_c}\right]
\bigg\{
\int_0^{\bar{u}'p^+} \frac{dk^+}{k^+} 
\left[- \delta(\bar{u}'-\bar{u} - k^+/p^+)\left(\frac{\bar{u}}{\bar{u}'}\right)\left(\bar{u}'+ u \right)
 +\delta(\bar{u}-\bar{u}')\right] 
 -\frac{3}{4}\delta(\bar{u}-\bar{u}') 
\nonumber\\
&\ & \hskip 0.7in  + 
\int_{-{u}'p^+}^0 \frac{dk^+}{k^+} 
\left[ \delta(u-u' - k^+/p^+) \left(\frac{{u}}{{u}'}\right)\left(u' + \bar{u} \right)
 - \delta(u-u') \right]
 -\frac{3}{4}\delta(u-u') 
 \bigg\} 
 \Bigg\}.
\nonumber
\end{eqnarray}
The virtual contribution from the complex conjugate diagrams 
of those in Fig.~\ref{fig:qqfrag-virtual1} is the same as that in Eq.~(\ref{eq:evo_vabc})
but with the momentum fractions $u$ and $v$ switched.

In Eq.~(\ref{eq:evo_vabc}), we recognize that $(\bar{u}/\bar{u}')(\bar{u}'+u)\to 1$ 
as $\bar{u}\to\bar{u}'$, and that  $(u/u')(u'+\bar{u})\to 1$ as $u\to u'$.
Thus the $k^+$-integration, 
\begin{eqnarray}
&\ &
\int_0^{\bar{u}'p^+} \frac{dk^+}{k^+} 
\left[\delta(\bar{u}'-\bar{u} - k^+/p^+)\left(\frac{\bar{u}}{\bar{u}'}\right)\left(\bar{u}'+ u \right)
 -\delta(\bar{u}-\bar{u}')\right] 
 \nonumber\\
&\ & \hskip 0.2in 
= \int_0^{\bar{u}'p^+} \frac{dk^+}{k^+} 
\left[\delta(\bar{u}'-\bar{u} - k^+/p^+)
 -\delta(\bar{u}'-\bar{u})\right] \left(\frac{\bar{u}}{\bar{u}'}\right)\left(\bar{u}'+ u \right)
\label{eq:ubar-indentity}
\end{eqnarray}
is finite.  We now define a plus-distribution for $(\bar{u}'-\bar{u})$ by
\begin{eqnarray}
\frac{\theta\left(\bar{u}'-\bar{u}\right)}{\left(\bar{u}'-\bar{u}\right)_+} 
&\equiv & 
\int_0^{\bar{u}'p^+} \frac{dk^+}{k^+} 
\left[\delta(\bar{u}'-\bar{u} - k^+/p^+)
 -\delta(\bar{u}'-\bar{u})\right]
 \nonumber\\
& = &
 \int_0^{1} \frac{dx}{1-x} 
 \left[\delta(x\bar{u}'-\bar{u}) - \delta(\bar{u}'-\bar{u})\right] \, ,
\label{eq:plus-new}
\end{eqnarray}
which becomes the standard ``$+$''-function of $(1-z)$ defined 
in Eq.~(\ref{eq:def-plus}) if we let $\bar{u}'\to 1$.
When deriving Eq.~(\ref{eq:plus-new}), we change variables from $k^+$ to $x$, through
$k^+ = (\bar{u}'-x)p^+$.
In this notation,
we write the total virtual contribution to the order of $\alpha_s$ heavy
quark pair fragmentation function as
\begin{eqnarray}
{\mathcal D}^{(1,V)}_{[Q\bar{Q}(v8)]\to [Q\bar{Q}(v8)]}(z, u, v; u',v';\mu^2)
&=& 
\int^{\mu^2} \frac{dk_\perp^2}{k_\perp^2} 
\left(\frac{\alpha_s}{2\pi}\right)\, \delta(z-1)\, 
\nonumber\\
&\ & \hskip -1.8in  \times
\Bigg\{
\left[\frac{N_c}{2}\right]
\delta(u-u')\,\delta(v-v')
\Bigg[3+{\rm Re}\left(
\int_{-up^+}^0 - \int_0^{\bar{u}p^+} + \int_{-vp^+}^0 - \int_0^{\bar{v}p^+}
\right) \frac{dk^+}{k^+}\Bigg]
\nonumber\\
&\ & \hskip -1.65in
-
\left[\frac{1}{2}\right]
\bigg[
\delta(v-v') [4u(1-u)] + \delta(u-u') [4v(1-v)] 
\bigg]
\nonumber\\
&\ & \hskip -1.65in
+
\left[\frac{-1}{2N_c}\right]
\left[
\delta(v-v')\left(
\frac{\theta\left(\bar{u}'-\bar{u}\right)}{\left(\bar{u}'-\bar{u}\right)_+} 
\left(\frac{\bar{u}}{\bar{u}'}\right)\left(\bar{u}'+ u \right)
+\frac{3}{4}\delta(\bar{u}-\bar{u}')
\right. \right.
\nonumber\\
&\ & \hskip -0.3in 
\left.
+ 
\frac{\theta\left(u'-u\right)}{\left(u'-u\right)_+} 
\left(\frac{u}{u'}\right)\left(u'+ \bar{u} \right)
+\frac{3}{4}\delta(u-u')
\right)
\nonumber\\
&\ & \hskip -1.0in
+
\delta(u-u')\left(
\frac{\theta\left(\bar{v}'-\bar{v}\right)}{\left(\bar{v}'-\bar{v}\right)_+} 
\left(\frac{\bar{v}}{\bar{v}'}\right)\left(\bar{v}'+ v \right)
+\frac{3}{4}\delta(\bar{v}-\bar{v}')
\right. 
\nonumber\\
&\ & \hskip -0.2in \left.\left.
+ 
\frac{\theta\left(v'-v\right)}{\left(v'-v\right)_+} 
\left(\frac{v}{v'}\right)\left(v'+ \bar{v} \right)
+\frac{3}{4}\delta(v-v')
\right)\right]\Bigg\}\, ,
\label{eq:evo-v}
\end{eqnarray}
where the generalized ``$+$''-function is defined in Eq.~(\ref{eq:plus-new}), 
and the divergence in the $k^+$ integration will cancel the divergences 
of the real-gluon contributions.  Combining this virtual contribution in Eq.~(\ref{eq:evo-v}) 
with the real contribution in Eq.~(\ref{eq:evo_r3}), and using Eq.~(\ref{eq:evo_sol}), 
we obtain the evolution kernel as
\begin{eqnarray}
&\ & 
{\Gamma}^{(1)}_{[Q\bar{Q}(v8)]\to [Q\bar{Q}(v8)]}(z, u, v; u',v')
\nonumber\\
&\ & \hspace{5mm} =
\left(\frac{\alpha_s}{2\pi}\right) 
\Bigg\{
\left[\frac{1}{2N_c}\right]\bigg\{
\frac{1}{2}\frac{z}{(1-z)_+}\,S_+\,\Delta_{-}^{[8]}
-\delta(z-1)
\nonumber\\
&\ & \hskip 1.1in
\times
\bigg[
\delta(v-v')\left(
\frac{\theta\left(\bar{u}'-\bar{u}\right)}{\left(\bar{u}'-\bar{u}\right)_+} 
\left(\frac{\bar{u}}{\bar{u}'}\right)\left(\bar{u}'+ u \right)
+
\frac{\theta\left(u'-u\right)}{\left(u'-u\right)_+} 
\left(\frac{u}{u'}\right)\left(u'+ \bar{u} \right)
\right) 
\nonumber\\
&\ & \hskip 1.2in
+
\delta(u-u')\left(
\frac{\theta\left(\bar{v}'-\bar{v}\right)}{\left(\bar{v}'-\bar{v}\right)_+} 
\left(\frac{\bar{v}}{\bar{v}'}\right)\left(\bar{v}'+ v \right)
+
\frac{\theta\left(v'-v\right)}{\left(v'-v\right)_+} 
\left(\frac{v}{v'}\right)\left(v'+ \bar{v} \right)
\right)
\nonumber\\
&\ & \hskip 1.2in
+
3\,\delta(v-v')\,\delta(u-u') \bigg] \bigg\}
\nonumber\\
&\ & \hskip 0.8in
- \left[\frac{1}{2}\right] \delta(z-1)
\bigg[
\delta(v-v') [4u(1-u)] + \delta(u-u') [4v(1-v)] 
\bigg]
\nonumber\\
&\ & \hskip 0.8in
+\left[\frac{N_c}{2}\right] \delta(z-1)\, \delta(u-u')\, \delta(v-v')
\left[3-\ln\left(u\bar{u}v\bar{v}\right) \right]
\Bigg\}\, ,
\label{eq:Gamma_f}
\end{eqnarray}
where $S_+$ and $\Delta_{-}^{[8]}$ are given in Eq.~(\ref{eq:def-sdelta}).
This result is consistent with the kernel derived in Ref.\ \cite{Fleming:2013qu}
except for the logarithmic term and the contact term.  
The logarithmic term is a consequence of the different integration limits of 
$k^+$-integration between the real contribution in Eq.~(\ref{eq:evo_r3})
and the virtual contribution in Eq.~(\ref{eq:evo-v}).  
As shown in Eqs.~(\ref{eq:evo_r3}) and (\ref{eq:evo-v}), both the real and virtual
contribution have infrared divergences from the $k^+$-integration, and 
the infrared divergences are exactly canceled when the real and virtual contributions
are combined.  More specifically, the cancelation takes place in 
the following $k^+$-integration, 
\begin{eqnarray}
4\int_0^{p^+} \frac{dk^+}{k^+}
+{\rm Re}\left(
\int_{-up^+}^0 - \int_0^{\bar{u}p^+} + \int_{-vp^+}^0 - \int_0^{\bar{v}p^+}
\right) \frac{dk^+}{k^+}
= - \ln\left(u\,\bar{u}\,v\,\bar{v}\right)\, .
\label{eq:IR-cancelation}
\end{eqnarray}
As required by the factorization, the evolution kernel in Eq.~(\ref{eq:Gamma_f}) 
is indeed free of any singularity.  

Evolution kernels between different spin-color states of heavy quark pairs 
can be calculated by using the corresponding spin-color projection operators
derived in Sec.~\ref{sec:frag}.
At first non-trivial order in $\alpha_s$, some evolution kernels vanish. 
To make the evolution (or change) between various spin-color states of heavy quark pairs 
clearer, we rewrite the evolution equation for heavy quark pair fragmentation functions 
in Eq.~(\ref{eq:evo_2p}) and corresponding evolution kernels in a matrix form,
\begin{eqnarray}
\frac{\partial}{\partial{\ln\mu^2}}
\left(\begin{array}{c} 
{\mathcal D}_{[Q\bar{Q}(v8)]\to H} \\ 
{\mathcal D}_{[Q\bar{Q}(v1)]\to H} \\ 
{\mathcal D}_{[Q\bar{Q}(a8)]\to H} \\
{\mathcal D}_{[Q\bar{Q}(a1)]\to H} \\
{\mathcal D}_{[Q\bar{Q}(t8)]\to H} \\
{\mathcal D}_{[Q\bar{Q}(t1)]\to H} 
\end{array} 
\right)
&=&
\left(\frac{\alpha_s}{2\pi}\right)
\left(\begin{array}{cccccc} 
{\mathcal K}_v     & {\mathcal R}      & {\mathcal T}_1  & {\mathcal T}_2 & 0 & 0 \\ 
\widetilde{\mathcal R} & {\mathcal S} & \widetilde{\mathcal T}_2  & 0 & 0 & 0  \\
{\mathcal T}_1  & {\mathcal T}_2  & {\mathcal K}_a  & {\mathcal R} & 0 & 0  \\
\widetilde{\mathcal T}_2 &  0 & \widetilde{\mathcal R} & {\mathcal S} & 0  & 0 \\
0 & 0 & 0 & 0  & {\mathcal K}' & {\mathcal R}' \\
0 & 0 & 0 & 0  & \widetilde{\mathcal R}' & {\mathcal S}'
\end{array}
\right)
\otimes
\left(\begin{array}{c} 
{\mathcal D}_{[Q\bar{Q}(v8)]\to H} \\ 
{\mathcal D}_{[Q\bar{Q}(v1)]\to H} \\ 
{\mathcal D}_{[Q\bar{Q}(a8)]\to H} \\
{\mathcal D}_{[Q\bar{Q}(a1)]\to H} \\
{\mathcal D}_{[Q\bar{Q}(t8)]\to H} \\
{\mathcal D}_{[Q\bar{Q}(t1)]\to H} 
\end{array} \right)\, ,
\nonumber\\
\label{eq:evo-matrix}
\end{eqnarray}
where $\otimes$ represents the convolution over momentum fractions, $z'$, $u'$ and $v'$, 
as defined in Eq.~(\ref{eq:evo_2p}).  
The elements of the matrix form of the evolution kernels are defined as
\begin{eqnarray}
&{\mathcal K}_v = P_{v8\to v8}\, ,  \hskip 1.1in
&{\mathcal K}_a = P_{a8\to a8}\, , 
\nonumber\\
&{\mathcal K}' =  P_{t8\to t8} \, , \hskip 1.1in
\nonumber\\
&{\mathcal S} = P_{v1\to v1}=P_{a1\to a1} \, ,  \hskip 0.4in
&{\mathcal S}' = P_{t1\to t1} \, ;
\nonumber\\
&{\mathcal R} = P_{v8\to v1}=P_{a8\to a1}\, ,  \hskip 0.4in
&\widetilde{\mathcal R} = P_{v1\to v8}=P_{a1\to a8} \, , 
\nonumber\\
&{\mathcal R}' = P_{t8\to t1}\, ,  \hskip 1.2in
&\widetilde{\mathcal R}' = P_{t1\to t8}\, ,
\nonumber\\
&{\mathcal T}_1 = P_{v8\to a8}=P_{a8\to v8} \, , \hskip 0.5in
&{\mathcal T}_2  = P_{v8\to a1}=P_{a8\to v1} \, ,
\nonumber\\
&\widetilde{\mathcal T}_2 = P_{v1\to a8}=P_{a1\to v8} \, , 
\hskip 0.5in &
\label{eq:kernel-def2}
\end{eqnarray}
with the relation to the kernels,
\begin{equation}
\Gamma_{[Q\bar{Q}(\kappa')]\to [Q\bar{Q}(\kappa)]} \equiv 
\left(\frac{\alpha_s}{2\pi} \right)
P_{\kappa \to \kappa'}\, ,
\end{equation}
where $\kappa=sI$ with $s=v,a,t$ and $I=1,8$ (the same for $\kappa'$).  
All kernels $P_{\kappa \to \kappa'}$ are given in Appendix~\ref{sec:appendix-qqb}. 


%% file: sec5.tex
\section{Summary and conclusions}
\label{sec:conclusions}

We have presented a perturbative QCD factorization formalism for the inclusive production of heavy quarkonia 
at large $p_T$, which provides a systematic approach to study their production 
at collider energies beyond leading power.  The factorization formalism is organized in terms of a power expansion 
of $1/p_T$, which is equivalent to an organization in terms of the characteristic times at which the 
heavy quark pair is produced, before it transforms into a physical quarkonium.  
The leading power contribution comes from partonic subprocesses in which a single parton 
is produced at the hard collision, of distance scale  $1/p_T$, followed by 
single parton evolution and hadronization into an observed heavy quarkonium at a much later time.
The subleading power term, which we have discussed in this paper, describes the 
production of a heavy quark pair, either directly at the distance scale of $1/p_T$, at which the hard collision
takes place, or at any intermediate scale $1/\mu$ up to $1/m_Q$.  Although the rate to produce the pair 
at intermediate time $1/\mu$ is suppressed by $1/\mu^2$ 
in comparison with the production of single parton, the probability for the pair to become 
a heavy quarkonium is  larger than that for a single parton to form a heavy
quarkonium by fragmenting into a heavy quark pair at large times. 
We have shown in this paper that both the leading power and next-to-leading power contributions 
to the production cross section can be factorized in terms of 
perturbatively calculable short-distance partonic coefficient functions and 
 non-perturbative, but universal, fragmentation functions for  partons to evolve
into observed heavy quarkonia.

We identified operators for  heavy quark pair fragmentation functions, and 
corresponding projection operators for calculating the factorized leading and 
next-to-leading power short-distance partonic hard parts.  
We derived a closed set of evolution equations for both single parton and heavy quark pair 
fragmentation functions. We pointed out that once we work beyond the leading power, 
QCD evolution of fragmentation functions with respect to the variation of the factorization scale 
mixes the heavy quark pair fragmentation functions with single parton fragmentation functions.  
Such mixing in evolution corresponds to a resummation 
of the probability for the single fragmenting parton to generate a heavy quark pair 
from the distance scale of the hard collision, $\sim 1/p_T$, to a scale, 
$1/\mu_0 \sim 1/m_Q$ at which the fragmentation process becomes non-perturbative.
We calculated perturbatively the lowest order evolution kernels for all channels 
of heavy quark pair fragmentation functions, and also derived the first order 
evolution kernels for a single parton to evolve into a heavy quark pair.  
As expected from the factorization, all calculated evolution kernels are infrared finite.

The predictive power of this new factorization formalism relies on the infrared safety of 
short-distance coefficient functions, and the universality of the process-independent 
fragmentation functions.  The short-distance hard parts reflect  partonic dynamics 
at a distance scale of $1/p_T$, and are the same for the production of all heavy quarkonium 
states.  The leading order short distance functions for the production of a heavy quark pair 
in all perturbative color-spin states are presented in a companion paper \cite{KMQS-hq2}. 

In order to compare our calculations with experimental data, 
we need fragmentation functions at the input factorization scale, $\mu_0 \gtrsim 2 m_Q$,
so that the evolution equations can evolve these input fragmentation functions to 
generate the fragmentation functions at any other scales.
In principle, input fragmentation functions are non-perturbative and 
should be extracted from fitting experimental data, just as one derives 
parton fragmentation functions to light hadrons through QCD global analysis. 
However, as pointed out in our companion paper \cite{KMQS-hq2}, it may also
be a very reasonable conjecture to use the NRQCD factorization formalism 
to calculate all input fragmentation functions.  With the calculated/estimated 
input fragmentation functions, and perturbatively calculated hard parts and evolution kernels, 
our new factorization formalism could provide predictions with absolute normalization, 
which can be tested by data from the LHC and other colliders \cite{Ma:2014xxx}.

%% file: appendixa.tex
\section{Another set of spin projection operators}
\label{sec:appendix-spin}

The explicit ``end-point" singularities of the short-distance partonic hard parts, as discussed in Sec.~\ref{sec:frag},
reflect the possibility that the momentum of the produced active quark or antiquark can vanish, even though the total momentum of the pair remains finite.  This kind of  singularity is only possible when  more than one active parton is produced, and the projection operators for the production, such as those in Eq.~(\ref{eq:spin-pj}), are independent of momenta of active partons (or the spinors of produced quark and antiquark).  

The apparent end point singularities could be systematically removed from the partonic parts, if we modify the $\gamma\cdot p$ in the spin projection operators in Eq.~(\ref{eq:spin-pj}) as follows,
\begin{equation}
\left(\gamma\cdot p\right)_{ji} 
\rightarrow 
\left(\frac{\gamma\cdot\hat{P}_{\bar{Q}}\,
\gamma\cdot n\, \gamma\cdot \hat{P}_{Q}}{2p\cdot n}\right)_{ji}
=\left( \frac{u\bar{u}}{z^2}\right)\, \left(\gamma\cdot p\right)_{ji}\, ,
\label{eq:spin-pj_new}
\end{equation}
where $\hat{P}_{Q}^\mu=(u/z)\, p^\mu$, and 
$\hat{P}_{\bar{Q}}^\mu=(\bar{u}/z)\, p^\mu$.
With this choice, the spin projection operators for the partonic hard parts are explicitly proportional to the
momenta of the produced heavy quark and antiquark, 
via $\gamma\cdot \hat{P}_Q=\sum_s {u}_s(\hat{P}_Q)\, \overline{u}_s(\hat{P}_Q)$,
with quark spinor $u_s$ and $\bar{u}_s$ 
(or $\gamma\cdot \hat{P}_{\bar{Q}}=\sum_s {v}_s(-\hat{P}_Q)\, \overline{v}_s(-\hat{P}_Q)$ for antiquark),
where we have neglected the quark mass for the partonic hard parts.
The explicit dependence on the momenta of the active quark and antiquark in Eq.~(\ref{eq:spin-pj_new}) 
cancels the endpoint singularity when the momentum of the produced heavy quark 
or antiquark vanishes.  Correspondingly, we adjust the spin projection operators for the cut-vertices of 
heavy quark pair fragmentation functions in Eq.~(\ref{eq:spin-cv}) by the following replacement,
\begin{equation}
\frac{1}{4p\cdot n} \left(\gamma\cdot n\right)_{ij} 
\rightarrow 
\frac{p\cdot n}{(2\hat{P}_{Q}\cdot n)(2\hat{P}_{\bar{Q}}\cdot n)} 
\left(\gamma\cdot n\right)_{ij} 
= \left(\frac{z^2}{u\bar{u}}\right)\, \frac{1}{4p\cdot n} \left(\gamma\cdot n\right)_{ij} \, .
\label{eq:spin-cv_new}
\end{equation}
From Eqs.~(\ref{eq:spin-pj_new}) and (\ref{eq:spin-cv_new}), 
it is clear that the modification of the spinor projection operators is effectively to move
a spin-independent factor: $z^4/(u\bar{u}v\bar{v})$ from the partonic hard part 
to the definition of the corresponding fragmentation functions.  In this paper, we present our results 
calculated by using the spinor projection operators in Eqs.~(\ref{eq:spin-pj}) and (\ref{eq:spin-cv}), without this replacement.

%% file: appendixb.tex
\section{Single-parton to double-parton evolution kernels}
\label{sec:appendix-mix}

In this appendix, we summarize all evolution kernels for a single parton to evolve into a heavy quark pair at order  $\alpha_s^2$, which appear in the evolution equation in Eq.~(\ref{eq:evo_1p}).  The detailed calculation of $\gamma^{(2)}_{[Q\bar{Q}(v8)]/q}(z,u,v)$ was given in subsection~\ref{sec:frag_mixing}.
Like $\gamma^{(2)}_{[Q\bar{Q}(v8)]/q}(z,u,v)$ in Eq.~(\ref{eq:evo_q2QQ2}), all kernels below are derived 
by using the special gluon propagator to remove the mass singularity analytically, and by choosing the factorization scale as a cutoff of the invariant mass of fragmenting parton.

\bigskip
\noindent{1) Light quark case:}
\ben
\gamma^{(2)}_{q\to [Q\bar{Q}(v8)]} &=& 
\alpha_s^2 \left[\frac{N_c^2-1}{4N_c} \right] 
\frac{64(1-z)}{z^2}
\\
\gamma^{(2)}_{q\to [Q\bar{Q}(v1)]} 
&=& \gamma^{(2)}_{q\to [Q\bar{Q}(a1)]} 
= \gamma^{(2)}_{q\to [Q\bar{Q}(a8)]}
= \gamma^{(2)}_{q\to [Q\bar{Q}(t1)]} 
= \gamma^{(2)}_{q\to [Q\bar{Q}(t8)]} = 0
\een

\bigskip
\noindent{2) Light antiquark case:}
\ben
\gamma^{(2)}_{\bar{q}\to [Q\bar{Q}(v8)]}  &=&
\gamma^{(2)}_{q\to [Q\bar{Q}(v8)]}  =
\alpha_s^2 \left[\frac{N_c^2-1}{4N_c} \right] 
\frac{64(1-z)}{z^2}
\\
\gamma^{(2)}_{\bar{q}\to [Q\bar{Q}(v1)]} 
&=& \gamma^{(2)}_{\bar{q}\to [Q\bar{Q}(a1)]} 
= \gamma^{(2)}_{\bar{q}\to [Q\bar{Q}(a8)]}
= \gamma^{(2)}_{\bar{q}\to [Q\bar{Q}(t1)]} 
= \gamma^{(2)}_{\bar{q}\to [Q\bar{Q}(t8)]} = 0
\een

\bigskip
\noindent{3) Gluon case:}
\ben
\gamma^{(2)}_{g\to [Q\bar{Q}(v1)]} 
&=& \alpha_s^2 \left[\frac{1}{4N_c}\right]
\frac{4(u-\bar u)(v - \bar v)}{u\bar u v\bar v}
[z^2+(1-z)^2]
\\
\gamma^{(2)}_{g\to [Q\bar{Q}(v8)]} 
&=& \alpha_s^2 \frac{1}{2u\bar u v \bar v}
\left\{\frac{N_c}{z^2} \left[4(1-z)^2 - 4 (1-2u\bar u-2v\bar v) (1-z)^2(z+2) \right.\right.
\nnu
&\ &
\left.
+(u-\bar u)^2(v - \bar v)^2 (2z^4+2z^3-3z^2-4z+4) \right] 
\nonumber\\
&\ &
\left.
+ \frac{N_c^2-4}{N_c} (u-\bar u)(v - \bar v) [z^2+(1-z)^2]\right\}
\\
\gamma^{(2)}_{g\to [Q\bar{Q}(a1)]} 
&=& \alpha_s^2 \left[\frac{1}{4N_c}\right]
\frac{4}{u\bar u v \bar v}\, [z^2+(1-z)^2]
\\
\gamma^{(2)}_{g\to [Q\bar{Q}(a8)]} 
&=& \alpha_s^2 \frac{2}{u\bar u v \bar v}
\left[\frac{N_c}{2}(\bar u\bar v + u v)-\frac{1}{N_c}\right]
[z^2+(1-z)^2]
\\
\gamma^{(2)}_{g\to [Q\bar{Q}(t1)]} 
&=& \gamma^{(2)}_{g\to [Q\bar{Q}(t8)]} = 0 
\een

\bigskip
\noindent{4) Heavy quark case:}
\ben
\gamma^{(2)}_{Q\to [Q\bar{Q}(v1)]} 
&=& \alpha_s^2 
\left[\frac{C_F^2}{N_c}\right]
\frac{4(1-z)(1+z \bar u) (1+z\bar v)}
{\bar u\bar v(1-z u)(1-z v)}
\\
\gamma^{(2)}_{Q\to [Q\bar{Q}(v8)]} 
&=& \alpha_s^2 
\left[ \frac{N_c^2-1}{N_c^3}\right]
\frac{1-z}{z^2}\,
\frac{1}{\bar u\bar v}\,
\frac{4N_c \bar u (1-z u)+z(1+z\bar u)}{1-z u}
\nonumber \\
&\ & \hskip 1in
\times 
\frac{4N_c \bar v (1-z v)+z(1+z\bar v)}{1-z v}
\\
\gamma^{(2)}_{Q\to [Q\bar{Q}(a1)]} 
&=&  \gamma^{(2)}_{Q\to [Q\bar{Q}(v1)]}
=  \alpha_s^2 
\left[\frac{C_F^2}{N_c}\right]
\frac{4(1-z)(1+z \bar u) (1+z\bar v)}
{\bar u\bar v(1-z u)(1-z v)}
\\
\gamma^{(2)}_{Q\to [Q\bar{Q}(a8)]} 
&=& \alpha_s^2 
\left[\frac{N_c^2-1}{4N_c^3}\right]
\frac{4(1-z)(1+z \bar u) (1+z\bar v)}
{\bar u\bar v(1-z u)(1-z v)}
\\
\gamma^{(2)}_{Q\to [Q\bar{Q}(t1)]} 
&=& \alpha_s^2 
\left[\frac{C_F^2}{N_c}\right] 
\frac{8(1-z)z^2}{\bar{u}\bar{v}(1-z u)(1-z v)} 
\\
\gamma^{(2)}_{Q\to [Q\bar{Q}(t8)]} 
&=& \alpha_s^2 
\left[\frac{N_c^2-1}{4N_c^3}\right]
\frac{8(1-z)z^2}{\bar{u}\bar{v}(1-z u)(1-z v)} 
\een

\bigskip
\noindent{5) Heavy antiquark case:}
\ben
\gamma^{(2)}_{\bar{Q}\to [Q\bar{Q}(v1)]} 
&=& \alpha_s^2 
\left[\frac{C_F^2}{N_c}\right]
\frac{4(1-z)(1+z u) (1+z v)}
{u v (1-z \bar{u})(1-z \bar{v})}
\\
\gamma^{(2)}_{\bar{Q}\to [Q\bar{Q}(v8)]} 
&=& \alpha_s^2\left[
\frac{N_c^2-1}{N_c^3}\right] 
\frac{1-z}{z^2}\,
\frac{1}{uv}\,
\frac{4N_c u (1-z \bar u)+z(1+z u)}{1-z \bar u}
\nonumber \\
&\ & \hskip 1in
\times 
\frac{4N_c  v (1-z \bar v)+z(1+z v)}{1-z\bar v}
\\
\gamma^{(2)}_{\bar{Q}\to [Q\bar{Q}(a1)]} 
&=&  \gamma^{(2)}_{\bar{Q}\to [Q\bar{Q}(v1)]}
=  \alpha_s^2 
\left[\frac{C_F^2}{N_c}\right]
\frac{4(1-z)(1+z u) (1+z v)}
{u v (1-z \bar{u})(1-z \bar{v})}
\\
\gamma^{(2)}_{\bar{Q}\to [Q\bar{Q}(a8)]} 
&=& \alpha_s^2 
\left[\frac{N_c^2-1}{4N_c^3}\right]
\frac{4(1-z)(1+z u) (1+z v)}
{u v (1-z \bar{u})(1-z \bar{v})}
\\
\gamma^{(2)}_{\bar{Q}\to [Q\bar{Q}(t1)]} 
&=& \alpha_s^2 
\left[\frac{C_F^2}{N_c}\right] 
\frac{8(1-z)z^2}{{u}{v}(1-z \bar{u})(1-z \bar{v})} 
\\
\gamma^{(2)}_{\bar{Q}\to [Q\bar{Q}(t8)]} 
&=&
\alpha_s^2 \left[\frac{N_c^2-1}{4N_c^3}\right]
\frac{8(1-z)z^2}{{u}{v}(1-z \bar{u})(1-z \bar{v})} 
\een

%% file: appendixc.tex
\section{Heavy quark pair to heavy quark pair evolution kernels}
\label{sec:appendix-qqb}

In this appendix, we summarize all evolution kernels for a heavy quark pair to evolve into
a heavy quark pair at the order of $\alpha_s$, which are derived in both light-cone and
Feynman gauge.  We present these kernels in connection with the evolution equations in
the matrix form in Eq.~(\ref{eq:evo-matrix}). 

{\noindent 1)\ Diagonal kernels:}
\ben
{\mathcal S}
= P_{v1\to v1}&=&P_{a1\to a1}=C_F\, \delta(1-z) \Bigg\{3 \delta(u-u^\prime)\delta(v-v^\prime)
\nnu
&\ & \hskip 0.4in
+\delta(v-v') \left[\frac{\theta(\bar{u}'-\bar{u})}{(\bar{u}'-\bar{u})_+} \frac{\bar u}{\bar u'}\left(\bar{u}'+u\right)
+\frac{\theta(u'-u)}{(u'-u)_+} \frac{u}{u'}\left(u'+\bar{u}\right)\right]
\nnu
&\ & \hskip 0.4in
+\delta(u-u') \left[\frac{\theta(\bar{v}'-\bar{v})}{(\bar{v}'-\bar{v})_+} \frac{\bar v}{\bar v'}\left(\bar{v}'+v\right)
+\frac{\theta(v'-v)}{(v'-v)_+} \frac{v}{v'}\left(v'+\bar{v}\right)\right]\Bigg\},\\
{\mathcal K}_a
=P_{a8\to a8}&=&\frac{N_c}{2}\left[3-\ln(u\bar{u}v\bar{v})\right]\delta(1-z)\delta(u-u^\prime)\delta(v-v^\prime)
\nnu
&\ & \hskip 0.4in
- \frac{{\mathcal S}}{N_c^2-1}+ \frac{1}{2N_c}\frac{z}{2(1-z)_+}S_+\Delta_{-}^{[8]},
\\
{\mathcal K}_v
=P_{v8\to v8}&=&
{\mathcal K}_a - \left[\frac{1}{2}\right]
\delta(1-z)
\bigg\{
\delta(v-v') [4u(1-u)] + \delta(u-u') [4v(1-v)] 
\bigg\}
\een

\ben
{\mathcal S}'
= P_{t1\to t1}&=&C_F\, \delta(1-z) \Bigg\{3 \delta(u-u^\prime)\delta(v-v^\prime)
\nnu
&\ &  \hskip 0.8in
+\delta(v-v') \left[\frac{\theta(\bar{u}'-\bar{u})}{(\bar{u}'-\bar{u})_+} \frac{\bar u}{\bar u'}
+\frac{\theta(u'-u)}{(u'-u)_+} \frac{u}{u'}\right]
\nnu
&\ &  \hskip 0.8in
+\delta(u-u') \left[\frac{\theta(\bar{v}'-\bar{v})}{(\bar{v}'-\bar{v})_+} \frac{\bar v}{\bar v'}
+\frac{\theta(v'-v)}{(v'-v)_+} \frac{v}{v'}x\right]\Bigg\},\\
{\mathcal K}'
=P_{t8\to t8}&=&\frac{N_c}{2}\left[3-\ln(u\bar{u}v\bar{v})\right]\delta(1-z)\delta(u-u^\prime)\delta(v-v^\prime)
\nnu
&\ &  \hskip 0.8in
-\frac{{\mathcal S}'}{N_c^2-1}+
\frac{1}{2N_c}\frac{z}{2(1-z)_+}\left(S_+\Delta_{-}^{[8]}+S_-\Delta_{+}^{[8]}\right).
\een

\bigskip
{\noindent 2)\ Off diagonal kernels:}
\ben
{\mathcal R}
= P_{v8\to v1}=P_{a8\to a1}&=&
\left[\frac{1}{2N_c}\right] \frac{z}{2(1-z)}S_+\Delta_-^{[1]},\\
{\mathcal R}'
= P_{t8\to t1}&=&
\left[\frac{1}{2N_c}\right] 
\frac{z}{2(1-z)}\left(S_+\Delta_-^{[1]}+S_-\Delta_+^{[1]}\right),\\
{\mathcal T}_1
=P_{v8\to a8}=P_{a8\to v8}&=&
\left[\frac{1}{2N_c}\right] \frac{z}{2(1-z)}S_-\Delta_{-}^{[8]},\\
{\mathcal T}_2
= P_{v8\to a1}=P_{a8\to v1}&=&
\left[\frac{1}{2N_c}\right] 
\frac{z}{2(1-z)}S_-\Delta_-^{[1]},
\een
\ben
P_{X1\to Y8}&=& \left(N_c^2-1\right) P_{X8\to Y1},\\
P_{tI\to vJ}&=&P_{tI\to aJ}=P_{vJ\to tI}=P_{aJ\to tI}=0
\een
with $X, Y= v, a, t$ and $I,J=1,8$. We have introduced the following symmetric notations,
\ben
S_{\pm}&=&\left(\frac{u}{u'}\pm\frac{\bar{u}}{\bar{u}'}\right)
\left(\frac{v}{v'}\pm\frac{\bar{v}}{\bar{v}'}\right),\\
\Delta_\pm^{[1]}&=&\left[\delta(u-zu^\prime)\pm
\delta(\bar{u}-z\bar{u}^\prime)\right]\left[\delta(v-zv^\prime)\pm
\delta(\bar{v}-z\bar{v}^\prime)\right],\\
\Delta_\pm^{[8]}&=&\left\{(N_c^2-2)\left[\delta(u-zu^\prime)\delta(v-zv^\prime)+
\delta(\bar{u}-z\bar{u}^\prime)\delta(\bar{v}-z\bar{v}^\prime)\right]\right.\nonumber\\
& &\left.\mp\, 2\left[\delta(u-zu^\prime)\delta(\bar{v}-z\bar{v}^\prime)+
\delta(\bar{u}-z\bar{u}^\prime)\delta(v-zv^\prime)\right]\right\}.
\een
In the limit that $z\to 1$, we have the  power behaviors
\ben
S_{-}\,\Delta_-^{[1]} &\to& O\left((1-z)^4\right),\\
S_{-}\,\Delta_+^{[1]} &\to& O\left((1-z)^2\right),\\
S_{-}\,\Delta_\pm^{[8]} &\to& O\left((1-z)^2\right),\\
S_{+}\,\Delta_-^{[1]} &\to& O\left((1-z)^2\right),\\
S_{+}\,\Delta_+^{[1]} &\to& O\left(1\right),\\
S_{+}\,\Delta_\pm^{[8]} &\to& O\left(1\right),
\een
which is the reason that we do not need ``+" prescription for off diagonal kernels.  